\documentclass[fleqn,usenatbib,twocolumn]{mnras}

\usepackage[T1]{fontenc}
\usepackage{ae,aecompl}

\usepackage{graphicx}	
\usepackage{amssymb}	
\usepackage{longtable}
\usepackage{pdflscape}
\usepackage{lscape}
\usepackage{booktabs}
\usepackage{csquotes}
\usepackage{color, colortbl}
\usepackage{amstext}
\usepackage{wasysym}
\usepackage{multirow}
\usepackage{natbib}
\usepackage{ulem}

\definecolor{LightGray}{gray}{0.9}
\definecolor{LightGray1}{gray}{0.8}

\definecolor{DarkGreen}{RGB}{22, 141, 22}
\definecolor{pad}{RGB}{242,12,242}

\def\degr{\hbox{$^\circ$}}

\title[Short title, max. 45 characters]{MNRAS \LaTeXe\ template -- title goes here}

\title[Oort spike comets with large perihelion distances]{Oort spike comets with large perihelion distances}

\author[M. Kr\'{o}likowska \& P. A. Dybczy\'{n}ski]{Ma{\l}gorzata Kr\'{o}likowska ,$^{1}$\thanks{E-mail: mkr@cbk.waw.pl}
Piotr A. Dybczy\'{n}ski,$^{2}$\thanks{E-mail: dybol@amu.edu.pl}
\\
$^{1}$Space Research Centre of the Polish Academy of Sciences, Bartycka 18A, 00-716 Warsaw, Poland\\
$^{2}$Astronomical Observatory Institute, Faculty of Physics, A.~Mickiewicz University, S{\l}oneczna 36, 60-286 Pozna\'{n}, Poland
}

\date{Accepted XXX. Received YYY; in original form ZZZ}

\pubyear{2015}

\begin{document}
\label{firstpage}
\pagerange{\pageref{firstpage}--\pageref{lastpage}}
\maketitle

\begin{abstract}

The complete sample of large-perihelion nearly-parabolic comets discovered during the period 1901-2010 is studied starting from their orbit determination.  Next, an orbital evolution that includes three perihelion passages (previous-observed-next) is investigated where a full model of Galactic perturbations and perturbations from passing stars have been incorporated.

We show that the distribution of planetary perturbations suffered by actual large-perihelion comets during their passage through the Solar system has a deep, unexpected minimum around zero which indicates a lack of ,,almost unperturbed'' comets. By a series of simulations we show that this deep well is moderately resistant to some diffusion of orbital elements of analysed comets. It seems reasonable to state that the observed stream of these large-perihelion comets experienced a series of specific planetary configurations when passing through the planetary zone.

An analysis of the past dynamics of these comets clearly shows that dynamically new comets may appear only when their original semimajor axes are greater than 20\,000\,au. However, only for semimajor axes longer than  40\,000\,au dynamically old comets are completely not present. We demonstrated that the observed $1/a_{\rm ori}$-distribution exhibits a local minimum separating dynamically new from dynamically old comets.

Long-term dynamical studies reveal a large variety of orbital behaviour. Several interesting examples of action of passing stars are also described, in particular the impact of Gliese~710 which will pass close to the Sun in the future. However, none of the obtained stellar perturbations is sufficient to change the dynamical status of  analysed comets. 

\end{abstract}




\section{Introduction}\label{sec:intro}

The Oort Cloud hypothesis understood as a rich population of comets, containing billions of cometary-type objects and  forming a huge, spherically symmetrical cloud on the outskirts of the Solar system is still waiting for a convincing observational evidence. The latest and comprehensive overview of our knowledge of the Oort Cloud and ideas related to its origin and evolution is given by \citet{Dones:2015}. Investigations presented here contribute to the observational basis on which all Oort Cloud theories are built or should be in harmony. 

\noindent The detailed analysis of orbits of the actual long-period comets (hereafter LPCs) having original semimajor axes greater than a few thousand au is essential to search for their source region or regions. It is commonly believed that the distribution of the original $1/a$ has a single maximum, called the Oort spike, peaked somewhere between (30-60)$\cdot 10^{-6}$\,au$^{-1}$. This conclusion is based on the original $1/a$-distribution binned with a resolution of $10\cdot 10^{-6}$\,au$^{-1}$. Today, however, it is possible to construct $1/a$-distribution with the resolution of $5\cdot 10^{-6}$\,au$^{-1}$ if the $1/a$-uncertainties are taken into account. We present here such an analysis and obtain a detailed shape of Oort spike. Such a narrow bin histogram was recently introduced by \citet{krol-sit-et-al:2014} for a distribution of LPCs mostly composed of comets with small perihelion distances. Here, we show a distribution of a complete sample of LPCs with perihelia far from the Sun (hereafter in short: large-perihelion LPCs) and argue that this two-humped distribution can be interpreted as two populations of LPCs, partially overlapping each other: dynamically new and dynamically old comets.  

The orbit determination for nearly parabolic comets is a difficult task since we have observations limited to a unique, short part of a large eccentricity orbit with orbital periods of at least hundreds of thousands of years. Moreover, the non-gravitational effects are hard to determine in such cases.  Therefore we selected here only large-perihelion LPCs (having perihelia further than 3.1\,au from the Sun) since perihelion placed far from the Sun makes a motion of comet significantly less sensitive to non-gravitational forces (though some non-gravitational effects still can be detected as is described below). Additionally, the appearance of comets in large heliocentric distances makes nowadays their astrometric observations easier and more precise. Furthermore, many of these comets were observed over relatively 
long periods of several years. This is also the reason why the majority of large-perihelion comets discovered after 2010 is still observable.

\noindent All these arguments augmented with the homogeneous and very careful data treatment allow us to obtain a valuable sample of cometary orbits of high quality and precision. 


This paper deals with the similar issues as those discussed in \citet[hereafter Paper~I]{dyb-kroli:2011} and can be treated as a substantial extension and update of investigations presented there. 

First, we analysed here an updated sample of large-perihelion comets. Although we observe an ever-increasing rate of discoveries of LPCs with aphelia further than ten thousands of au from the Sun, less than three hundred LPCs having semimajor axes greater than 5\,000\,au were detected before 2010, while one hundred more such comets were discovered since 2010. However, in the entire sample of LPCs discovered so far only about 35 per cent are large-perihelion comets, and the majority of these discovered after 2010 is still observable. 

Second, the orbital evolution for one orbital period to the past and future is completely recalculated here for all considered comets since we decided to include perturbations from all known potential stellar perturbers into our dynamical model. Since Paper~I we also have changed slightly our definition of dynamically new and dynamically old comets, basing on their previous perihelion and started to use a modified orbit quality assessment method, introduced by \citet{kroli-dyb:2013}. 
On the other hand, some interesting issues were already discussed by us in Paper~I, so we devote a little space to these questions here. This includes, for example, an evolution of angular orbital elements in a Galactic frame and the concept of a Jupiter-Saturn barrier. Regarding this last issue, a few detailed studies of actual LPCs undertaken in the recent decade show that Jupiter-Saturn barrier is much less effective than some theoretical arguments suggest \citep[see for example:][]{kaib-quinn:2009,dyb-kroli:2015,fouchard:2014a}.

In the next section, we describe a complete sample of large perihelion LPCs having original semimajor axes larger than 5\,000\,au which were discovered in the years 1901--2010; this sample is enriched with a few from many comets detected after 2010. This is followed by two sections describing in detail the observational material used for new orbital solutions, our strategy applied for determination of non-gravitational (hereafter: NG) effects in the cometary motion, and resulting NG-solutions for 16~comets from the sample where NG-accelerations were detectable in their motion. Statistics of orbital elements of investigated LPCs is presented in Section~\ref{sec:sample_statistics}. Next two sections  analyse the $1/a$-distributions of original and future barycentric orbits (sect.~\ref{sec: original_orbits}) and the resulting distribution of observed planetary perturbations acting on these LPCs during their passage through the planetary zone (sect.~\ref{sec:perturb_observed}). In Section~\ref{sec:perturb_sim} we attempt to explain the atypical shape of this distribution of planetary perturbations derived here for actual comets, in which the clear deficit of comets suffering extremely small perturbations is visible.

Second part of this investigation (Sections~\ref{sec:past_next}--\ref{sec:prev-obs-next}) is devoted to studies of various aspects of a long-term dynamical evolution of LPCs during their three consecutive perihelion passages: from previous through observed to the next. In Section~\ref{sec:past_next} we describe results of numerical integrations for one orbital period to the past and to the future of full swarms of 5001~virtual comets (hereafter: VCs) representing each comet in our sample. Galactic perturbations as well as perturbations from all known potential stellar perturbers were taken into account. 
A study of past motion allows us to discriminate between dynamically new and old comets basing on their previous perihelion distances. Future motion analysis confirms a well-known feature that comets are either ejected on hyperbolic orbits or captured into more tightly bound orbits.  Section~\ref{sec:prev-obs-next} presents an overall picture of LPCs' dynamical evolution during three consecutive perihelia. We also present there a detailed analysis of several interesting cases of dynamical evolution for individual comets.
Section~\ref{sec:summary} summarizes our results and discusses the implications of our findings.

To the possible extent we have tabulated all our orbital results which are presented  as supplementary material due to their large sizes. Tables~\ref{tab:observational_material}--\ref{tab:orbit_future}  contain characteristics of an observational material and all new orbital solutions for 31~comets: observed (heliocentric) orbits at the epoch close to perihelion passage, and original and future (barycentric) orbits at  a   distance of 250\,au from the Sun. Moreover, results obtained for long-term dynamical evolution to the previous and next perihelia are given in Tables~\ref{tab:previous-returning}--\ref{tab:all-next} where we included orbital evolution results for all LPCs considered here since in addition to the Galactic perturbations we have taken into account perturbations from all currently known nearby stars for the first time in our research. 
\newline Further material for individual comets is also available at {\tt ssdp.cbk.waw.pl/LPCs} and {\tt  apollo.astro.amu.edu.pl/WCP}.

\section{The studied sample of LPCs}\label{sec:sample}

\begin{center}
\begin{table*}

\caption{\label{tab:comet_list} A list of a full sample of large-perihelion comets having
$1/a_{\rm ori} \leq 2\cdot 10^{-4}$\,au$^{-1}$; 69~comets and their orbital solutions were taken from \citep{krol-sit-et-al:2014} and \citep{krolikowska:2014}. Comets with new orbital solutions presented here are shown in bold; ballistic orbital solutions are marked by 'GR' and non-gravitational orbital solutions are marked by 'NG', where NG$_{\rm CO}$ indicates solutions dedicated to the CO-sublimation; an orbit quality assessment for each comet is also presented.}

\setlength{\tabcolsep}{1.5pt} 
\begin{tabular}{llllllllll}
\hline 
\\
\multicolumn{10}{c}{\bf All large perihelion comets discovered in the years 1901-1950;} \\
\multicolumn{10}{c}{taken from \citet{krol-sit-et-al:2014}, C/1906~E1 was excluded due to its splitting} \\
\\
Comet                 & solution,                & Comet                  & solution,             & Comet                     & solution,         & Comet                  & solution,              & Comet                  & solution, \\
 &  orb. qual.  &  & orb. qual. &  & orb. qual. &  & orb. qual. &  & orb. qual. \\
     C/1914 M1        &        GR, 1b        &       C/1925 F1        &        GR, 1b     &        C/1935 Q1          &      GR, 1a   &      C/1942 C2         &      GR, 1b        &      C/1948 T1         &      GR, 1b\\
\\ \hline 
\\
\multicolumn{10}{c}{\bf All large-perihelion comets discovered in the years 1951-2010} \\
\multicolumn{10}{c}{where 64 orbital solutions were taken from \citet{krolikowska:2014}, and still observable comet C/2010~U3  is omitted} \\
\\
Comet                 & solution,                & Comet                  & solution,             & Comet                     & solution,         & Comet                  & solution,              & Comet                  & solution, \\
 &  orb. qual.  &  & orb. qual. &  & orb. qual. &  & orb. qual. &  & orb. qual. \\
{\bf C/1954 O2}       &   {\bf GR, 1a}       &  {\bf C/1954 Y1}       &   {\bf GR, 1a}    &   {\bf C/1955 G1}         & {\bf GR, 1b}  & {\bf C/1960 M1}        & {\bf GR, 1b}       &      C/1972 L1         &      GR, 1a\\
     C/1973 W1        &        GR ,1b        &       C/1974 V1        &        GR, 1b     &        C/1976 D2          &      GR, 1a   &      C/1976 U1         &      GR, 1b        &      C/1978 A1         &      GR, 1a\\
{\bf C/1977 D1}       &   {\bf GR, 1b}       &       C/1978 G2        &        GR, 1b     &        C/1979 M3          &      GR, 1b   &      C/1980 E1         &      NG, 1a+       & {\bf C/1981 G1}        & {\bf GR, 1a}\\
     C/1983 O1        &        NG, 1a        &       C/1984 W2        &        NG, 1a     &        C/1987 F1          &      GR, 1a   &      C/1987 H1         &      GR, 1a+       &      C/1987 W3         &      GR, 1a\\
     C/1988 B1        &        GR, 1a        &  {\bf C/1991 C3}       &   {\bf GR, 1b}    &        C/1993 F1          &      GR, 1a   &      C/1993 K1         &      GR, 1a        &      C/1997 A1         &      GR, 1b\\
     C/1997 BA$_{6}$  &        NG, 1a+       &  {\bf C/1997 P2}       &   {\bf GR, 2b}    &   {\bf C/1998 M3}         & {\bf GR, 1b}  &      C/1999 F1         &      GR, 1a+       &      C/1999 F2         &      GR, 1a\\
     C/1999 H3        &        NG, 1a        &       C/1999 J2        &        GR, 1a+    &        C/1999 K5          &      GR, 1a+  &      C/1999 N4         &      GR, 1a        &      C/1999 S2         &      GR, 1a\\
     C/1999 U1        &        GR, 1a        &       C/1999 U4        &        GR, 1a+    &        C/2000 A1          &      GR, 1a   &      C/2000 CT$_{54}$  &      NG, 1a+       & {\bf C/2000 H1}        & {\bf GR, 2a}\\
     C/2000 K1        &        GR, 1a        &       C/2000 O1        &        GR, 1a     &   {\bf C/2000 SV$_{74}$}  & {\bf NG, 1a+} &      C/2000 Y1         &      GR, 1a        & {\bf C/2001 B2}        & {\bf GR, 1a}\\
     C/2001 C1        &        GR, 1a        &       C/2001 G1        &        GR, 1a     &        C/2001 K5          &      GR, 1a+  &      C/2002 A3         &      GR, 1a        &      C/2002 J4         &      GR, 1a+\\
     C/2002 J5        &        GR, 1a+       &       C/2002 L9        &        GR, 1a+    &        C/2002 R3          &      NG, 1a   & {\bf C/2003 A2}        & {\bf GR, 1a}       &      C/2003 G1         &      GR, 1a\\
{\bf C/2003 O1}       &   {\bf GR, 1a+}      &       C/2003 S3        &        GR, 1a     &   {\bf C/2003 WT$_{42}$}  & {\bf GR, 1a+} &      C/2004 P1         &      GR, 1a        &      C/2004 T3         &      GR, 1a\\
     C/2004 X3        &        GR, 1a        &       C/2005 B1        &        NG, 1a+    &        C/2005 EL$_{173}$  &      NG, 1a+  &      C/2005 G1         &      GR, 1a        &      C/2005 K1         &      NG, 1a\\
     C/2005 L3        &        GR, 1a+       &       C/2005 Q1        &        GR, 1a     &        C/2006 E1          &      GR, 1a   &      C/2006 K1         &      GR, 1a+       &      C/2006 S2         &      NG, 1b\\
{\bf C/2006 S3}       &   {\bf NG$_{\rm CO}$,1a+} &  {\bf C/2006 X1}       &   {\bf GR, 2b}    &        C/2006 YC            &      GR, 2a   &      C/2007 D1         &      GR, 1a+       &      C/2007 JA$_{21}$  &      GR, 1a\\
{\bf C/2007 U1}       &   {\bf NG, 1a }      &       C/2007 VO$_{53}$ &        GR, 1a+    &        C/2007 Y1          &      GR, 2a   & {\bf C/2008 FK$_{75}$} & {\bf NG$_{\rm CO}$,1a+}   &      C/2008 P1         &      GR, 1a+\\
{\bf C/2008 S3}       &   {\bf GR, 1a+}      &  {\bf C/2009 F4}       &   {\bf GR, 1a+}   &        C/2009 P2          &      GR, 1a   &      C/2009 U5         &      GR, 1a        & {\bf C/2009 UG$_{89}$} & {\bf GR, 1a+}\\
     C/2010 D3        &        GR, 1a        &  {\bf C/2010 L3}       &   {\bf GR, 1a}    &   {\bf C/2010 R1}         & {\bf GR, 1a+} & {\bf C/2010 S1}        & {\bf GR, 1a+}      & {\bf C/2011 L6}        & {\bf GR, 1b}\\
{\bf C/2012 A1}       &   {\bf GR, 1a+}      &  {\bf C/2012 B3}       &   {\bf GR, 1b}    &   {\bf C/2013 B2}         & {\bf NG, 1a}  & {\bf C/2013 E1}        & {\bf GR, 1a}       & {\bf C/2013 L2}        & {\bf GR, 2a}\\
 &  &  &  &  &  &  &  &  & \\
\hline 
\end{tabular}
\end{table*}
\end{center}

We constructed the sample of near-parabolic comets with $q > 3.1$\,au and $1/a_{\rm ori} < 0.000200$\,au$^{-1}$, that is, large-perihelion comets of original semi-major axes larger than 5\,000\,AU. The majority of comets satisfying both conditions (74 objects) were taken from \citet{krolikowska:2014}  and \citet{krol-sit-et-al:2014}.  However, for five of these 74 comets we decided to repeat an orbit determination because the previous orbits were based on a shorter arc of observations and, additionally, we found non-gravitational effects in the motion of two of them. To complete the sample of large-perihelion comets with original semi-major axes greater than 5\,000\,au, we analysed 26 more comets discovered since 1950 and previously not considered by us. As a result we present new orbital solutions for 31~LPCs.
\newline 

\noindent Therefore, our final sample consists of 100~large-perihelion comets having $1/a_{\rm ori}$ less than $2\cdot 10^{-4}$\,au$^{-1}$ and represents the complete sample of all such objects discovered in the years 1901--2010 (94~comets)\footnote{Since two Oort spike objects of large perihelion distances, C/1906~E1~Kopff and C/1956~F1~Wirtanen, are among the group of split comets (see for example \cite{boehnhardt:2004}, and {\tt www.icq.eps.harvard.edu/ICQsplit.html}) they are not included here. Furthermore, comet C/2010~U3~Boattini was not taken into a consideration because it is still observable.}; additionally six more comets detected after 2010 are also included. It is worth mentioning that in the years 1901-1960 only nine large-perihelion comets were discovered, while as many as 45~comets were detected in the period 2001-2010.

A full list of comets is given in Table~\ref{tab:comet_list}, where these with new solutions are highlighted in bold. For each comet a relevant type of dynamical model (GR -- gravitational, NG -- non-gravitational) and orbit quality class are also presented there. It can be seen that cometary orbits of analysed comets are generally of the first orbital quality class, only six orbits are of a second quality class (C/1997~P2 Spacewatch, C/2000~H1 LINEAR, C/2006~X1 LINEAR, C/2006~YC Catalina-Christensen, C/2007~Y1 LINEAR, and C/2013~L2 Catalina).

\section{Positional data and new orbital solutions}\label{sec:sample_new_solutions}

The general descriptions of observational material taken here for orbit determination 
for each of 31~comets (given in bold in Table~\ref{tab:comet_list}) are presented in Table~\ref{tab:observational_material}, whereas orbital solutions resulting from these observational data are shown in Table~\ref{tab:orbit_osculating}. 

In Table~\ref{tab:observational_material} we have nine comets discovered after a perihelion passage (see 
'post' in column [8], and Cols.~[3]--[4]). Two of them, C/1955~G1 Abell and C/2013~L2 Catalina were detected later than a year after their perihelion passages. It shows that LPCs with large perihelion distances are also now discovered when they are moving away from the Sun. In the whole sample of 100~comets discussed here as many as 24~comets were first spotted on the outgoing leg of their orbits.

The shortest arc, when orbital elements can be determined with reasonably small uncertainties is about 1.5 months, as in the case of C/1997~P1 (Col.~[6]). The orbit derived turned out to be of 2b-class for this comet; for more details about quality classes used here see the last paragraph of this section. On the other hand, in Table~\ref{tab:observational_material} are some examples of a very long period of observations, and as many as 14~of 31~near-parabolic comets listed there were observed longer than 3\,yrs. 

A long-observed group of objects contains C/2006~S3 LONEOS which was observed systematically  during eleven consecutive oppositions in the period of 9.7\,yr. This comet was discovered on 19~September 2006 at the heliocentric distance of 14.4\,au and was followed through perihelion (5.1\,au) up to 11.4\,au from the Sun. Moreover, ten prediscovery astrometric positions were found, which extended the arc of data back to 13~October 1999 (heliocentric distance of 26.1\,au) giving the period of observation as long as 16.6\,yr (Table~\ref{tab:observational_material}). 

Another example of a long-observed comet among those listed in Table~\ref{tab:observational_material} is 
C/2008~S3 Boattini which was followed during ten oppositions for 8.6\,yr from the heliocentric distance of 
12.4\,au through perihelion (8.0\,au) till 11.4\,au from the Sun. The third comet, followed longer than 
6\,years is C/2009~F4 McNaught observed in the range of heliocentric distances: 9.0\,au -- 5.5\,au 
(perihelion) -- 10.5\,au.

Other comets observed longer than 6\,years are C/1980~E1 Bowell (6.9\,yrs), 
C/1983~O1~\v{C}ernis (7.8\,yrs), C/1997~BA$_6$ Spacewatch (7.7\,yrs), C/1999~F1 Catalina (6.5\,yrs), 
C/2005~L3 McNaught (8.7\,yrs), and  C/2007~D1 LINEAR (6.2\,yrs); their orbits are given in \citet{krolikowska:2014}. C/2005~L3 was discovered on 3~June 2005 about 2.5\,yrs prior to perihelion and next was followed until 15~March 2013. Later, prediscovery detections by Siding Spring Survey were found (16~July and 16~August 2004), extending the period of data to 8.7\,yrs in a range of heliocentric distances: 10.3\,au -- 5.593\,au (perihelion) -- 13.4\,au. C/2005~L3 and C/2006~S3 were among the brightest comets beyond 5\,au from the Sun, and  sizes of their nuclei can be expected to lie somewhere between size of comet 1P/Halley and comet C/1995~O1 Hale Bopp \citep{sarneczky-et-al:2016}.

Generally,  modern techniques allow discovery of near-parabolic comets far outside Jupiter's orbit. 
As was already mentioned we have 45~comets discovered in the first decade of the 21$^{\rm st}$ century in the sample of 100~comets considered here. Among them as many as 36 (80 per cent)  were  discovered further than 5\,au from the Sun, 28 (62 per cent)-- further than 6\,au from the Sun, and 21 (47 per cent) -- further than 7\,au from the Sun. 

During our numerical calculations leading to orbit determination, the equations of motion have been integrated using the recurrent power series method \citep{sitarski:1989}, taking into account perturbations by all planets and including relativistic effects. We applied the selection and weighting procedure simultaneously with the orbit determination from the data. It was earlier found that the weighting procedure is crucial for the orbit fitting \citep[see for more details][]{krolikowska-sit-soltan:2009,kroli-dyb:2010}. We decided, however, not to weight observations for the least numerous data sets, that is for comets C/1955~G1 Abell, C/1977~D1 Lovas, C/1981~G1 Elias and C/1991~C3 McNaught-Russell.

The number of positional observations, root-mean-square residuals (rms) and number of residuals resulting from the final selection and weighting procedure are given in Cols.\,$[5]$ and $[13]$ of Table~\ref{tab:observational_material}.  
\noindent We made an attempt to determine the NG~effects in the motion of each comet analysed here. However, only in a small number of comets these effects turned out to be detectable from positional data  for a rather obvious reason: all comets discussed here   have perihelion distances further than 3.1\,au from the Sun, where the outgassing is relatively less effective than in comets of small perihelion distances.
The type of model of motion used for final orbit determination is shown in Table~\ref{tab:comet_list} and Col.~[9] of Table~\ref{tab:observational_material}, where 'GR' -- means ballistic solution, 'NG' -- model where NG-effects  were determined. In NG-cases, each individual set of astrometric data has been selected and weighted simultaneously with an iterative process of NG-orbit determination. 
\newline Details about the NG-model of motion and resulted solutions are described and 
discussed in the next section. Osculating elements of all new orbital solutions (in a heliocentric reference 
frame) are listed in Table~\ref{tab:orbit_osculating}.

Column~[12] of Table~\ref{tab:orbit_osculating} gives the new quality assessment according to a recipe given by  
\citet{kroli-dyb:2013}. This new method of orbit quality estimation is based on the original method 
introduced by \citet[hereafter MSE]{mar-sek-eve:1978}, but is slightly more restrictive and seems to give a better diversification than the MSE~method, especially concerning recently discovered objects that are observed using modern techniques. In the following discussion, we refer to this new orbit quality estimation, however in Cols~[10]--[12], values of $Q^*$ are given, that can be directly used to calculate the MSE~orbital quality assessment according to the recipe described in \citet{krolikowska:2014}. The same quality assessment tags can be found in Table~\ref{tab:comet_list}.


\begin{center}
\begin{table}
\caption{\label{tab:gr-like_functions}Parameters used in Eq.~\ref{eq:g_r}}

\setlength{\tabcolsep}{5.0pt} 
\begin{tabular}{lllll}
\hline 
\\
\multicolumn{5}{c}{standard g(r)-function (water sublimation) }\\
\\
$\alpha$    & $r_0$ & $m$      & $n$   & $k$           \\
0.1113      & 2.808 & $-$2.15  & 5.093 & $-$4.6142     \\
\\
\multicolumn{5}{c}{g(r)-like function (CO sublimation)} \\
\\
$\alpha$    & $r_0$ & $m$      & $n$   & $k$           \\
0.01003     & 10.0  & $-$2.0   & 3.0   & $-$2.6        \\
\hline
\end{tabular}
\end{table}
\end{center}

\section{Non-gravitational orbits}\label{sec:NG_model}

Sixteen large-perihelion comets investigated here exhibit some  measurable traces of NG-acceleration in their orbital motion. For all these comets, we noticed slight decrease of rms and reduction of some trends (if any were visible in ballistic solution) in time variations of residuals, i.e. ({\it Observed minus Calculated} )-time variations in right ascension and declination (hereafter: O-C time variations) when NG~model of motion were used to orbit determination.

To determine a NG-orbit we applied  a standard formalism proposed by \citet{marsden-sek-ye:1973} where three orbital components of the NG~acceleration acting on a comet are  proportional to the $g(r)$-function which is symmetric relative to perihelion,

\begin{eqnarray}
F_{i}=A_{\rm i} \> g(r),& A_{\rm i}={\rm ~const~~for}\quad{\rm i}=1,2,3,\nonumber\\
 & \quad g(r)=\alpha(r/r_{0})^{m}[1+(r/r_{0})^{n}]^{k},\label{eq:g_r}
\end{eqnarray}

\noindent where $F_{1},\, F_{2},\, F_{3}$ represent radial, transverse, and normal components of the 
NG~acceleration, respectively, and the radial acceleration is defined as positive outward along the Sun-comet line. For a water sublimation, the exponential coefficients $m,\, n,\, k$, the distance scale, $r_{0}$, and the normalization constant, $\alpha$ (fulfilling the condition: $g(1\,{\rm au})=1$) are given in Table~\ref{tab:gr-like_functions}. 

\begin{figure}

\includegraphics[width=8.8cm]{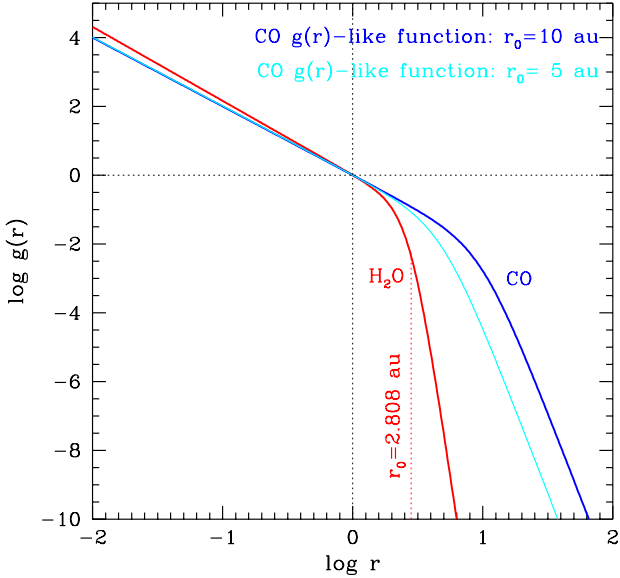} 
\protect\caption{\label{fig:gr_functions}Standard g(r)-function (red curve) in comparison to g(r)-like function adapted here for CO-sublimation (blue curve), also g(r)-like function used by Farnocchia for C/2006~S3 is shown (cyan); his $g(r)$-like function is published only on JPL/SSD web page (see text).} 
\end{figure}

\begin{figure}

\includegraphics[width=8.8cm]{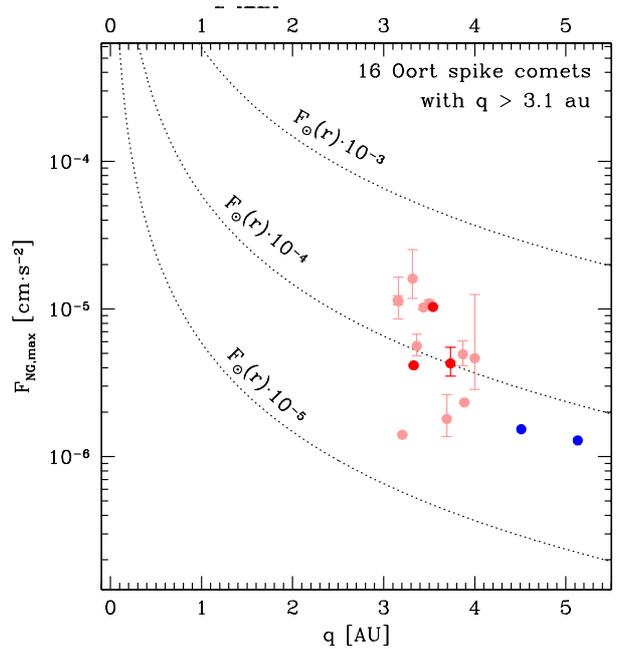} 
\protect\caption{\label{fig:16comets_gr_forces} The dependence of the maximum of the NG-force on a perihelion distance in large-perihelion comets with NG~solutions. Solutions obtained in Paper~I are shown by light-red dots, whereas solutions presented here are shown by red dots (standard g(r)-function) and blue dots (g(r)-like function for CO-sublimation). Three dotted curves are lines of constant ratio ($10^{-3},10^{-4}$, and $10^{-5}$, respectively) between NG-force and a solar attraction.}
\end{figure}

We used here two different sets of $m,\>n,\>k,\>r_{0}$ parameters depending on the perihelion distance of a comet. Namely, for $q \leq 4$\,au we applied the standard g(r)-function dedicated to water sublimation, whereas for $q \geq 4$\,au we rely on g(r)-like function more adequate for a CO-sublimation. This second set of parameters $\alpha, n, m, k$ and $r_0$ is also presented in Table~\ref{tab:gr-like_functions}. We adopted here exactly the same exponential coefficients $m,\, n,\, k$ as Farnocchia used for C/2006~S3 LONEOS, however we took $r_0=10$\,au instead of $r_0=5$\,au (see: {\tt http://ssd.jpl.nasa.gov/}). By testing values of $r_0$ in the range of 5--15\,au, we noticed slightly better NG~solution when value of $r_0\ge 10$\,au was assumed for C/2006~S3 LONEOS. Therefore, we decided to adopt $r_0 = 10$\,au for all comets with perihelion distances larger than 4\,au. The differences in shapes of g(r)-like functions used here are shown in Fig.~\ref{fig:gr_functions} by red (dedicated to water sublimation) and blue (CO-sublimation) curves.

Eleven satisfactory NG-models of the investigated near-parabolic comets were taken from Paper~I (NG~parameters were given in \citet{krolikowska:2014}). These solutions are based on the standard $g(r)$-function described above, and are listed in the first part of Table~\ref{tab:NG-parameters} and shown as light-red dots in Fig.~\ref{fig:16comets_gr_forces}. 

Here, we presented new NG-solutions for five comets. 
Three of them have perihelion distances below 3.8\,au and their NG-solutions are derived using standard g(r)-function.  These solutions are presented in the second part of Table~\ref{tab:NG-parameters} and displayed as red dots in Fig.~\ref{fig:16comets_gr_forces}. Next two comets, C/2006~S3 (mentioned above) and C/2008~FK$_{75}$ Lemmon-Siding, have perihelia farther than 4.5\,au from the Sun and their NG-solutions are based on g(r)-like function dedicated to CO-sublimation (blue curve in Fig.~\ref{fig:gr_functions}; third part of Table~\ref{tab:NG-parameters}) and are represented by blue dots in Fig.~\ref{fig:16comets_gr_forces}.  We have checked that for both these comets the g(r)-like function defined by CO-set of parameters shows better fitting to the positional data than a standard g(r)-function.
\newline We have also tested whether NG-solutions based on CO-driven formula give a better fitting to the data for two other comets with  $q \leq 4$\,au and previously found NG-solutions (Paper~I, \citet{krolikowska:2014}). 
The results were so similar that it cannot be resolved which formula is more appropriate. Thus, we decided to stay with  standard NG-solutions for all comets with detectable NG-effects and perihelion distances below 4\,au. 

The maximum value of the NG-acceleration which affected the cometary motion, that is the NG-force operating in perihelion,  $F_{\rm {NG,max}}$, could be calculated by using the relation:

\begin{displaymath}
F_{\rm {NG,max}}= g(q)\cdot \sqrt{A_1^2+A_2^2+A_3^2}.
\end{displaymath}

\noindent Fig.~\ref{fig:16comets_gr_forces} shows the $F_{\rm {NG,max}}/F_{\odot}$ as  a  function of the perihelion distance for all sixteen comets with detectable NG-effects,  where the solar gravitational acceleration   satisfies the equation:  $F_{\odot}(r)=0.59\cdot r^{-2}$\,[cm\,s$^{-2}$], and $r$ is in astronomical units. The three dotted black curves in Fig.~\ref{fig:16comets_gr_forces}  display $10^{-3}, 10^{-4}, 10^{-5}$ of $F_{\odot }(r)$, respectively. 

\noindent All values of $F_{\rm {NG,max}}$ are in the range (2--30)$\cdot 10^{-5}\cdot F_{\odot}(r)$ (last column of Table~\ref{tab:NG-parameters}) with the weighted mean value of $5.5\cdot 10^{-5}\cdot F_{\odot}(r)$. 
This is in  quite good agreement with the previous estimates based on LPCs with smaller perihelion distances. 
\citet{marsden-sek-ye:1973} concluded that the actual magnitude of the NG-forces is typically about $10^{-5}$~times the solar attraction at 1\,au for 23 short-period comets (SPCs), and (1--2)$\cdot 10^{-4}$ of the solar attraction for small sample of 7~LPCs. Generally a similar difference in NG-forces between SPCs and LPCs was obtained by \citet{krolikowska:2004} for these two cometary populations with perihelion distances well below 3\,au. Different NG~models of motion were considered there and the weighted mean value of NG-forces at perihelion of about (6--8)$\cdot 10^{-5}\cdot F_{\odot}$ was derived for 17~LPCs (depending  on   the NG-model) and about 1.1$\cdot 10^{-5}\cdot F_{\odot}$ for 22~SPCs for the NG-model of rotating spherical nucleus. Thus, a value of $5.5\cdot 10^{-5}\cdot F_{\odot}(r)$ derived here for 16~near-parabolic comets with large perihelion distances provide an interesting extension to this general picture. We can conclude that NG-forces operating at perihelia of near-parabolic comets are typically in the range of (1--20)$\cdot 10^{-5}\cdot q^{-2}$\,[cm$\cdot$ s$^{-2}$], where $q$ is in astronomical units.  

We thus confirmed the previous general conclusion resulting from \citet{marsden-sek-ye:1973} and \citet{krolikowska:2004} calculations, that we have typically greater NG~forces for LPCs compared to SPCs. An interpretation of this result is not obvious. There are many profound studies on this subject, see for example \citet{sosa_fernandez:2009,sosa_fernandez:2011}.

\par
\begin{center}
\begin{table*}
\caption{\label{tab:NG-parameters} NG-parameters derived in NG-orbital solutions given in 
Table~\ref{tab:orbit_osculating}.  Last column shows the $F_{\rm NG}/F_{\odot}$, that is NG-acceleration 
operating in perihelion in units of solar attraction acting on this distance}
\setlength{\tabcolsep}{6.0pt}{
\begin{tabular}{@{}llc@{$\pm$}cc@{$\pm$}cccc@{}}
\hline 
Comet    & $q$ &\multicolumn{5}{c}{NG parameters defined by Eq.~\ref{eq:g_r} in units of 10$^{-8}\,$AU\,day$^{-2}$ } &   g(r) &  $F_{\rm NG,max}/F_{\odot}$  \\
         & [au] & \multicolumn{2}{c}{A$_1$}  & \multicolumn{2}{c}{A$_2$}  & \multicolumn{1}{c}{A$_3$} & in perihelion  & in perihelion \\
 $[1]$   & $[2]$ & \multicolumn{2}{c}{$[3]$}  & \multicolumn{2}{c}{$[4]$}  & \multicolumn{1}{c}{$[5]$} & $[6]$  & $[7]$     \\
\hline
\\
\multicolumn{9}{c}{ I.~~ NG solutions taken from \citep{krolikowska:2014} for the standard  g(r)-function} \\
\\
C/1980 E1       & 3.364 &  1095    &   181   &  535.89  &  93.1   & --       & 2.302$\cdot 10^{-4}$ & 1.08$\cdot 10^{-4}$    \\ 
C/1983 O1       & 3.318 &  2683    &   942   &  158.10  & 677     & --       & 2.982$\cdot 10^{-4}$ & 3.00$\cdot 10^{-4}$    \\
C/1984 W2       & 4.000 & 36844    & 23157   & \multicolumn{2}{c}{--} & --   & 6.289$\cdot 10^{-6}$ & 1.26$\cdot 10^{-4}$   \\  
C/1997 BA$_6$   & 3.436 &  3341    &   118   &   24.3   &  54.1   &~$\,-$29.8 $\pm$ 11.7         & 1.529$\cdot 10^{-4}$ & 2.05$\cdot 10^{-4}$    \\ 
C/1999 H3       & 3.501 &  4112    &   193   & 3007     & 228     &$-$509.0 $\pm$ 72.0        & 1.061$\cdot 10^{-4}$ & 2.26$\cdot 10^{-4}$    \\
C/2000 CT$_{54}$& 3.156 &   778.0  &    53.6 &   51.5   &  25.9   & --       & 7.325$\cdot 10^{-4}$ & 1.93$\cdot 10^{-4}$    \\
C/2002 R3       & 3.869 & 17850    &  2640   & 5810     &3510     & --       & 1.309$\cdot 10^{-5}$ & 1.25$\cdot 10^{-4}$    \\
C/2005 B1       & 3.205 &    74.7  &    12.6 &$-$77.7   &   9.17  &$-$63.94 $\pm$ 4.66          & 5.596$\cdot 10^{-4}$ & 2.45$\cdot 10^{-5}$    \\
C/2005 EL$_{173}$& 3.886 & 6602    &   773   &$-$7175   &  496    & --       & 1.191$\cdot 10^{-5}$ & 5.96$\cdot 10^{-5}$    \\
C/2005 K1       & 3.693 &  2515    &   741   &   184    &  762    & --       & 3.559$\cdot 10^{-5}$ & 4.16$\cdot 10^{-5}$    \\ 
C/2006 S2       & 3.161 &   772    &   299   &$-$167    &  199    & --       & 7.111$\cdot 10^{-4}$ & 1.91$\cdot 10^{-4}$    \\
\\
\multicolumn{9}{c}{ II.~~ NG solutions derived in this investigation for the standard g(r)-function} \\
\\
C/2000 SV$_{74}$ & 3.542 & 6064     & 76      &   551     &  60     & $-$596 $\pm$ 22      &  8.413$\cdot 10^{-5}$ &  2.19$\cdot 10^{-4}$  \\  
C/2007 U1        & 3.329 &  685     & 45      &  $-$222   &  89     & $-$172 $\pm$ 18      &  2.799$\cdot 10^{-4}$ &  7.80$\cdot 10^{-5}$ \\
C/2013 B2        & 3.734 &  403     & 1978    &  7577     & 1582    & --                   &2.816$\cdot 10^{-5}$ &  1.01$\cdot 10^{-4}$   \\ \\
\multicolumn{9}{c}{ III.~~ NG solutions derived in this investigation for g(r)-like function dedicated to the CO-sublimation} \\
\\
C/2006 S3        & 5.131 & 1.589   &  0.036   & 1.714     &  0.041  & 0.1796 $\pm$ 0.0066   & 2.739$\cdot 10^{-2}$  &  5.74$\cdot 10^{-5}$ \\ 
C/2008 FK$_{75}$ & 4.511 & 1.861   &  0.048   & $-$0.5222 &  0.060  & 0.2751 $\pm$ 0.0111   & 3.922$\cdot 10^{-2}$  &  5.29$\cdot 10^{-5}$ \\
\end{tabular}}
\end{table*}
\end{center}

\section{General statistics of orbital elements}\label{sec:sample_statistics}

\begin{figure}
 
\includegraphics[width=8.6cm]{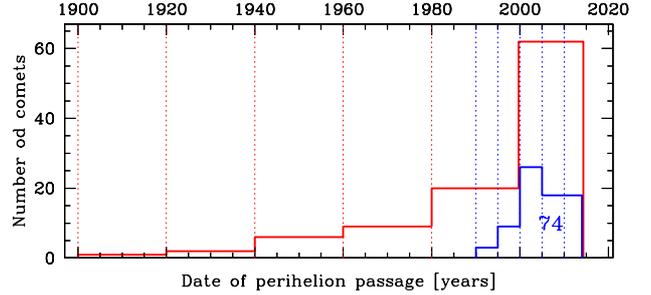} 
\protect\caption{\label{fig:hist_perihelia}  Time distribution of perihelia passages of analysed comets. The red histogram shows perihelia distribution in  20-year intervals (the last bin starting from 2000 is obviously incomplete), whereas the blue histogram shows the same distribution since 1990 in the 5-year periods.}
\end{figure}

\begin{figure}

\includegraphics[width=8.6cm]{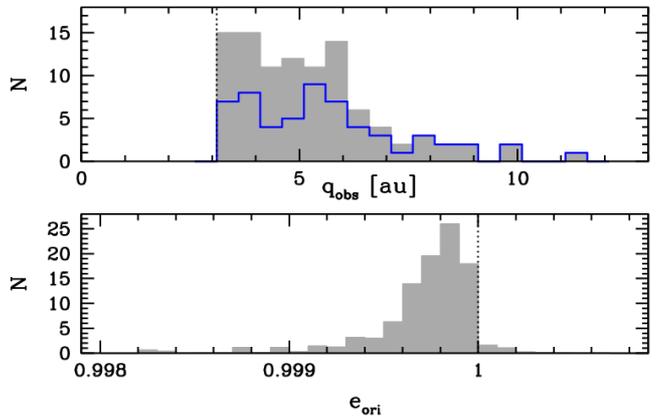} 
\protect\caption{\label{fig:hist_q_e} Distribution of the observed perihelion distances (upper panel, grey histogram) and original eccentricities (lower panel) of the investigated sample of 100 large-perihelion LPCs. The blue histogram shows a distribution of perihelion distances for 58~comets discovered since 2000.}
\end{figure}

\begin{figure}

\includegraphics[width=8.6cm]{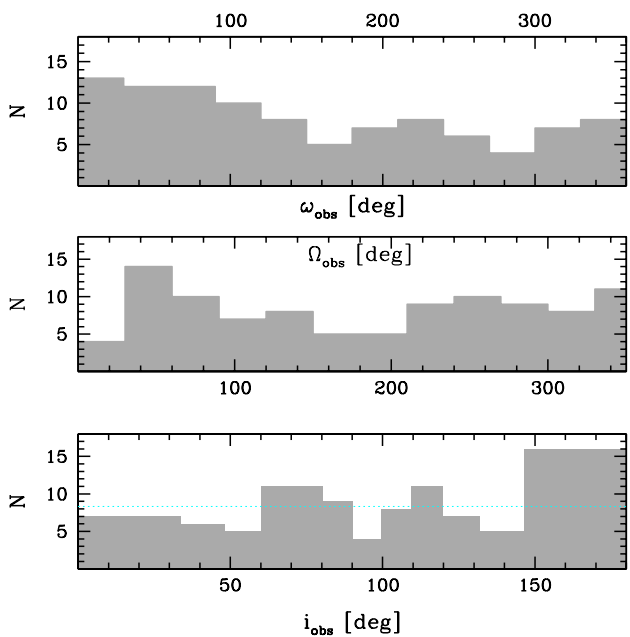} 
\protect\caption{\label{fig:hist_angles} Observed distributions of angular orbital elements (referred to  the ecliptic J2000) of the investigated sample of 100 large-perihelion LPCs. Bins in two upper panels have a width of 30\degr , whereas in the lowest panel the widths of the $i$-bins result from dividing the sky into equal-area strips. Cyan dotted horizontal line in the lowest panel indicates the mean number of comets for the assumption of uniform distribution of $i$.}
\end{figure}

In this section distributions of the observed orbital elements for an investigated sample of large-perihelion LPCs with original semi-major axes greater than 5\,000\,au are discussed. 
Therefore, starting with this section, we will distinguish between heliocentric orbital elements determined at the osculation epoch close to observed perihelion passage (and marked with subscript $_{\rm obs}$), and barycentric orbital elements, dynamically evolved backward and forward to a distance of 250\,au from the Sun (called original and future, subscripts $_{\rm ori}$ and $_{\rm fut}$), and barycentric orbital elements evolved to the previous or next perihelia (called previous and next, subscripts $_{\rm prev}$ and $_{\rm next}$).

Figure~\ref{fig:hist_perihelia}  presents the distribution of perihelion passages of the analysed sample. The red histogram displays the number of passages through the perihelion in 20-year periods, while the blue histogram shows exactly the same distribution for comets discovered after 1990 and in five-year time-intervals. The red distribution reveals a conspicuous increase in number of cometary discoveries  during the last hundred years, and it turns out that the majority of analysed comets (82 objects) were discovered after 1980. 

Upper panel of Fig.~\ref{fig:hist_q_e} shows the flat  distribution (grey histogram) of perihelion distances, $q_{\rm obs}$, in the range 3.1--6.1\,au, giving in average 13~comets in each of 0.5\,au-wide bins. The number of observed comets with more distant perihelia ($q_{\rm obs}>6.1$\,au) decreases drastically, and there are only 22~comets in the next six bins ($6.1 < q_{\rm obs} < 9.1$\,au). 
Taking into account only comets discovered since 2000 (blue histogram) we can see that the differences in numbers between bins on the left from $q_{\rm obs}=6.1$\,au and on the right from this limit are less dramatic. However, it is difficult to speculate to what extent this drop of observed comets  may be dominated by observational bias. In addition, a small local decrease in the number of comets of 4.1\,au\,$<q_{\rm obs}\,<\,5.1$\,au is visible in blue histogram, however a much richer statistics is needed to decide whether it is real and can be attributed to Jupiter action.

The distribution of original eccentricities (at 250\,au before penetrating into the planetary zone) is displayed in the lower panel of Fig~\ref{fig:hist_q_e}. The uncertainties of eccentricities were included in this particular plot since they can give spread over a few bins for individual comet. Thus, we took here 5001\,clones for each comet, constructed according to \citet{sitarski:1998} method, instead of taking only the nominal orbital solution (see Section~\ref{sec: original_orbits}). When we cut-off the outermost 10\,per cent of VCs on both sides of the wings, we get the original eccentricity range of 0.999408--0.999941. 

The distributions of angular orbital elements are shown in Fig~\ref{fig:hist_angles}. We can notice  more or less uniform distributions in the argument of perihelion, $\omega _{\rm obs}$ (upper panel) and in the ascending node, $\Omega _{\rm obs}$ (middle panel). However, some deviations from uniform distribution can be observed there, the more prominent is for $0 < \omega < 90$\degr where we have as many as 37 per cent of analysed comets. Using a $\chi ^2$ test and assuming the significance level of 0.05 we obtained that $\omega _{\rm obs}$-distribution statistically differs from the  homogeneous distribution.

The lowest panel of Fig~\ref{fig:hist_angles} shows distribution of orbital inclinations. We observe here evident overpopulation for the inclination greater than 150\degr. However, the almost perfect balance between the prograde and retrograde orbits is realized: 51~comets are moving on retrograde orbits and as many as 49~comets have prograde orbits. Three comets on prograde orbits with the smallest inclinations to the ecliptic plane in our sample ($i<15$\degr) are discussed in Section~\ref{sec:perturb_observed}.

\section{ Original and future orbits}\label{sec: original_orbits}

\begin{figure*}
\includegraphics[width=8.6cm]{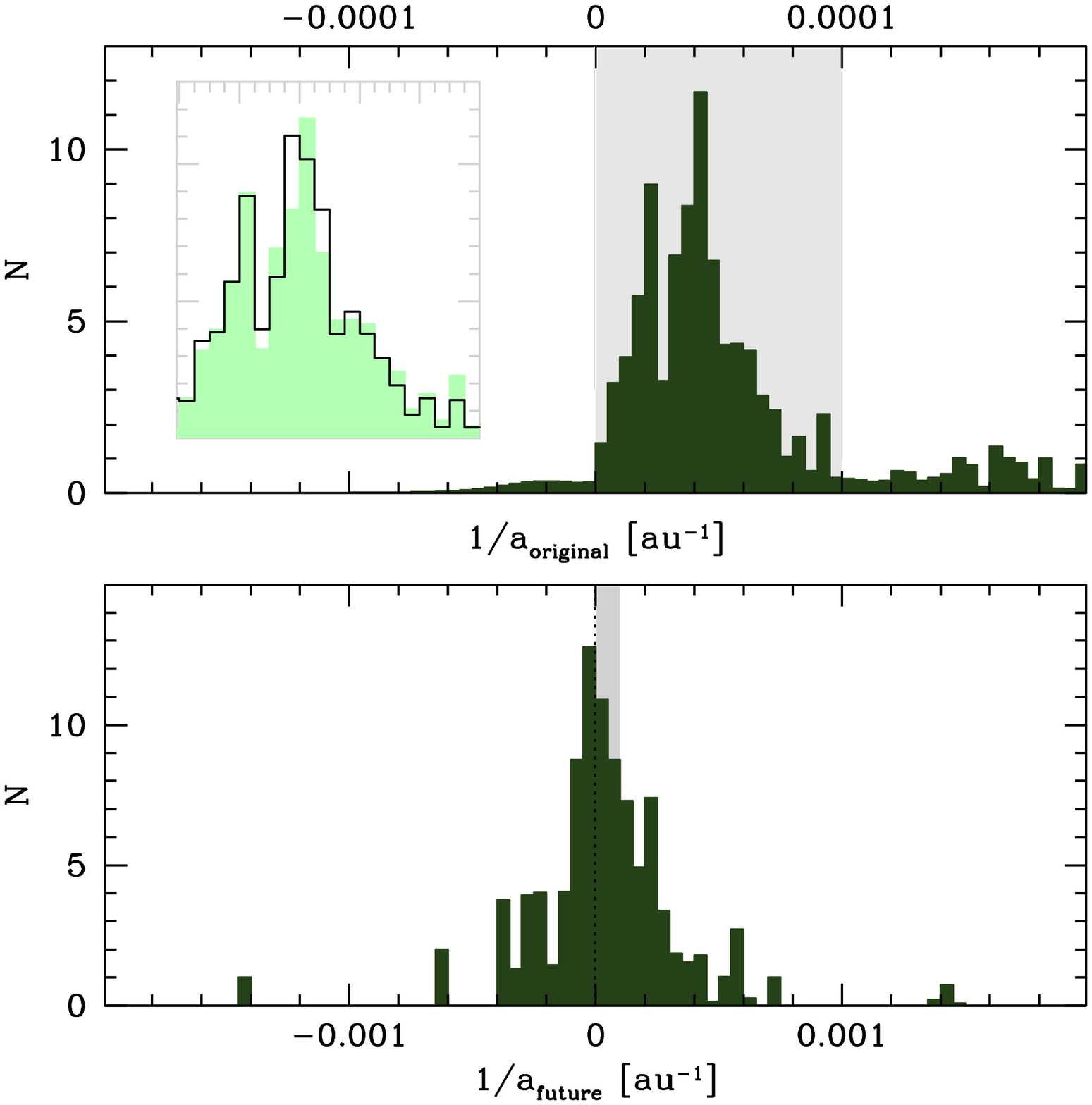} 
\includegraphics[width=8.6cm]{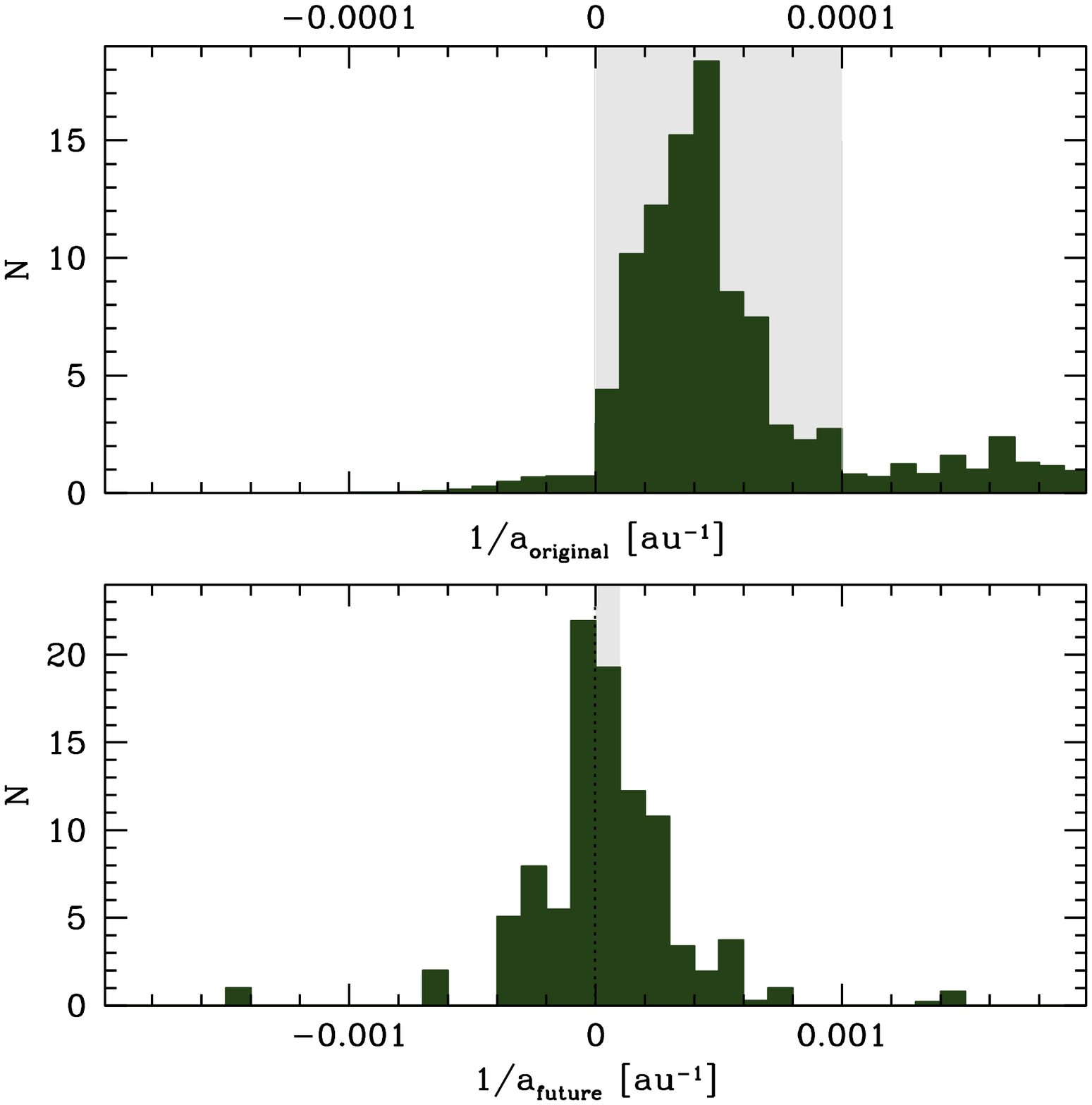} 

\protect\caption{\label{fig:original_future} Original (upper panel) and future (lower panel) distributions of $1/a$ for the investigated sample of 100~near-parabolic comets, where for 16 of them the NG~solutions were obtained. The left panels show distributions given for two-fold narrower bins than those given in the right panels. Oort spike range of 0--100$\times 10^{-6}$\,au$^{-1}$ is highlighted in grey to show the different horizontal scale in upper and lower panels. The analysed $1/a_{\rm ori}$-distribution represents a complete sample of large-q near-parabolic comets in the wider range of 0--200$\times 10^{-6}$\,au$^{-1}$.
The dark-green histogram from the main part of upper-left panel is copied using light-green colour to the inset, while  black curve displayed in this inset represents distribution based on 100 ballistic solutions.}
\end{figure*}

To be able to reliably follow the uncertainties of orbital elements during the dynamical evolution of cometary orbit, we constructed a swarm of 5001\,VCs, including the nominal orbit for each individual comet. These swarms were constructed according to a Monte Carlo method proposed by \citet{sitarski:1998}, for more details see also \citet{krolikowska-sit-soltan:2009}. This method allowed us to determine the uncertainties at any epoch covered by the numerical integration.
The dynamical calculations of each swarm of VCs were performed backwards and forwards in time until 
each VC reached 250\,au from the Sun, that is, a distance where planetary perturbations are already negligible. These swarms of orbits are called original and future, respectively. Further dynamical evolutions to the previous and next perihelion passages are described in Section~\ref{sec:past_next}. 

Original and future barycentric orbital elements are given in Tables~\ref{tab:orbit_original} and~\ref{tab:orbit_future} for all comets whose orbits were determined in this investigation (31 comets given in bold in Table~\ref{tab:comet_list}), while respective orbital elements for the remaining 69~comets can be found in \citet{krolikowska:2014} and \citet{krol-sit-et-al:2014}.

Figure~\ref{fig:original_future} shows distributions of original and future $1/a$ for the whole sample considered here. Solid dark-green histograms show  distributions of 100~orbits, where 16 of them are NG-solutions. Right panel gives the distributions of  $1/a_{\rm ori}$ (top plot) and  $1/a_{\rm fut}$  (bottom plot) in a standard way, that is in intervals of a width of $10\times 10^{-6}$\,au$^{-1}$ and $100\times 10^{-6}$\,au$^{-1}$, respectively. Both  distributions are clearly asymmetrical with respect to their maxima. However, when we apply twofold narrower bins in the $1/a_{\rm ori}$-distribution, then some local minimum appears which covers three consecutive bins between two local maxima located in $1/a_{\rm ori}$-intervals of (20--25)$\cdot 10^{-6}$\,au$^{-1}$ and (40--45)$\cdot 10^{-6}$\,au$^{-1}$, respectively. Such a two-humped shape can be easily explained by two populations of comets. Dynamically new comets can form the first local maximum of the global $1/a_{\rm ori}$-distribution, while dynamically old comets can be responsible for the second local maximum as it was clearly shown in \cite{dyb-kroli:2015} (Fig.10 therein); this is also discussed in Section~\ref{sec:past_next}. 
\newline To compare the extent to which a small part of NG-solutions (16 per cent in the entire sample) changes the overall picture we also show distributions for 100~purely ballistic orbits using  a black curve shown in the inset in the left-upper panel of Fig.~\ref{fig:original_future} where only the Oort spike part of the distribution is shown (range of a horizontal axis in the inset is: $0 <(1/a)_{\rm ori}<0.000100$\,au$^{-1}$). For a comparison we copied to the inset this dark-green distribution from the main part of panel using light-green colour for better visualisation.
One can see that the pure ballistic distribution also reveals a two-humped signature in the Oort spike region. Thus, this minimum is not a product of two types of solutions in which one (NG-type of solutions) causes systematic shifts of $1/a_{\rm ori, NG}$ to the right in comparison to a ballistic solution. For the analysed sample of large-perihelion LPCs, differences in NG- and GR-solutions, $1/a_{\rm ori,NG}-1/a_{\rm ori,GR}$, are generally small, that is, these differences are less than $10\cdot 10^{-6}$\,au$^{-1}$ for nine comets with detectable NG~solutions. On the other hand, for C/1999~H3 LINEAR and C/2000~SV$_{74}$ LINEAR these differences are greater than $40\cdot 10^{-6}$\,au$^{-1}$.

\section{ Observed planetary perturbations}\label{sec:perturb_observed}

\begin{figure*}
\includegraphics[width=5.8cm]{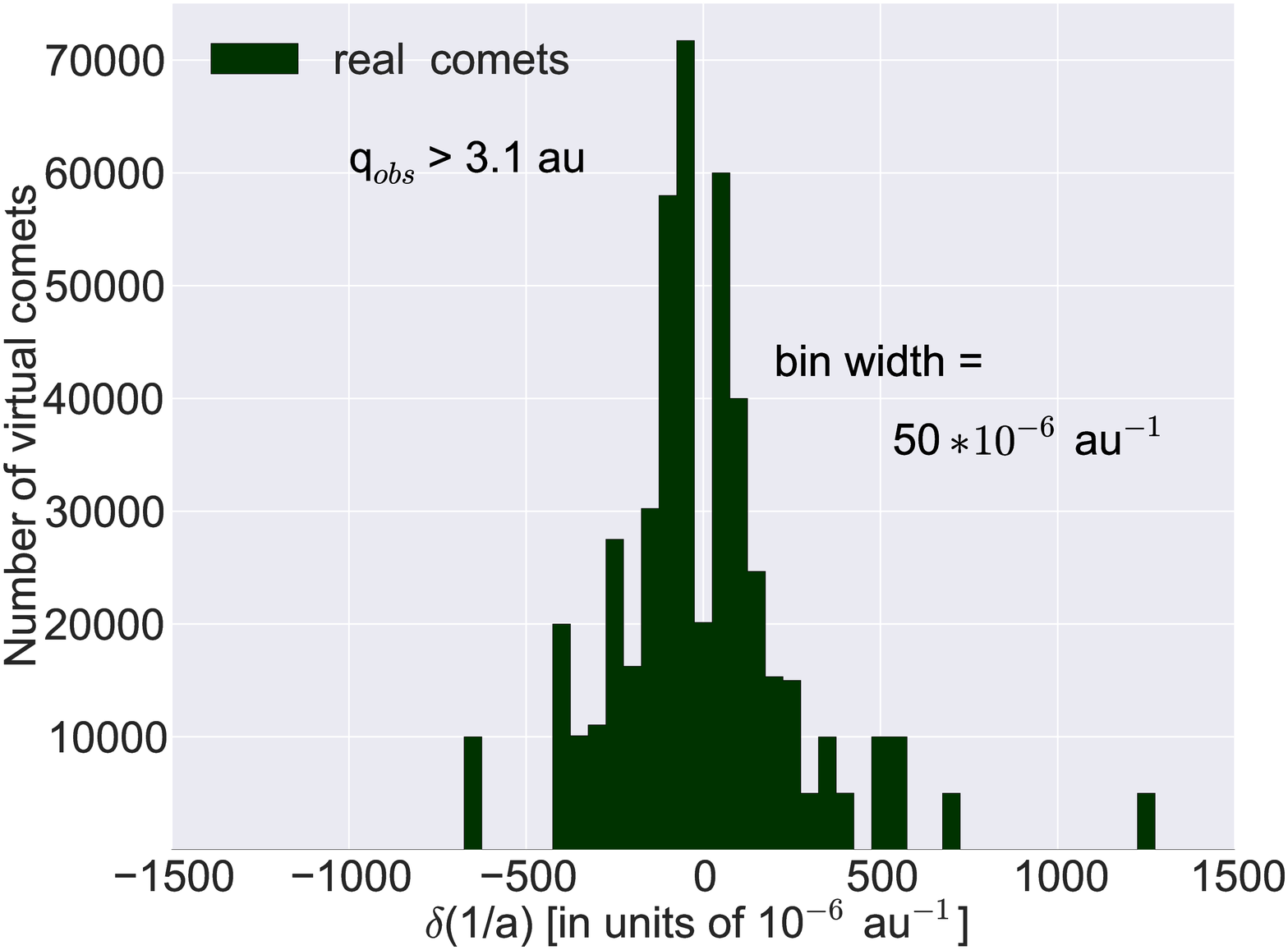} 
\includegraphics[width=5.8cm]{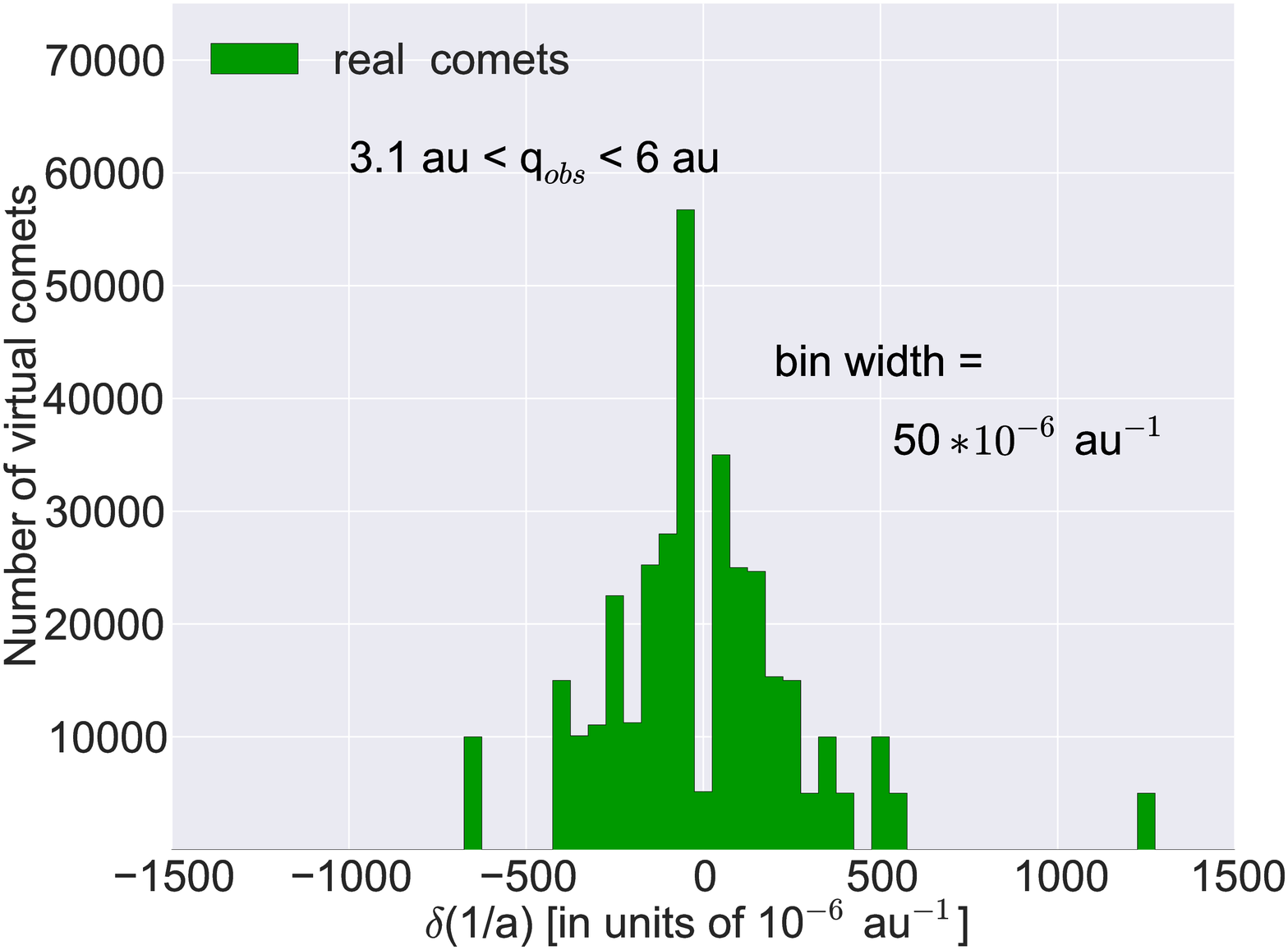} 
\includegraphics[width=5.8cm]{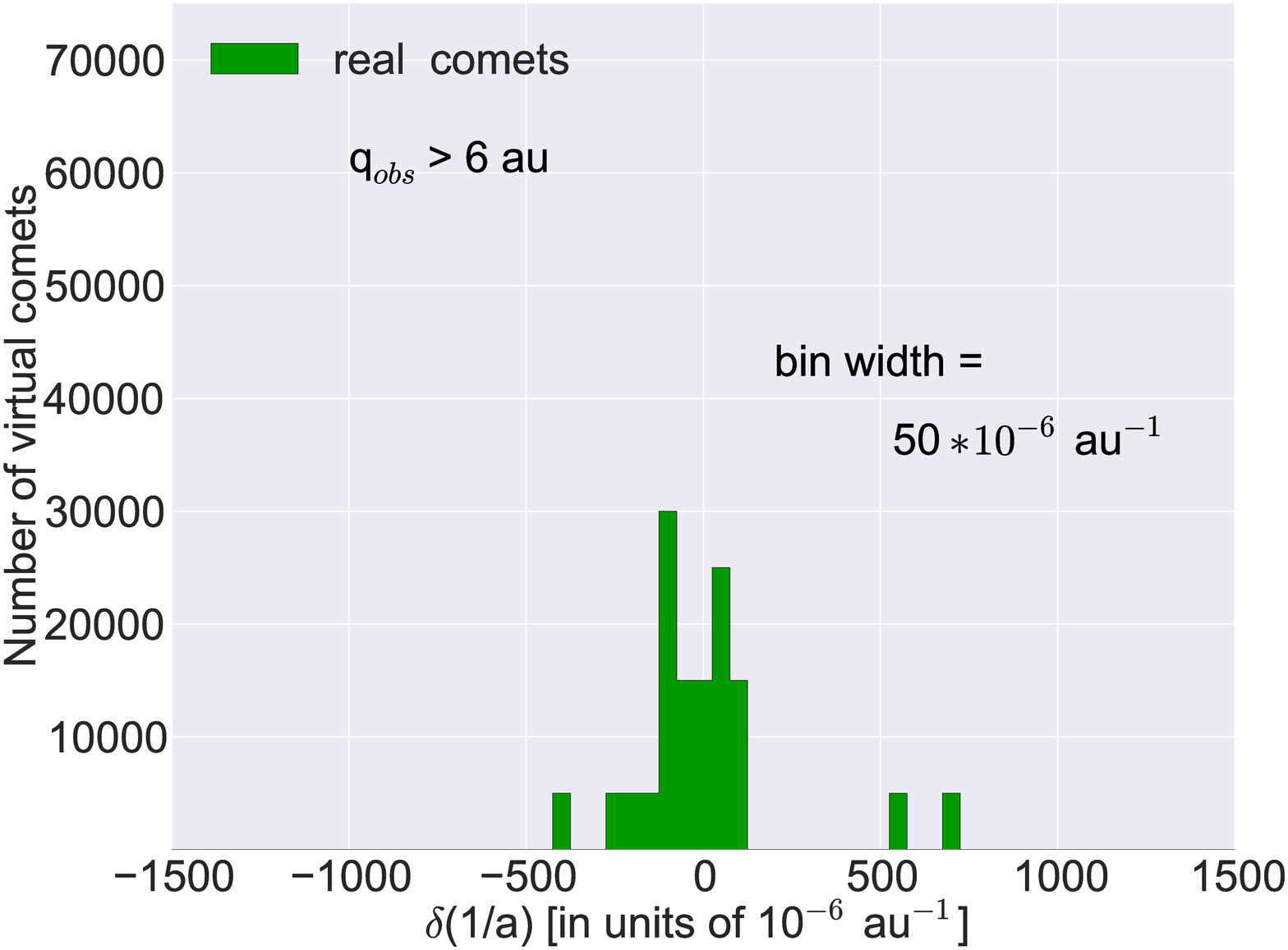} 
\protect\caption{Histograms of planetary perturbations measured by $\delta (1/a) = 1/a_{\rm fut} - 1/a_{\rm ori}$.
Dark-green histogram in the left panel shows the $\delta (1/a)$-distribution of the whole sample. Middle and right panels display similar histograms for comets of perihelion distances between  $3.1 < q_{\rm obs} <6.$\,au from the Sun and greater than 6\,au from the Sun, respectively. Single bin width is $50\cdot 10^{-6}$\,au$^{-1}$.}\label{fig:perturbations_01} 
\end{figure*}

\begin{figure}
\includegraphics[clip,width=1.0\columnwidth]{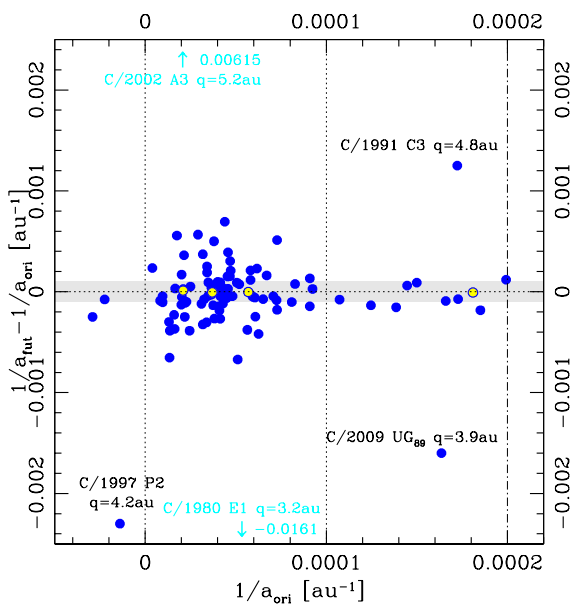} 
\protect\caption{Planetary perturbations in function of the original $1/a$. 
Each blue point represents the nominal orbit and four blue circles with yellow interior represent comets with extremely small planetary perturbations; more explanation in the text. 
}\label{fig:perturbations_02} 
\end{figure}

It turns out that the observed distribution of planetary perturbations suffered by analysed comets during their passage through the Solar system has a spectacular decrease in the range of $-0.000025$\,au$^{-1} < \delta (1/a) <  +0.000025$\,au$^{-1}$, where  $\delta (1/a) = 1/a_{\rm fut} - 1/a_{\rm ori}$. 
Dark-green histogram in the left-side panel of Fig.~\ref{fig:perturbations_01} shows the observed $\delta (1/a)$-distribution for all considered comets, whereas light-green histograms in the middle and right panels represent the  same distribution but for comets with perihelia up to 6\,au and with $q_{\rm obs}\,>\,6$\,au, respectively.  Vertical axes in all three panels in this figure counts VCs, where each swarm of a real comet consists of 5001\,VCs. 

In the deep well we have VCs representing only four comets: C/1976~D2 ($q_{\rm obs}=6.88$\,au, $i_{\rm obs}=112$\degr, MOID\footnote{Minimum Orbit Intersection Distance} with respect to Jupiter equal  to  1.54\,au, $1/a_{\rm ori}=57\cdot 10^{-6}$\,au$^{-1}$), C/1999~F1 (5.79\,au, 92\degr, MOID = 2.93\,au, 37$\cdot 10^{-6}$\,au$^{-1}$), C/2003~O1 (6.85\,au, 118\degr, 4.62\,au, 181$\cdot 10^{-6}$\,au$^{-1}$) and C/2008 S3 (8.02\,au, 163\degr, 3.3\,au, 21$\cdot 10^{-6}$\,au$^{-1}$), while neighbouring bins in the left-side panel contain VCs from about 14 and 12 comet swarms. 
\newline It is obvious that in NG-type of model both planetary perturbations as well as NG-accelerations can contribute to the value $1/a_{\rm fut,NG}-1/a_{\rm ori,NG}$. Therefore we have checked that the sample of 100~purely ballistic solutions  gives congruous distributions to these shown in Fig.~\ref{fig:perturbations_01}.

Statistical significance  of the comet deficit in the central bin of the observed distribution (a dark-green histogram in Fig.~\ref{fig:perturbations_01}) is difficult to estimate because it would require  us  to fit the multi-parameter function to the observed distribution. Such analysis should be based on  a  substantially larger sample. Therefore, in the next section we compare the overall shapes of the observed distribution and the simulated one  calculated for a random mixture of   LPC orbits resembling the observed  element distributions, see Section \ref{sec:perturb_sim}.

It is clear that a deep minimum in this distribution is mainly formed by a lack of ,,almost unperturbed'' comets having 3.1\,au\,$<q_{\rm obs}<$\,6\,au (middle panel in Fig.~\ref{fig:perturbations_01}). Therefore, it makes sense to suspect that mainly Jupiter is responsible for this deep minimum in the $\delta (1/a)$-distribution. Further on, we may conclude that the distribution of orbits of analysed comets (discovered over the past 100 years) is such  that Jupiter does not allow them to pass through our planetary system gaining  perturbations $\mid\delta (1/a)\mid\,<\,25\cdot 10^{-6}$\,au$^{-1}$. 
Let's assume that we are not dealing with some additional perturbations of unknown massive bodies located somewhere beyond Neptune and which we did not take into account in our research. If so, we have at least two basic possibilities. 
Either this is an unknown feature of the distribution of orbital elements of cometary objects forming the Oort spike, and then it can be a permanent phenomenon in time,  or it is a transient phenomenon. To bring us closer to a specific answer about the source responsible for this minimum, in the next section we discuss a  series of simple numerical simulations based on different Monte Carlo ways of dispersing the observed distribution of orbital elements of analysed comets.

Figure~\ref{fig:perturbations_02} shows planetary perturbations  as a   function of $1/a$-original, where
horizontal grey stripe represents a range of perturbations that are smaller  than  or comparable to the width of the Oort spike.
We can see, that the majority of analysed comets (95 per cent of the sample) have planetary perturbations  $\mid \delta (1/a) \mid \, < \, 800\cdot 10^{-6}$\,au$^{-1}$.
Four comets discussed above, with extremely small planetary perturbations, are shown by four blue circles with yellow   interiors.
\newline We also observed that for comets with  $1/a_{\rm ori} > 100\cdot 10^{-6}$\,au$^{-1}$ (original semimajor shorter than 10\,000\,au) planetary perturbations are typically less than $\mid \delta (1/a) \mid \, < \, 200\cdot 10^{-6}$\,au$^{-1}$. 

Figure~\ref{fig:perturbations_02} reveals  also that three of the   analysed comets, C/1942~C2 Oterma, C/1978~G2 McNaught-Tritton and C/1997~P2 Spacewatch, have formally negative $1/a_{\rm ori}$. However, their $1/a_{\rm ori}$-uncertainties are relatively large, that is  $1/a_{\rm ori}$ are  $(-29.1\pm 13.5)\cdot 10^{-6}$\,au$^{-1}$, $(-22.4\pm 37.6)\cdot 10^{-6}$\,au$^{-1}$, and $(-13.9\pm 13.8)\cdot 10^{-6}$\,au$^{-1}$, respectively (poor quality orbits). Therefore, there is an extremely small chance that these comets are interstellar. 

We note in Fig.~\ref{fig:perturbations_02} that only five comets (5 per cent of the sample) have suffered planetary perturbations larger than $800\cdot 10^{-6}$\,au$^{-1}$. Comet C/1980~E1~Bowell ($q_{\rm obs}=3.2$\,au) is the well-known case with perturbations of $-16\,100\cdot 10^{-6}$\,au$^{-1}$, and none amongst known large-perihelion comets had suffered such  large perturbation (MOID with respect to Jupiter equals 0.01\,au). Orbit of comet Bowell is inclined at a negligible angle of $i_{\rm obs}=1.8$\degr ~to the ecliptic plane. In the sample analysed here, just only two more objects have small inclinations.  Comet C/2003~A2~Gleason has an orbital inclination of 8.1\degr , but its perihelion distance is far from Jupiter ($q_{\rm obs}=11.4$\,au and MOID $= 6.1$\,au) and consequently this comet suffers small planetary perturbations of $93\cdot 10^{-6}$\,au$^{-1}$ (one of the blue dots in the main concentration in Fig.~\ref{fig:perturbations_02}). The second comet with a small inclination is C/1997~P2 ($q_{\rm obs}=4.22$\,au, $i_{\rm obs}=14.5$\degr , MOID = 0.47\,au), that also suffered large perturbations ($\delta (1/a)=-2300\cdot 10^{-6}$\,au$^{-1}$)\footnote{Despite a weak knowledge of $1/a_{\rm ori}$ for this comet, the value of planetary perturbations is relatively well estimated.}. The remaining three comets with large perturbations are: C/1991~C3 McNaught-Russell ($q_{\rm obs}=4.8$\,au, $i_{\rm obs}=113.4$\degr, MOID = 0.1\,au, $\delta (1/a)=1300\cdot 10^{-6}$\,au$^{-1}$), C/2002~A3 LINEAR ($5.2$\,au, $48.1$\degr, 0.22\,au, $6150\cdot 10^{-6}$\,au$^{-1}$) and C/2009~UG$_{89}$ Lemmon ($3.9$\,au, $130.1$\degr, 0.23\,au, $-1600\cdot 10^{-6}$\,au$^{-1}$). 

\begin{figure}
\vspace{-0.4cm}
\includegraphics[width=8.8cm]{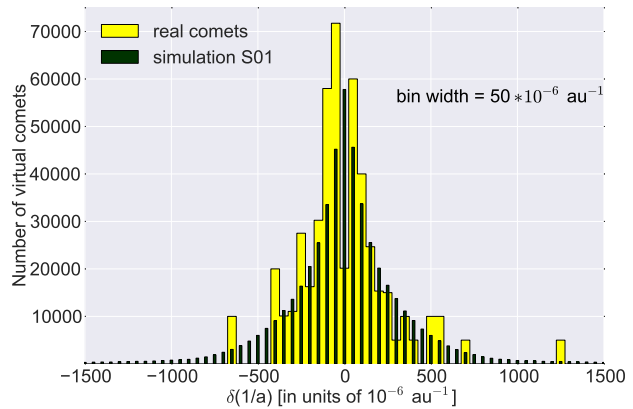} 
\protect\caption{\label{fig:simI_comp} The reference simulation S01 of planetary perturbations (dark green histograms with bars located in the middle of each bin) in comparison with the distribution of planetary perturbations for actual sample of comets represented by $100\times 5001$ VCs (full yellow histogram). In this simulation, both perihelion passage epochs and  Galactic ascending node longitudes were completely randomized.} 
\end{figure}

\begin{figure*}
\vspace{-0.4cm}
\includegraphics[width=8.8cm]{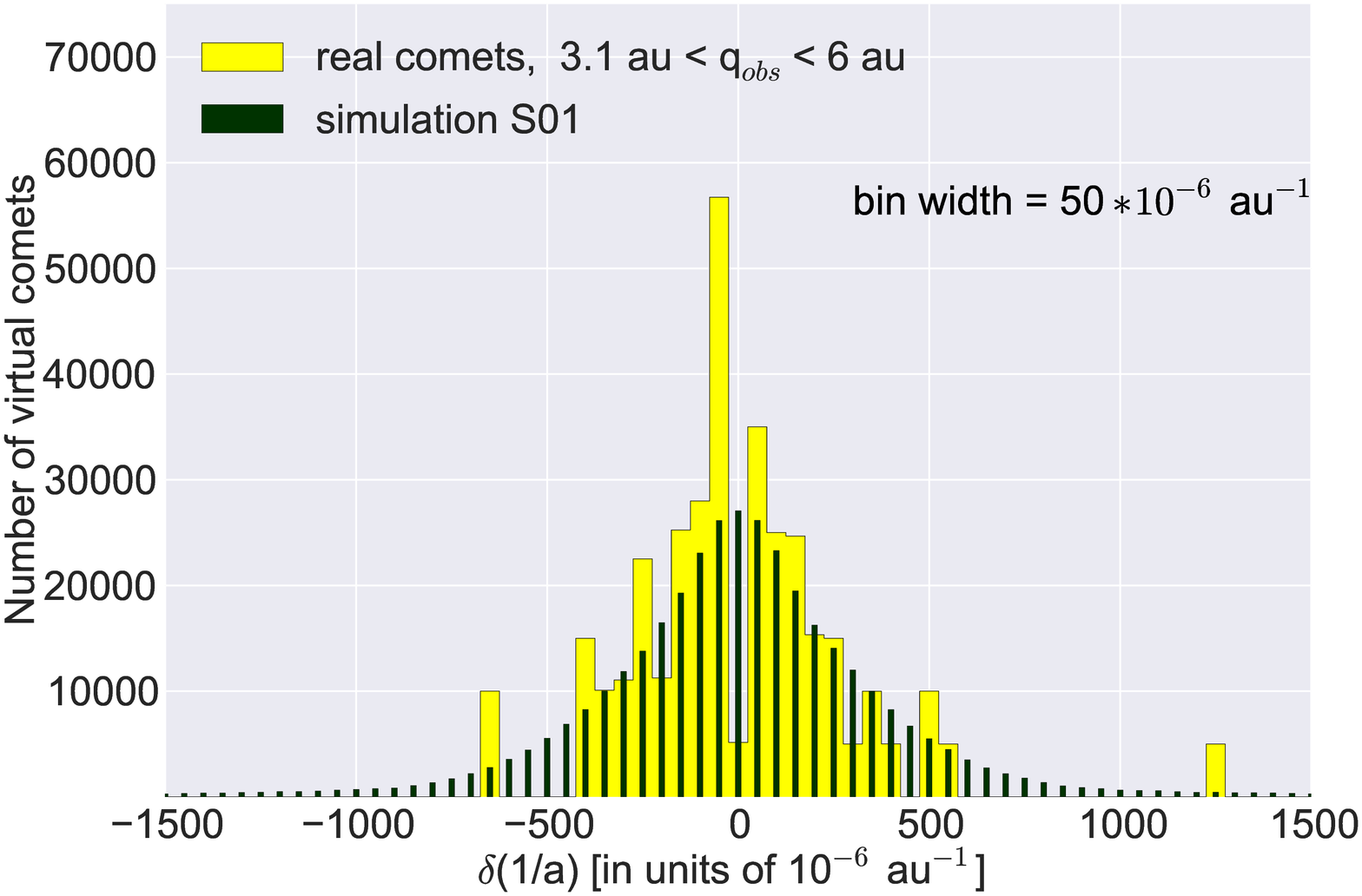} 
\includegraphics[width=8.8cm]{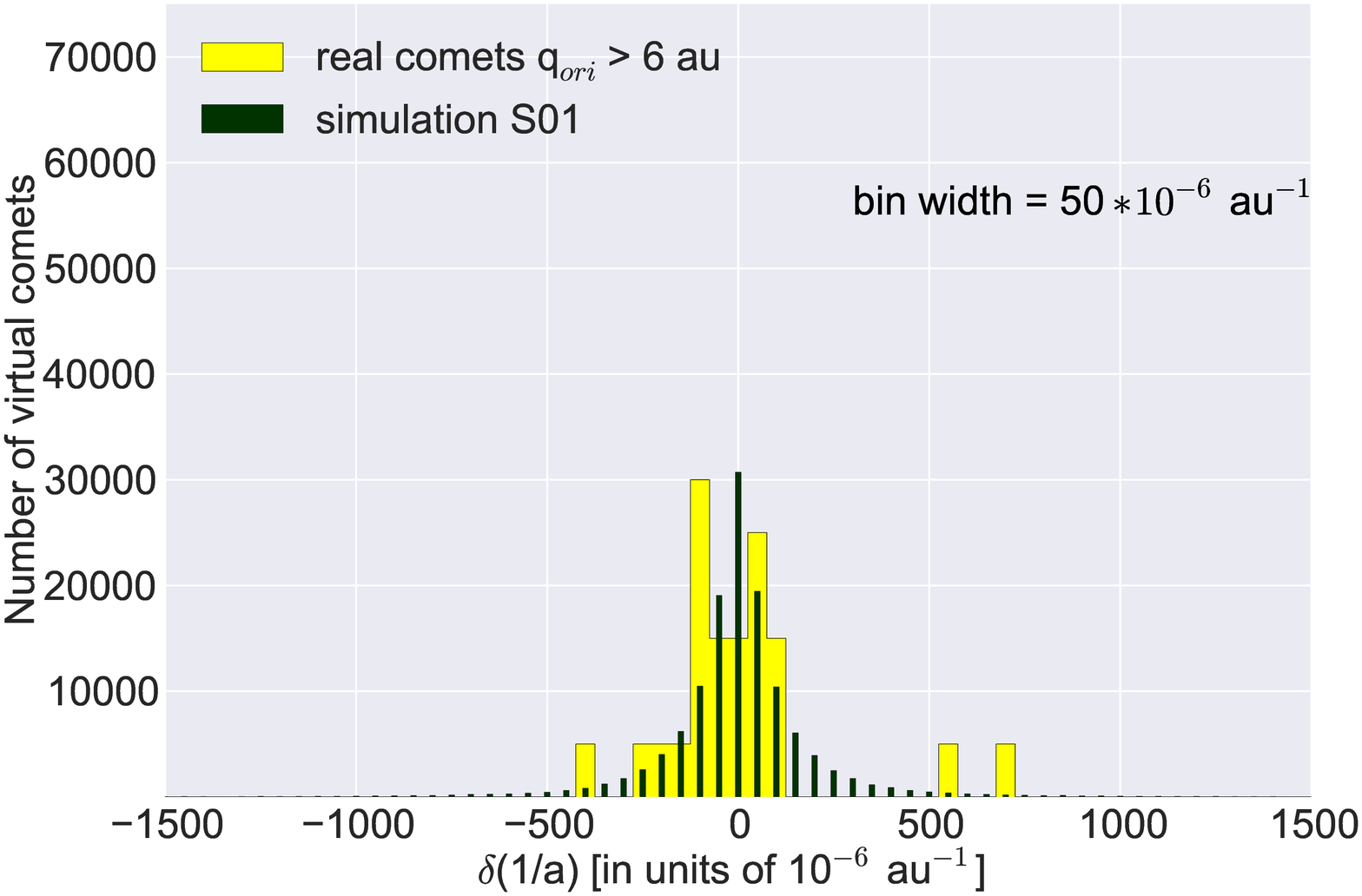} 
\protect\caption{\label{fig:simI_two_distibutions}A comparison between actual VCs  distributions (solid yellow histograms) and the reference simulation S01 (dark green histograms with bars located in the middle of each bin), where the left-side panel shows distributions for comets of perihelion distances of 3.1\,au\,$< q_{\rm obs}\,<\,6$\,au (77 per cent  of comets, flat  part of the distribution of $q_{\rm obs}$ in Fig~\ref{fig:hist_q_e}) and the right-side panel displays the same for comets of perihelion distances of $q_{\rm obs}\,>\,6$\,au (23 per cent of comets, decreasing branch of the $q_{\rm obs}$-distribution).}
\end{figure*}

\section{Simulated planetary perturbations of LPCs based on their observed distribution of \lowercase{$q_{\text{obs}}$ and $e_{\text{obs}}$.}}\label{sec:perturb_sim}
 
\begin{figure*}
\vspace{-0.4cm}
\includegraphics[width=8.8cm]{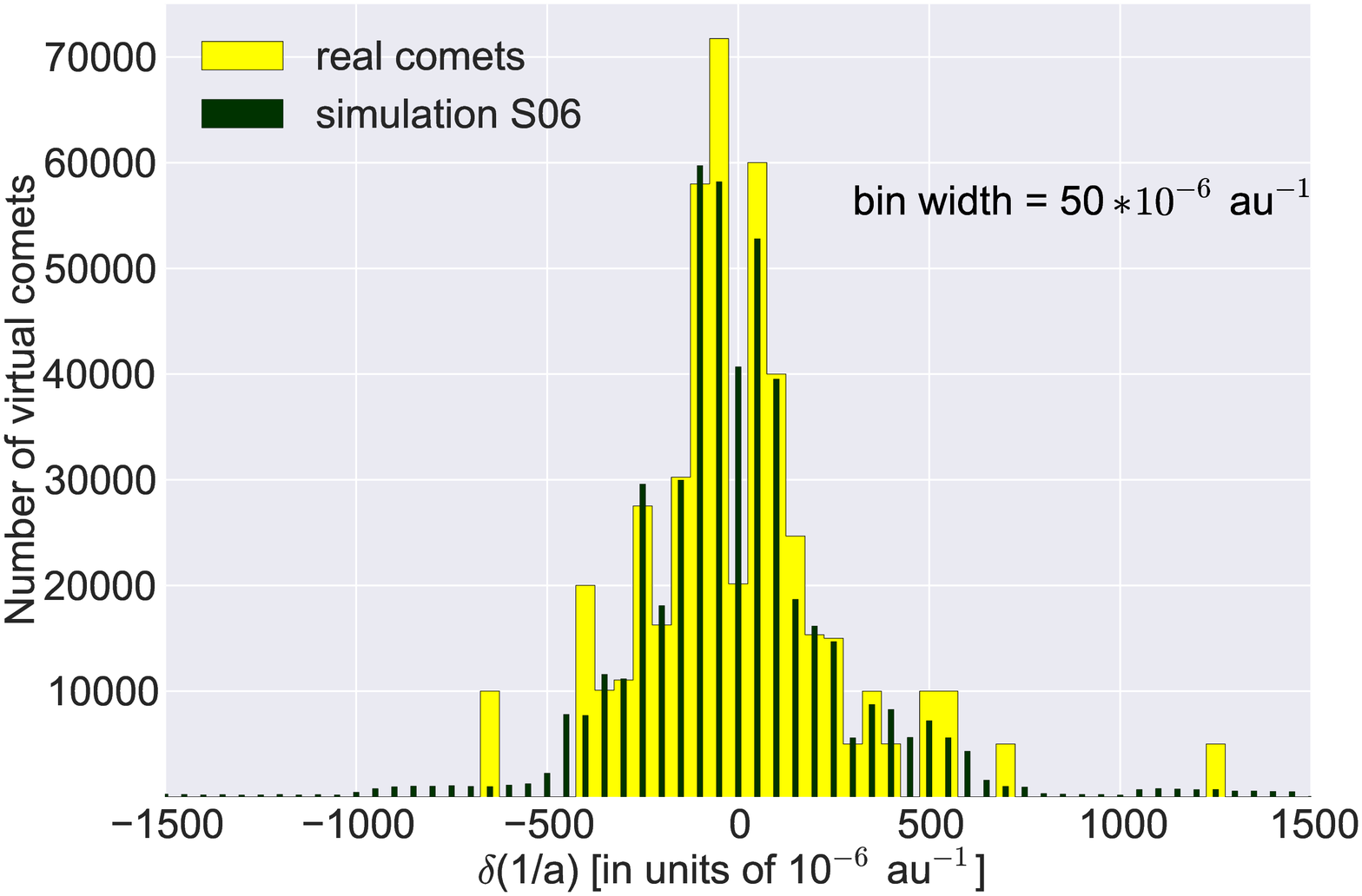} 
\includegraphics[width=8.8cm]{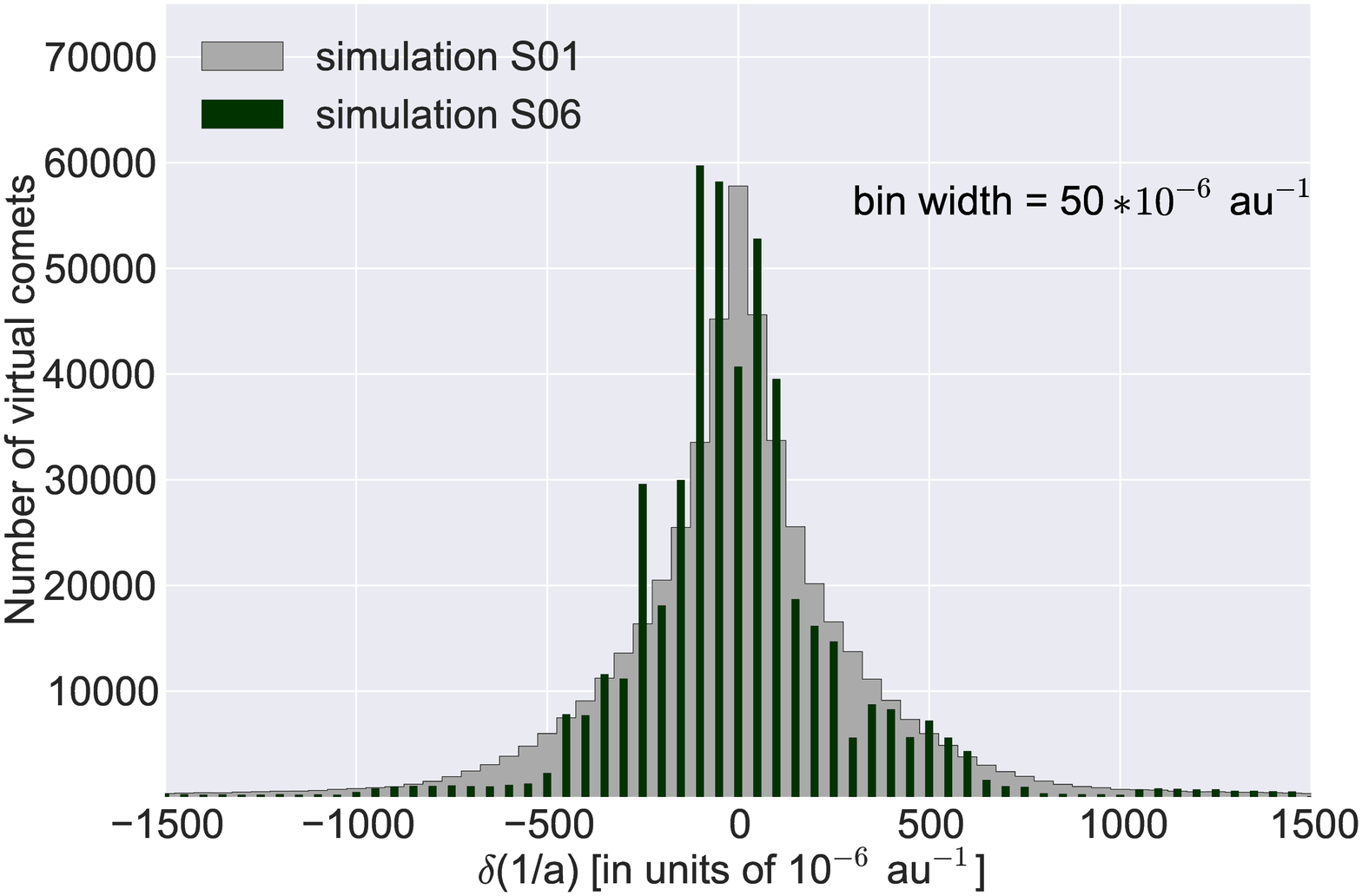} 
\includegraphics[width=8.8cm]{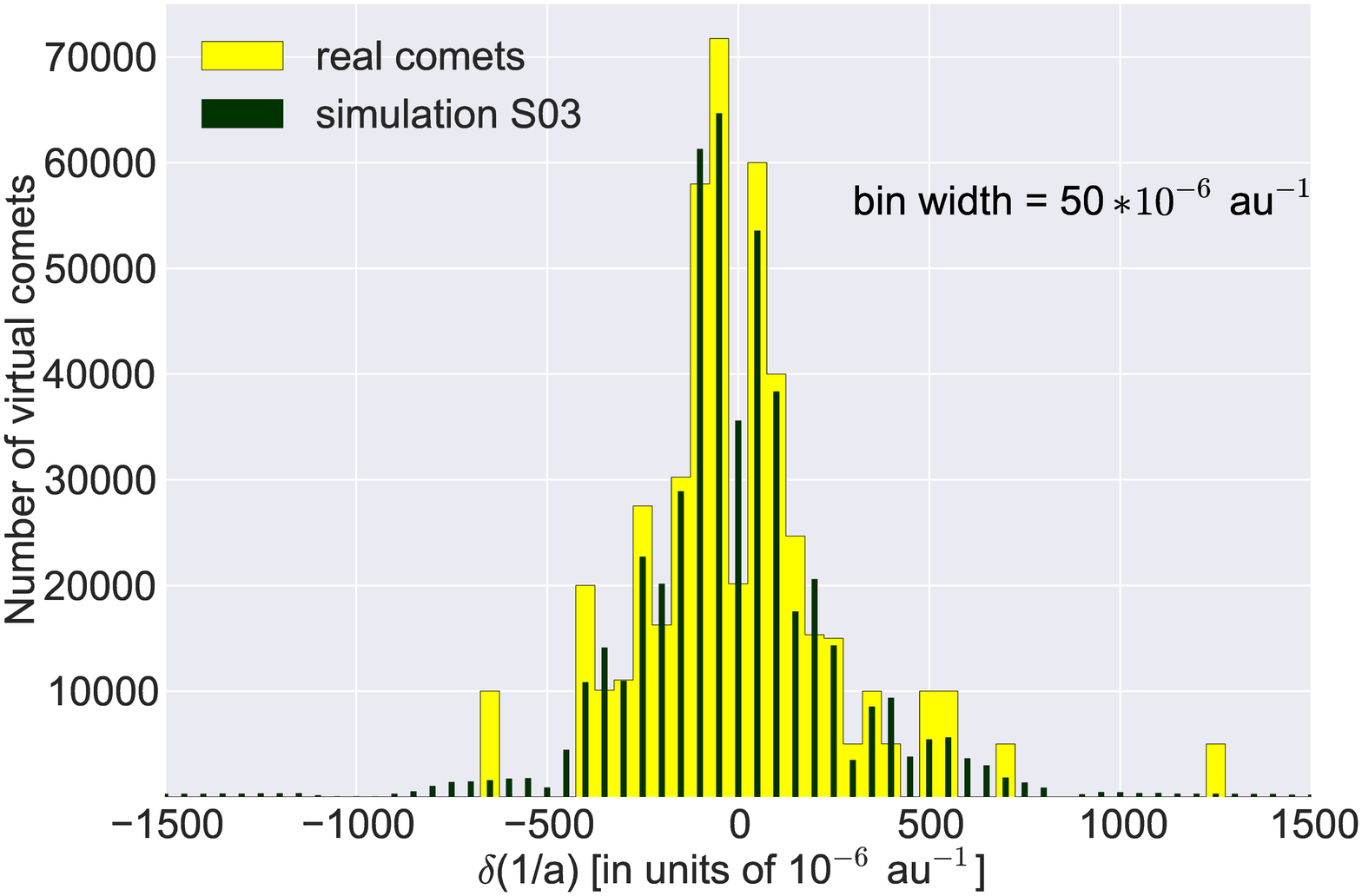} 
\includegraphics[width=8.8cm]{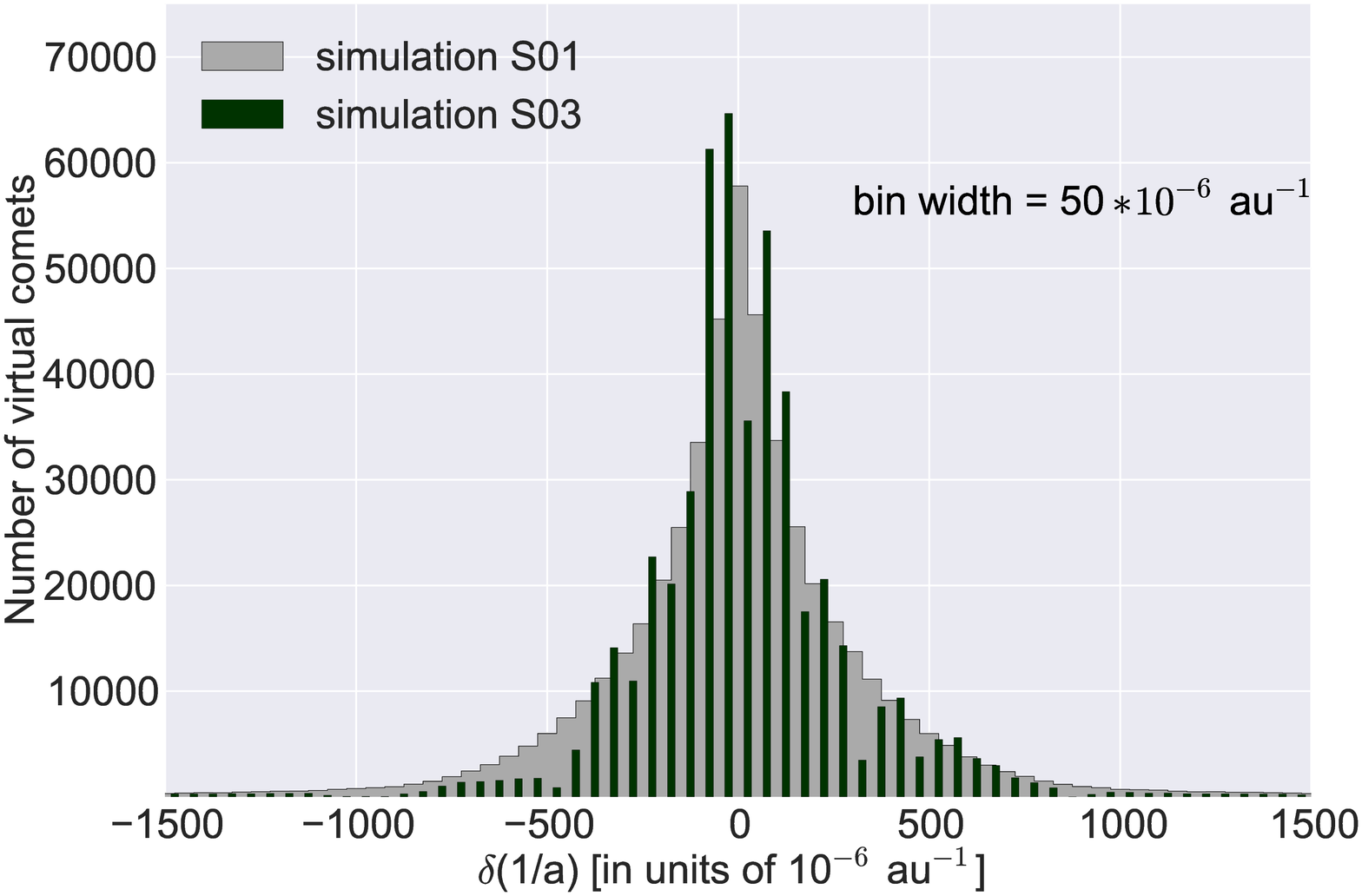} 
\caption{\label{fig:sim06_sim03}{A comparison between simulation S06 (upper panels, dark-green histogram with bars located in the middle of bins) and simulation S03 (lower panels). Yellow histogram shown as the background in the left-side panels represents the actual distribution of VCs for analysed comets while a grey  histogram in the right-side panels shows the distribution of planetary perturbations obtained from the reference simulation S01.}}
\end{figure*}

\begin{figure*}
\vspace{-0.4cm}
\includegraphics[width=8.8cm]{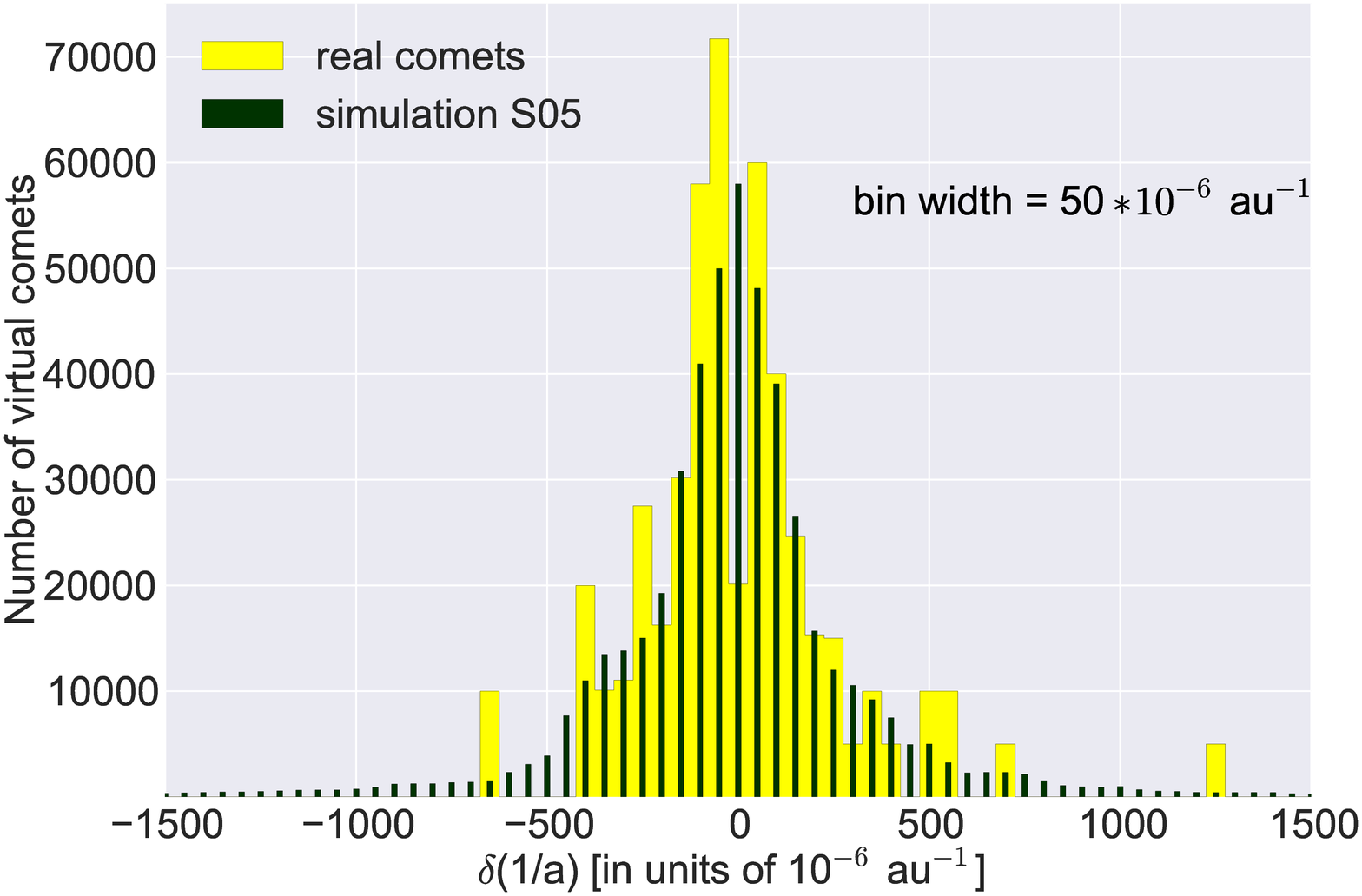} 
\includegraphics[width=8.8cm]{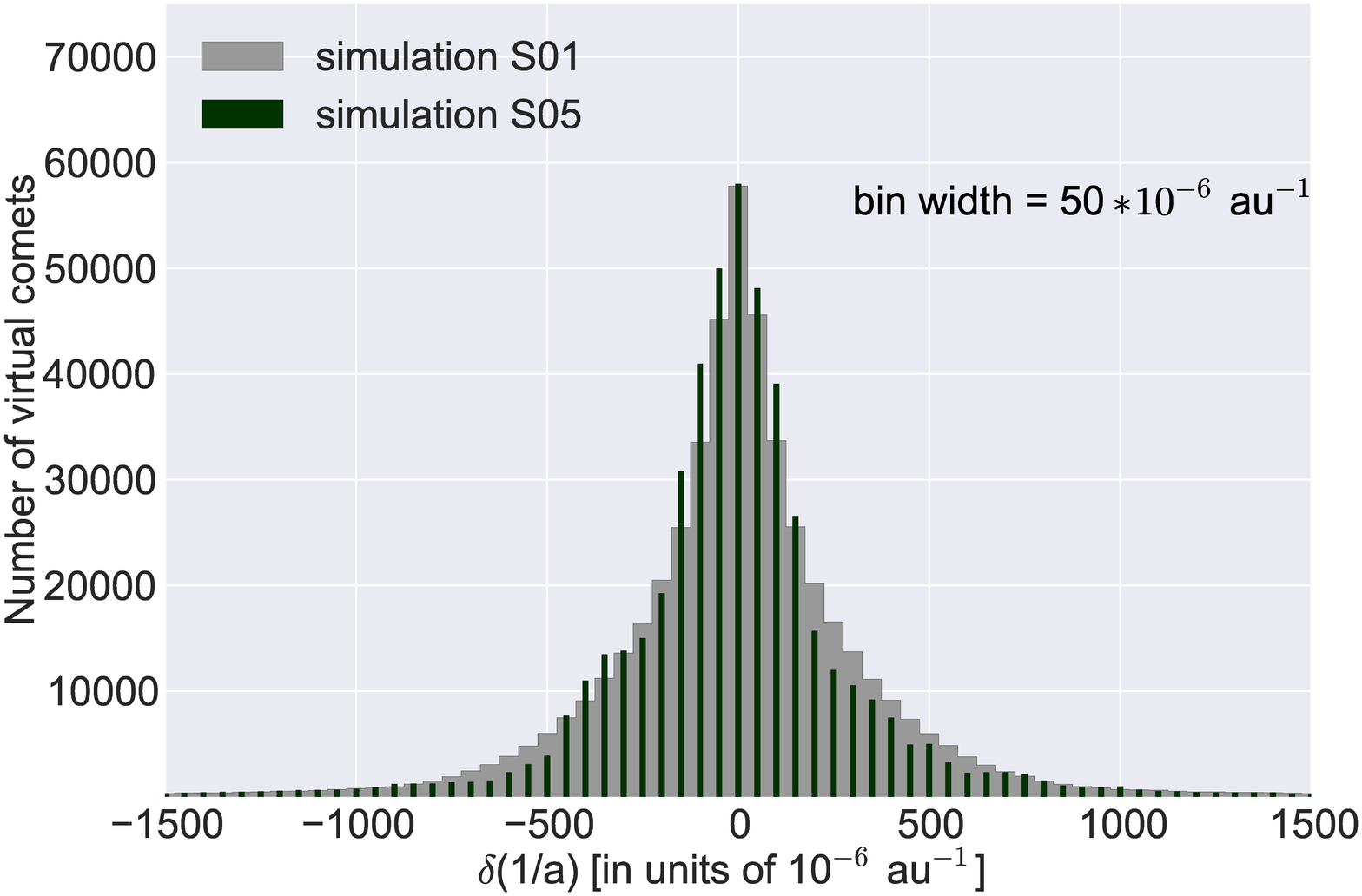} 
\protect\caption{\label{fig:sim05}Simulation S05 (dark green bars) on a background of an actual distribution of VCs (yellow histogram shown in the left-side panel) and on a background of simulation S01 (grey histogram presented in the right-side panel).} 
\end{figure*}

An unexpected but striking lack of comets with very small planetary perturbations depicted in Fig.~\ref{fig:perturbations_01} motivated us to search for its reason. It might be only a random, statistical fluctuation (see below) but we deal here with quite a large number of 100 cometary orbits. In order to check the dependence of this phenomenon on the geometry and timing of this specific stream of 100~LPCs we have performed several simulations, varying angular orbital elements and perihelion times separately and simultaneously at different levels of dispersion.
In this manner we show how the surprising deep minimum existing in the actual distribution of planetary perturbations (Fig.~\ref{fig:perturbations_01}) is resistant to a moderate diffusion of orbital elements of analysed comets. 

Numerical simulations of planetary perturbations acting on the motion of LPCs have a long-established tradition with traces pointing back to works of Hubert Anson Newton, e.g.~\cite{hubert_newton:1891} or later extensive studies by \cite{everhart:1968, everhart:1969}, \cite{fernandez:1981} and many others. To our best knowledge such a striking local minimum around zero was never observed.

In all simulation experiments discussed below we always started from the actual distribution of 100 swarms of comets, that is from 500\,100\,VCs, and we only changed two orbital elements:  the  epoch of the perihelion passage ($T$) and  the   longitude of the ascending node in the Galactic frame ($\Omega_{\rm Gal}$). The NG-forces were not taken into account in these simulations to increase the calculation speed. We are convinced (see also previous section) that this omission did not  influence the results.
A dispersion of perihelion passage epochs should make the simulated dynamics of large-perihelion LPCs independent  of   particular planetary configurations. To perform such a dispersion we used the JPL\,DE431 long term ephemeris \citep{folkner_et_al:2014} as a source for the planetary positions. Due to its large time span of over 30\,000 years we were able to spread simulated epochs of cometary perihelion passage over this interval. As it concerns the longitude of the ascending node with respect to the Galactic disc plane the principles of this approach are described in \citet{dyb-trans:2004}. In short, we consider the dominating perturbing force (the Galactic disc tide) to be axisymmetric and such a simple way of the orientation of orbits dispersion should not change qualitatively the overall dynamical evolution. In practice, to force more or less uniform distribution of $\Omega_{\rm Gal}$ we divided the whole range of 360\degr  ~into 50 intervals and then  drew   100\,different random values from each bin. Such a 'total' dispersion of these two orbital elements was used to perform the 'reference' simulation, marked as S01.  

The resulted distribution of planetary perturbations obtained from S01 is displayed in Fig.~\ref{fig:simI_comp} using a dark green histogram. It is a well-known fact that this is a symmetrical and heavy-tailed distribution, see for example discussions in \citet{stoica-et-al:2010} and \citet{rickman-2014}. Obviously this distribution is peaked around zero. Fig.~\ref{fig:simI_comp} also shows the difference between this distribution and the distribution of planetary perturbations acting on  the  actual sample of analysed comets (represented by solid yellow histogram in the background). The dark green histogram presents what is expected while the yellow one shows what we have obtained for our sample of 100 large-perihelion LPCs. 		

Differences between the observed distribution of planetary perturbations and the simulated one (S01) are tested here using a one-sample Kolmogorov test (simulated distribution is well represented by a continuous function). It appears that the low amplitude in the central bin of the observed distribution is compensated at both sides with the relative excess of comets, while in the simulated distribution these features are absent. A null 
hypothesis that the observed and simulated distributions are different is rejected at the level of $\alpha = 0.18$. It implies that the present material is insufficient to claim differences of both distributions.

In Fig.~\ref{fig:simI_two_distibutions} we additionally present a different contribution to the overall planetary perturbations from two distinct  parts of the studied sample: the left-side plot is for comets having $q_{\rm obs}\,<\,6$\,au while the right one  is  for $q_{\rm obs}\,>\,6$\,au. The same colour coding for distributions is used  here as in Fig.~\ref{fig:simI_comp}. Both  the  real and simulated distributions show remarkable differences. It can be noticed, that for comets with $3.1$\,au\,$<\,q_{\rm obs}\,<\,6$\,au the effect of a deep minimum around zero is evident whereas for comets with  $q_{\rm obs}\,>\,6$\,au a small minimum around zero is statistically unconvincing mainly due to a small total number of comets with such a large perihelion distances.

Further on, seven other simulations were performed according to the specification given in Table~\ref{tab:simulations}. The purpose of these simulations was to test at what level of dispersion of  perihelion epochs and/or geometry of orbits this striking local minimum will disappear. We noticed that the distribution of $\delta (1/a)$ in S02 ($\Omega_{\rm Gal}$  dispersed in the range of $\pm$1\degr) still shows a deep minimum, however in distributions  from S03  ($T$ scattered up to $\pm$60\,days) and S06 ($\Omega_{\rm Gal}$ dispersed in the range of $\pm$10\degr) the obtained minima are significantly more shallow than in the observed distribution (yellow histogram given in the background in the left-side panels of Fig.~\ref{fig:sim06_sim03}). Figure~\ref{fig:sim06_sim03} also reveals that the latter two simulations give similar distributions of planetary perturbations, even similarly exposing local structures of $\delta (1/a)$-distribution. In the right-side panels both simulations are compared  to S01. We concluded that $\Omega_{\rm Gal}$-dispersion up to $\pm$10\degr and/or $T$-dispersion up to 60~days effectively reduces the well in $\delta (1/a)$-distribution.

The minimum is completely filled-in in Fig.~\ref{fig:sim05}, where we changed the planetary configuration met by comets applying a dispersion in perihelion passage up to 600\,days, and in $\Omega_{\rm Gal}$ in the range of $\pm$10\degr . The distribution of planetary perturbations in this simulation (S05, dark green bars) shows only some small  deviations from the symmetrical distribution resulting from the simulation S01 (light grey histogram in the right panel), and its wings seem to be smoothed in comparison to wings in $\delta (1/a)$-distribution obtained using simulation S03 or S04 (compare right-side panels of Figs.~\ref{fig:sim06_sim03}~and~\ref{fig:sim05}). However, for a bit smaller dispersions in simulations S07 and S08 the deep minimum was not completely filled-in.

To conclude: the existence of a deep well in the planetary perturbations distribution obtained from the dynamics of 100\,LPCs with large perihelion distances studied here is remarkable, even if formally statistically possible. Simulations show that it vanishes after a moderate dispersion of cometary perihelion epochs and/or orbit orientation. It seems reasonable to state that this particular stream of 100~comets (lasting over a century) met such a series of planetary configurations that passing through a planetary zone without any changes in $1/a$ was almost impossible. 


\begin{center}
\begin{table}
\caption{\label{tab:simulations}Ranges of uniform scattering for simulated orbital elements}
\setlength{\tabcolsep}{5.0pt} 
\begin{tabular}{lll}
\hline 
simulation &  $T$              & $\Omega _{\rm Gal}$  \\
\hline
S01        &  30\,000 years    & $<0;2\pi)$           \\
S02        &   observed        & $\pm$1\degr          \\
S03        &  $\pm$60\,days    &  observed            \\
S04        &  $\pm$60\,days    & $\pm$1\degr          \\
S05        &  $\pm$600\,days   & $\pm$10\degr         \\
S06        &  observed         & $\pm$10\degr         \\
S07        &  $\pm$600\,days   &  observed            \\
S08        &  $\pm$60\,days    & $\pm$10\degr         \\
\hline
\end{tabular}
\end{table}
\end{center}


\section{Long term past and future dynamical evolution}\label{sec:past_next}
 
Having barycentric original and future orbits for all 100\,LPCs described in this paper we are able to study their past and future motion during the previous and next orbital revolution. To this purpose we used exactly the same dynamical model as in \citet{dyb-kroli:2015}. To obtain orbital elements at previous and next perihelion passage together with their uncertainties we followed numerically the motion of the whole swarm of 5001\,VCs for each comet in the studied sample. Galactic disk and centre perturbations as well as the influence of all known stellar perturbers were taken into account. It means that in addition to both Galactic tide terms, perturbations from 90~stars or stellar systems capable to pass closer than 3.5~pc from the Sun in the studied period of cometary dynamics were considered. These are all known stellar perturbers that might influence the orbital evolution of LPCs. The detailed procedure of selecting stellar perturbers as well as the methods of calculating their influence on cometary dynamics is fully described in \cite{dyb-kroli:2015}. Similarly to our previous papers we did not find (see below, this and next section) any strong stellar disturbance in the past dynamics of investigated comets and many of the recognised interactions have only a local (i.e. short in time) importance.

For each comet we stopped the numerical integration at a previous (when integrating backward) or next (going forward) perihelion passage epoch. For hyperbolic or very elongated elliptical orbits we applied the so-called escape limit (EL) of 120\,000~au and stopped the calculation when this threshold distance from the Sun was exceeded.  This threshold represents the maximum heliocentric distance of a comet up to which we can  reliably  follow its motion. Each individual VC is called \textit{returning} (R) if its maximum distance from the Sun is not greater than EL, otherwise it is called \textit{escaping} (E). Additionally,  we distinguished and counted hyperbolic orbits (H) inside \textit{escaping} VCs. This nomenclature was introduced and explained in detail in \cite{kroli-dyb:2010}.

The full description of previous and next orbital evolution for the studied sample of 100\,LPCs is presented in 
Tables~\ref{tab:previous-returning}--\ref{tab:previous-mixed} and Tables~\ref{tab:next-returning}--\ref{tab:next-mixed}, respectively; all  are  included as online material only. We also provide estimates of the uncertainty of all parameters given there. If a given parameter values follow the Gaussian distribution we present a mean value and its standard deviation. Otherwise we show three deciles: at 10, 50 (median) and 90 per cent.
\newline To easily search for data on a specific comet there are additional Tables~\ref{tab:all-previous} and \ref{tab:all-next} containing all 100 comets arranged chronologically with their previous and next orbit parameters.

\vspace{0.1cm}
\noindent

Please keep in mind during further discussion of our results that we are able to precisely apply the planetary perturbation only during the observed perihelion passage. For obvious reasons we can not do this for previous and next ones. Therefore strictly speaking, the \textit{previous} orbit describes an orbit after leaving planetary zone during previous passage through perihelion while the \textit{next} orbit describes an orbit just before the entrance to a planetary zone during the next passage through perihelion.

\subsection{Evolution to the previous perihelion}

Table~\ref{tab:previous-returning} presents 66\,comets (in order of decreasing $1/a_{\textrm{prev}}$) for which the entire swarm of VCs is returning (R=5001, E=0, H=0). In Cols.~[2]--[5] the statistics of orbital elements recorded at previous perihelion is shown. Columns~[6]--[8] show the percentage of the previous perihelion distance found in three intervals: $q_{\textrm{prev}}\leqslant$10\,au, 10\,au $<q_{\textrm{prev}}\leqslant$20\,au and $q_{\textrm{prev}}>$20\,au. 
The purpose of such statistics is to distinguish between dynamically old and dynamically new LPCs. Dynamical status resulting from these statistics is presented in the last column of the table, and was evaluated according to the following criteria: if more than 50~per cent of VCs are found in the first interval, we call that comet dynamically old with DO symbol. We mark it as DO+ in a case when 95 per cent or more VCs have  $q_{\textrm{prev}}$ smaller than 10\,au. In a similar way we define dynamically new comets: DN+ if 95~per cent or more VCs have $q_{\textrm{prev}}>$20\,au (or VCs are escaping) and most probably dynamically new (DN) when this percentage exceeds 50\,per cent. If the previous perihelion do not satisfy any of the above criteria we conclude that the status of such a comet is uncertain (DU).

Part~I of Table~\ref{tab:statistics_previous} summarizes a dynamical status statistics for comets having fully returning swarms of VCs during the past evolution to the previous perihelion (see also last column of Table~\ref{tab:previous-returning}). We find here 13~dynamically new comet including 10~comets classified as DN+, 37~comets as certainly dynamically old (DO+), three more dynamically old classified as DO, and 13~comets with an uncertain status. What might be surprising the latter 13~comets have orbits of a very good quality (exclusively 1a or 1a+) and compact (or very compact) swarms of VCs. They simply visited our planetary system having previous perihelion distance close to the assumed threshold value of $q_{\textrm{prev}}=$15\,au. In fact 10 of them have more than 90 per cent of VCs with $q_{\textrm{prev}}\leqslant$20\,au so they should rather be considered as dynamically old.

\begin{center}
\begin{table*}
\caption{\label{tab:statistics_previous}Statistics of the dynamical status for 66 fully returning swarms of VCs (Table~\ref{tab:previous-returning}) and 34~mixed or fully escaping swarms of VCs (Table \ref{tab:previous-mixed}) in the past evolution to the previous perihelion.}
\setlength{\tabcolsep}{8.0pt} 
\begin{tabular}{lrrrrrrr}
\hline 
Description of the subgroup   &  Number of & \multicolumn{6}{c}{D y n a m i c a l ~~~s t a t u s} \\
                              &  comets   & DO+ & DO & DN & DN+ & DU & in Fig.~\ref{fig:fig13} \\
\hline \\
\multicolumn{8}{l}{Table~\ref{tab:previous-returning}} \\
Fully returning VCs                              & 66  & 37  &  3  &  3  & 10  & 13   & black    \\
\hline \\
\multicolumn{8}{l}{Table~\ref{tab:previous-mixed}} \\
At least 95\% of returning VCs                   & 11  & 1   &  4  &  1  &  2  &  3   & grey     \\
Remaining comets with returning nominal clone    & 12  & --  &  5  &  3  &  3  &  1  & omitted   \\
Comets with escaping or hyperbolic nominal clone & 11  & --  &  -- & --  & 11  &  --  & omitted  \\
\hline
All of mixed or fully escaping VCs               & 34  & 1   &  9  &  4  & 16  &  4   &         \\
\hline   \\
All comets                                       &100  & 38  & 12  &  7  & 26  & 17   &         \\
\hline
\end{tabular}
\end{table*}
\end{center}

\begin{figure}
\includegraphics[clip,angle=270,width=1.0\columnwidth]{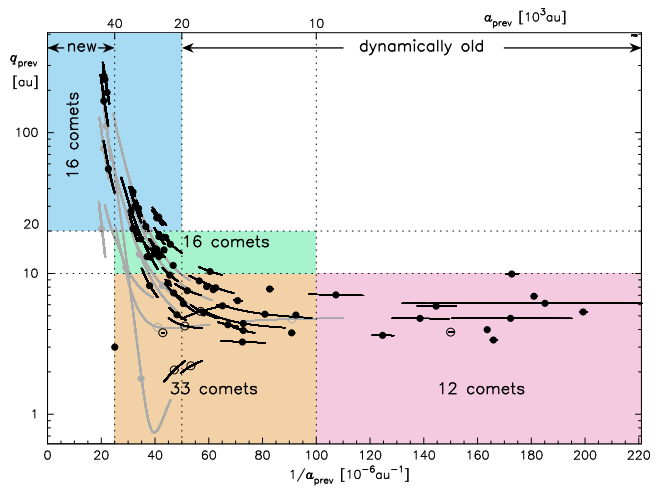}
\caption{\label{fig:fig13}{Perihelion distances versus semimajor axes for 77 comets returning in the past. We included here all 66 comets from Table~\ref{tab:previous-returning} and  the   first 11 comets from Table~\ref{tab:previous-mixed}. 
Dynamically old and dynamically new comets (marked as 'new' in the plot) are actually identified from their previous perihelion distances, which then turn out to correspond to previous semi-major axes of $< 20\,000$\,au
and $> 40\,000$\,au, respectively. See text for a detailed description.}}
\end{figure}

\begin{figure}
\vspace{-0.4cm}
\includegraphics[width=8.8cm]{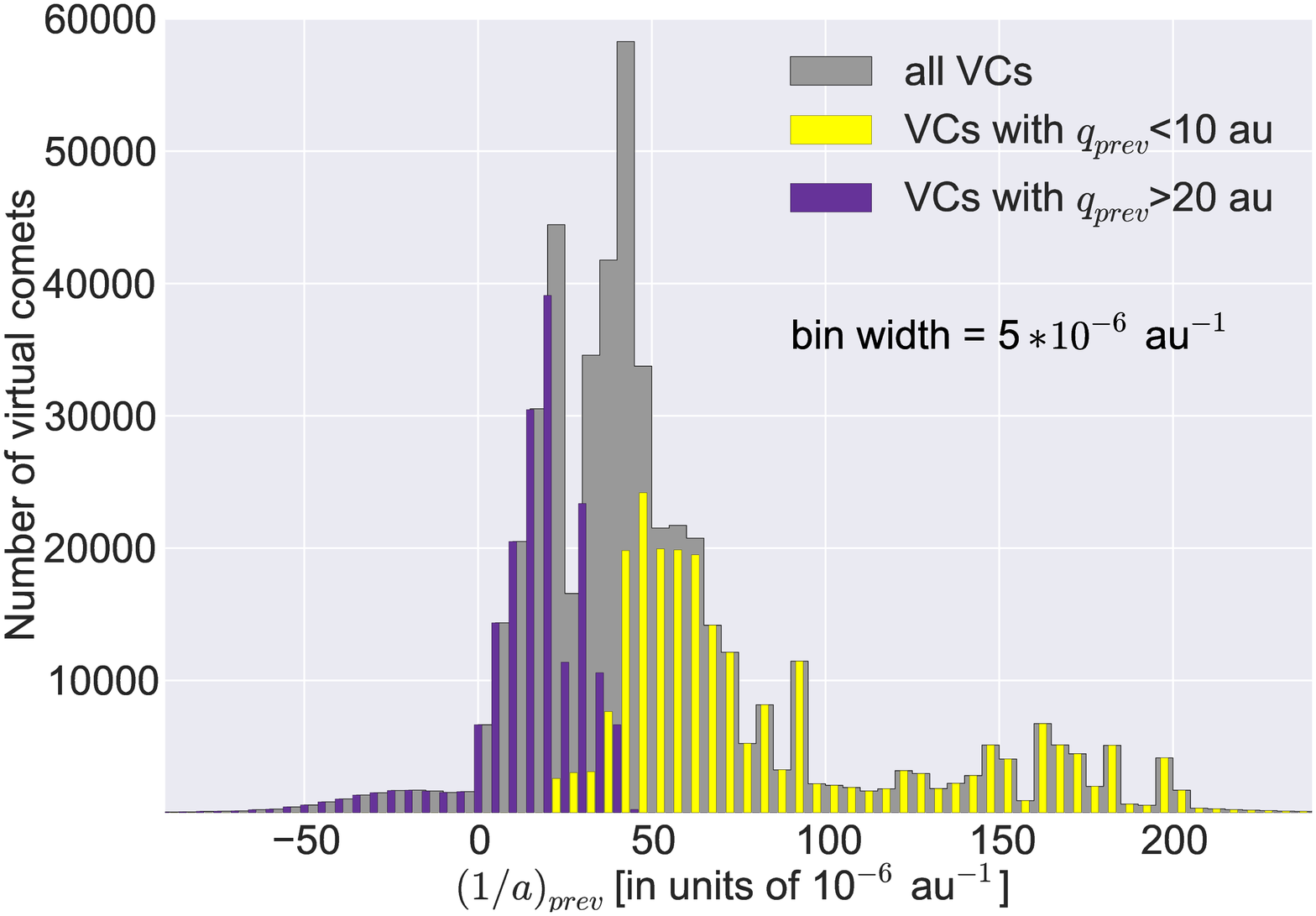} 

\includegraphics[width=8.8cm]{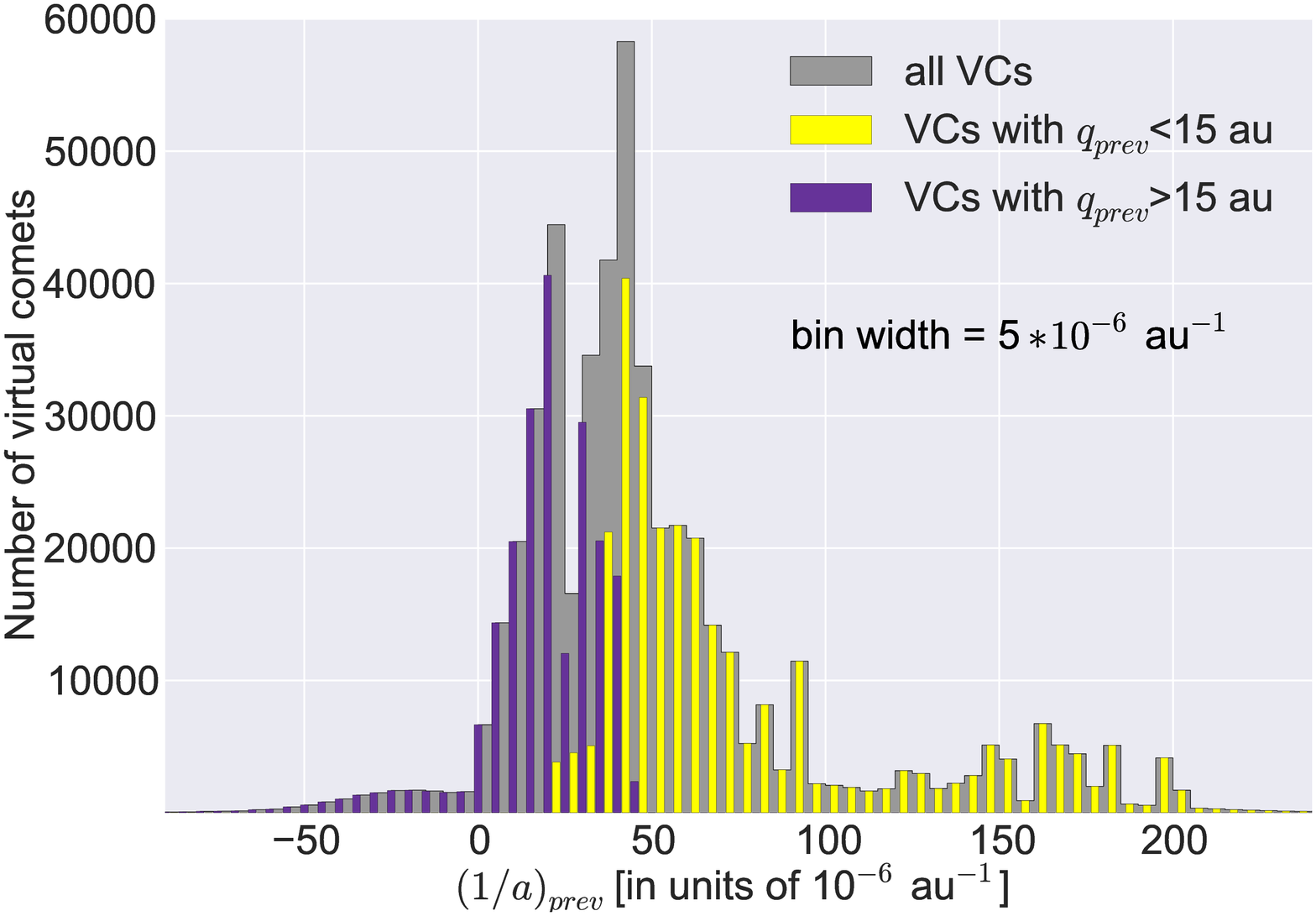} 

\includegraphics[width=8.8cm]{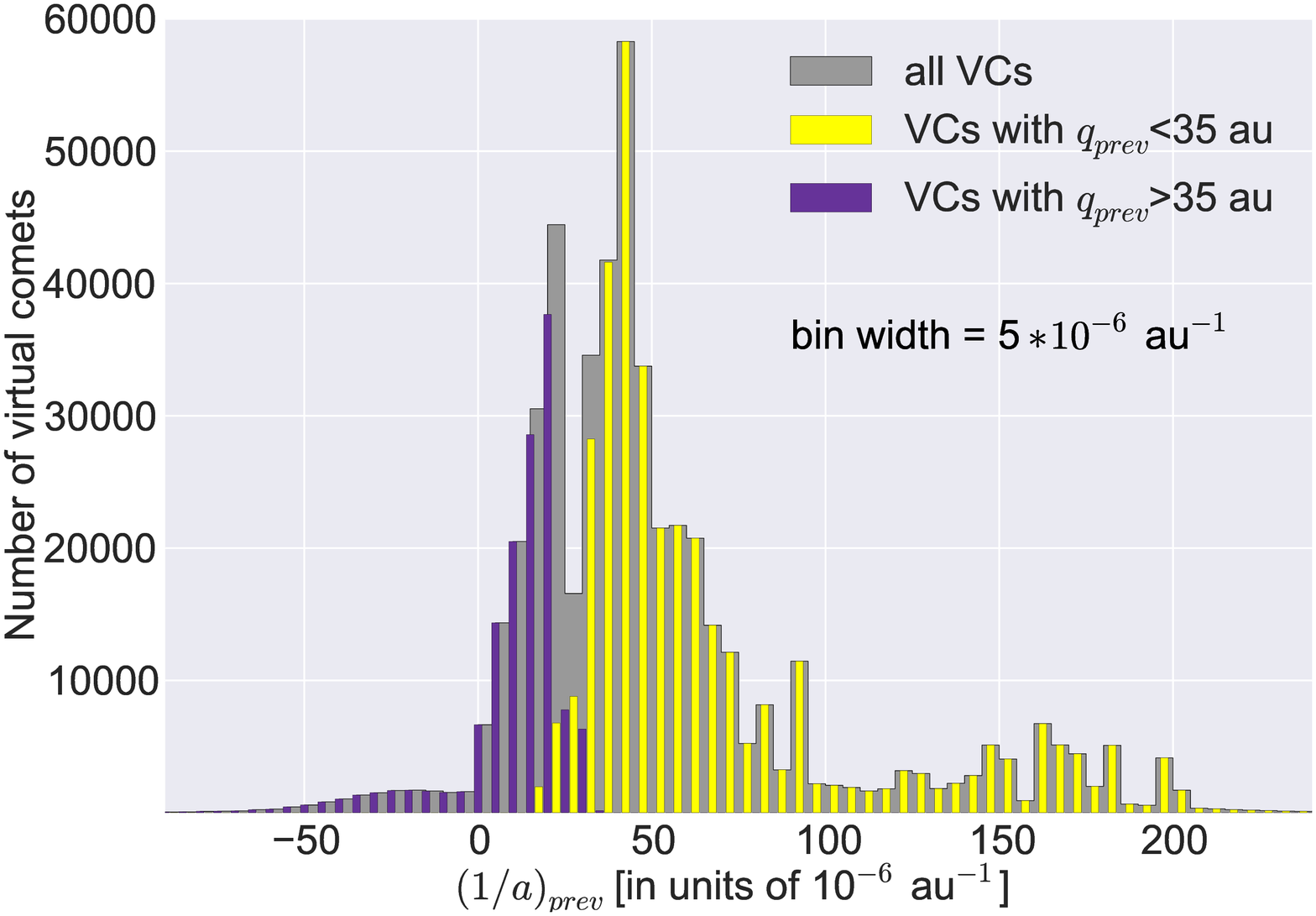} 
\protect\caption{\label{fig:previous_limit}
Distribution of dynamically new (violet histogram) and dynamically old (yellow histogram) of all VCs representing actual comets under consideration. The overall distribution of $1/a_{\rm prev}$ is shown by a grey histogram in all panels and is essentially equivalent to a dark green $1/a_{\rm ori}$-distribution given in the upper left panel in Fig.~\ref{fig:original_future}. Uppermost panel presents the situation when we keep the lower limit of 20\,au for dynamically new VCs and upper limit of 10\,au for dynamically old VCs, distribution of dynamically uncertain VCs are not shown in this plot. Middle panel illustrates the situation where both limits are united into a one border of 15\,au and in a consequence all comets are classified as dynamically old or dynamically new. In the lowermost panel this border between dynamically new/old is shifted to 35\,au from the Sun. Dividing numbers given in the vertical scale by 5001 we obtain numbers of comets in bins. } 
\end{figure}

The remaining 34\,comets in our sample have their swarms of VCs mixed or escaping at previous perihelion (note that none is fully escaping along a hyperbolic orbit).  We present their past dynamics in Table~\ref{tab:previous-mixed}. The structure of this table is similar to the previous one with one exception: after the name of a comet we inserted three additional columns containing the number of returning (R), the number of escaping (on elongated ellipses or hyperbolas, E) and separately the number of hyperbolic VCs (H). We show these comets in order of decreasing number of returning VCs. An asterisk mark ($^*$) appended to one of these numbers informs that this part of a swarm  contains the nominal solution. For the first 23~comets in Table~\ref{tab:previous-mixed} the nominal solution is inside the returning part of the swarm. For the remaining 11~comets we stopped the numerical integration synchronously with the fastest escaping VC to give  concise statistics of $1/a_{\textrm{prev}}$ and $q_{\textrm{prev}}$. Aphelion distance statistics is presented only for the elliptic part of the swarm, which is additionally marked by appending [R] symbol. 

These 11~comets from the bottom of Table~\ref{tab:previous-mixed} are all classified as certainly dynamically new according to our criterion but two cases should be commented. For C/2001~C1 as many as 1888 VCs are returning but the $q_{\textrm{prev}}$ statistics for this part (values recorded at previous perihelion) is described with deciles 35.7 -- 75.8 -- 117\,au. Similarly the distribution of a previous perihelion distance $q_{\textrm{prev}}$ for the returning part of the swarm of C/1935~Q1 (1306 VCs) is described as 23.0 -- 105 -- 249\,au. Therefore, in both cases the conclusion that these comets are certainly dynamically new seems to be fully legitimate. 
\newline Eight more comets from the upper part of Table~\ref{tab:previous-mixed} are  also marked as DN+ (five comets) or  DN (three comets) for the great majority of their $q_{\textrm{prev}}$values over 20\,au. Part~II of Table~\ref{tab:statistics_previous} gives  the   complete dynamical status statistics for comets with mixed or escaping swarms  of  VCs resulting from their evolution to the previous perihelion.

In contrast to Table~\ref{tab:previous-returning}, a large per cent of comets with an uncertain dynamical status in Table~\ref{tab:previous-mixed} (9 in total) are of a worse quality which is the main source of this inability to a dynamical status identification. Among comets with mixed swarms there are at least 10~dynamically old but only one of them is marked as DO+.

The distribution of the returning (in the past) part of our sample is additionally presented in detail in Fig.~\ref{fig:fig13}. We included here 77~comets: all 66 from Table~\ref{tab:previous-returning} and first 11~comets from Table~\ref{tab:previous-mixed}. 
The criterion here was to demand that more than 95~per cent of VCs are returning. Horizontal axis describes  $1/a_{\textrm{prev}}$ (and the corresponding semimajor axis itself at the top) while the vertical axis corresponds to $q_{\textrm{prev}}$, expressed in a logarithmic scale. For each comet we plot a single mark accompanied with a series of small dots representing the distribution of this comet VCs, limited to the 1$\sigma$ range. These serve as 'error bars' and they are invisible (covered with a mark) in case of a very compact VCs swarm. Comets from Table~\ref{tab:previous-returning} are plotted in black while the first 11~comets from Table~\ref{tab:previous-mixed} are plotted in grey (see also Table~\ref{tab:statistics_previous}). Full circles denote typical comets while open circles distinguish eight rare cases when an argument of perihelion of cometary orbit, expressed with respect to the Galactic disc plane and recorded at the previous perihelion falls into the second or fourth quarter. The Galactic disk tidal action is qualitatively different depending on the quarter of the argument of perihelion, see for example \citet{byl:1983,matese-w:1992,breiter-dyb:1996} for a detailed explanation. These eight LPCs are: C/1954~O2, C/1972~L1, C/1976~D2, C/1979~M3, C/1980~E1, C/2008~FK$_{75}$, C/2008~P1 and C/2013~L2. If a comet orbit is perturbed mainly by the Galactic centre tide the VCs run is in opposite direction, as it is clearly seen in four cases in Fig.~\ref{fig:fig13} (C/1954~O2, C/1976~D2, C/1980~E1 and C/2013~L2).

We have coloured four rectangles in Fig.~\ref{fig:fig13} plane. The magenta region includes 12~dynamically old comets outside the classical definition of the Oort spike. All other comets, satisfying the condition $1/a_{\textrm{ori}}<0.0001$au$^{-1}$ are traditionally (but erroneously) called 'new' comets (please keep in mind that all $1/a_{\textrm{ori}}$ values are practically identical to $1/a_{\textrm{prev}}$ ones). One can clearly distinguish an orange area, containing 33 dynamically old 	comets, green area with comets of the uncertain status and a blue area with only 16 dynamically new comets.
One should note, that in a group of 23~comets listed in Table~\ref{tab:previous-mixed} and omitted in this figure due to their highly mixed or fully escaping VC~swarms, there are  additional 14~dynamically new comets with highly elongated elliptic orbits (their aphelia are well above our escape limit of 120\,000\,au.) plus three more dynamically new comets with hyperbolic nominal previous orbits. Among the remaining six objects only one is marked as uncertain and five are dynamically old.

As pointed out above there are only three comets in our sample with slightly hyperbolic nominal previous orbits. These are C/1978~G2 ($1/a_{\textrm{prev}}=-21.39 \pm 37.71$ ), C/1997~P2  ($1/a_{\textrm{prev}}=-14.15 \pm 13.66$) and C/1942~C2 ($1/a_{\textrm{prev}}=-29.13 \pm 13.45$). However, the uncertainties of the derived $1/a_{\textrm{prev}}$ do not exclude that all these three comets are associated with the Solar system.

Summarising, an analysis of the past motion of 100~LPCs  shows that 50~of them are dynamically old, 32 comets are classified as dynamically new and  the remaining 18~comets cannot be clearly classified into one of the considered dynamical groups because their dynamical status seems to be uncertain in terms of the adopted definitions (see also Table~\ref{tab:statistics_previous}). It is important to stress that in our sample all comets with the original semimajor axis $a_{\textrm{prev}}\apprle$21\,000~au are dynamically old, with the only one small exception of C/2000~Y1~Tubbiolo. However, previous perihelion for the nominal orbit of  C/2000~Y1 is  10.3\,au and the entire swarm of VCs have $q_{\rm prev} < 15$\,au. Therefore, we can conclude that semimajor axes of dynamically new comets are above this limit of 21\,thousand au; it gives previous aphelia situated further than about 42\,000~au from the Sun. Comets with previous aphelia below this limit are exclusively dynamically old in the considered sample of large-perihelion comets. They are evidently immune to the so-called Jupiter-Saturn barrier, see~Paper~I for a critical discussion of this phenomenon. The exceptional case of a comet certainly dynamically old is C/2010~S1, with $a_{\textrm{prev}}$ as large as 40\,000~au. This comet has a very interesting and rather unusual past dynamics, see Section~\ref{sub:past-dynamics-C/2010 S1} for a detailed description. Furthermore, among comets with semimajor axes longer than 21\,thousand au there are 18~comets with an uncertain status but some of them have previous perihelia of the most of their VCs below 15\,au.

As stated above, the analysis of previous perihelion distances allowed us to divide our sample of 100~comets into 50 dynamically old, 32 dynamically new and 18 of an uncertain dynamical status. This makes a good opportunity to ask  how these groups of comets contribute to the overall histogram of $1/a_{\textrm{ori}}$ presented in the upper panels of Fig.~\ref{fig:original_future}. To answer this question we repeated a construction of this histogram using dynamically old and dynamically new parts of all swarms of VCs (500\,100 VCs in total), separately. It is worth to mention, that for all comets studied in this paper values of $1/a_{\textrm{ori}}$ and $1/a_{\textrm{prev}}$ are almost identical. This comes from the fact that during one orbital period and in the absence of strong stellar perturbation Galaxy itself modifies the orbital energy at almost an infinitesimal level. 
\newline The result is presented in Fig.~\ref{fig:previous_limit}. Uppermost panel presents a situation when we keep the lower limit of 20~au for dynamically new VCs and upper limit of 10~au for dynamically old VCs and completely omit dynamically uncertain VCs. Such an approach is fully consistent with the definition of a dynamical status that was used in a construction of Tables~\ref{tab:previous-returning}--\ref{tab:all-previous}. We noticed, that a striking local minimum in this histogram might be connected with our division into different dynamical states. In the middle and the lowermost parts of Fig.~\ref{fig:previous_limit} we applied a simplified dynamical status definitions, using one threshold value without any \textit{uncertainty margins}. In the middle panel we used a limiting value of 15\,au, and in the lowermost panel the threshold value is shifted to 35\,au.  This last attempt suggests that the local minimum in the $1/a_{\textrm{prev}}$ distribution might be better explained with a more distant threshold value, corresponding to the outer planetary zone radius.

\subsection{Evolution to the next perihelion}

Tables~\ref{tab:next-returning} -- \ref{tab:next-mixed}  describe future dynamics of comets investigated here after  their next orbital revolution. They were constructed in a similar way as Tables~\ref{tab:previous-returning} and \ref{tab:previous-mixed}. 
The most important difference here, comparing to the past motion description, is that in Tables~\ref{tab:next-returning} -- \ref{tab:next-mixed} we decided to omit 36 comets with whole swarms of VCs escaping along  an  hyperbolic orbit. Their $1/a_{\textrm{next}}$-values are almost identical to $1/a_{\textrm{fut}}$ and the remaining orbital parameters are not important here. These comets will be definitely lost from the Solar system. The remaining 64~comets are divided into two groups. The first group, presented in Table~\ref{tab:next-returning}, consists of 49~comets  with their swarms consisting of only returning VCs. The second group containing 15~comets with mixed swarms of VCs is listed in Table~\ref{tab:next-mixed}. Additional Table~\ref{tab:all-next} shows the full sample of 100~comets in a chronological order. The last column in Tables~\ref{tab:next-returning}--\ref{tab:all-next} repeats a dynamical status flag from tables describing previous orbits. The purpose of this repetition is to make an analysis of three consecutive perihelion passages of these comets easier for the reader.

Returning comets in Table~\ref{tab:next-returning} are presented in a decreasing order of their $1/a_{\textrm{next}}$, where the first 35~comets will return as observable LPCs outside the Oort spike. At the very top of this list is C/2002~A3. This comet has been captured by planetary perturbations to the shortest orbit with $a_{\textrm{next}}=162$\,au, see~Section~\ref{sub:particular-cases} for more details. In contrast to that, at the bottom of Table~\ref{tab:next-returning} we have 14~comets with the next semimajor axes longer than 10\,000\,au  ($1/a_{\textrm{next}}<$0.000100\,au$^{-1}$). These comets will return to the solar vicinity as the Oort spike members.  Nine of  these 14~comets will have their $q_{\textrm{next}}<$10\,au, two (C/1972~D2 and C/1999~N4) have their $q_{\textrm{next}}$ spread over the large interval of heliocentric distances and next two (C/2008~S3 and C/1999~S1) will have $q_{\textrm{next}}>$20\,au for their all VCs. 
\newline It is worth to mention that over 48 per cent of returning comets from Table~\ref{tab:next-returning} (24~per cent of all 100~LPCs studied in this paper) is visiting the interior of our planetary system at least during three consecutive perihelion passages analysed here. Next 19~comets from Table~\ref{tab:next-returning} will visit a zone of significant planetary perturbations twice, during the observed and next perihelion passages. Comet~C/2008~S3 is an interesting case of visiting the planetary zone only once. Its nominal previous and next perihelion distances are $q_{\textrm{prev}}=252$ and $q_{\textrm{next}}=24$~au, respectively. Thus, this comet certainly did not  experience any planetary perturbation before entering the observability zone and also will not suffer from strong planetary perturbations during the next passage. But it suffered moderate stellar perturbations - see Section~\ref{sub:particular-cases} for details.

Comets with mixed or escaping (elliptic or hyperbolic) swarms of VCs in their future motion are described in Table~\ref{tab:next-mixed}. Only the first comet in this table, C/2011~L6, can be regarded as returning -- the nominal orbit and 90~per cent of its clones will return but having next perihelion distances spread over a wide range of heliocentric distances. The remaining 14~comets should be treated as escaping on very elongated elliptic or hyperbolic orbits, up to seven of them rather definitely will leave our planetary system. C/2002~J5 is an interesting example of the escaping comet (in a sense of crossing the 120\,000~au escaping limit) but not necessarily leaving us. All VCs are escaping along the elliptic orbit but a swarm is rather compact and the greatest VC aphelion distance equals to only about 173\,000~au. Past and future motion of this comet was analysed in detail in our earlier paper \citep[see Fig.~5 therein]{dyb-kroli:2011}.

To summarise the above analysis of the future motion: 44~per cent of comets studied in this paper will leave the Solar system, 50~per cent will return in elliptic orbits and the remaining 6~per cent have their VCs swarms mixed and highly  dispersed (all having a slightly hyperbolic nominal orbit). Among the group of comets with returning orbits in the next perihelion as many as 15~comets will remain members of the classical Oort spike ($1/a_{\rm next}<0.000100$\,au$^{-1}$). However, only ten of them have $q_{\rm next}<11$\,au for the entire swarm of VCs.
We included here comet C/2011~L6 with a very disperse swarm (first comet in Table~\ref{tab:next-mixed}). 

\section{Long-term dynamical evolution through three consecutive perihelia (previous-observed-next)}\label{sec:prev-obs-next}

We notice that almost half of dynamically old comets and as many as two thirds of dynamically new comets will leave the Solar system in the future. Relative number of future hyperbolic  orbits drops to 25~per cent for dynamically old comets inside the range of $100 < 1/a_{\rm prev} < 200$ in units of $10^{-6}$\,au$^{-1}$, however  statistics for these comets are scarce.

It is worth noting   that as much as 22~comets from among the studied sample are certainly visiting  the  interior of our planetary system during all three consecutive perihelion passages investigated here (previous--observed--next). Two particular examples of such an evolution are discussed below, long-term evolution of C/2005~L3 is briefly described in Section~\ref{sub:2005l3_2006s3} and details of  the  past dynamics of C/2010 S1 are presented in Section~\ref{sub:past-dynamics-C/2010 S1}.

The striking coincidence is that exactly half of all comets studied here appear to be dynamically old and also exactly 50 comets (with only 23 ones in common) will leave our planetary system permanently in the future. Looking for a balance of planetary action we see that almost 50~per cent of dynamically old comets is ejected into interstellar space by planets while only 36~per cent of the observed stream of dynamically new comets is captured into a more tightly bound orbit.

Additionally, we count that as many as 23~comets have $1/a_{\rm next}\geqslant 200 \cdot 10^{-6}$\,au$^{-1}$, that is semimajor axes shorter than 5\,000\,au. The same number of 12~comets have $100 \leqslant 1/a_{\rm next}< 200$ (in the same units of $10^{-6}$\,au$^{-1}$) as in the observed perihelion passage. 
To draw a more precise evolution of observed $(1/a)$-distribution far outside the Oort spike, we should, however, take into account comets having original semimajor axes shorter than 5\,000\,au.
The dynamical evolution of comet C/2002~A3 with the shortest future semimajor axis of about 162\,au is discussed in Section~\ref{sub:particular-cases}. Six more comets, C/1991~C3, C/2007~D1, C/1993~K1, C/2000~CT$_{54}$, C/1974~V1 and C/1999~U1, have next semimajor axes smaller than 2\,000\,au (orbital periods shorter than 90\,thousand yrs).

\subsection{C/2005~L3 McNaught and C/2006~S3 LONEOS -- two probably large comets having very different dynamical evolution}\label{sub:2005l3_2006s3}

Both comets were detected further than 10\,au from the Sun and were among the brightest comets beyond the Jupiter orbit. C/2005~L3 ($q_{\rm obs}=5.59$\,au) was observed 8.7\,years whereas C/2006~S3 ($q_{\rm obs}=5.14$\,au) was followed 16.6\,yrs including pre-discovery detections. As was mentioned in Section~\ref{sec:sample_new_solutions}, it is expected that nuclei of both these comets are large, between sizes of two remarkable comets: 1P/Halley and C/1995~O1 Hale-Bopp \citep{sarneczky-et-al:2016}. In Table~1 of their paper both these comets are defined as 'dynamically new' according to the criterion that each comet having original semimajor axis greater than 10\,000\,au is a dynamically new one. As it was demonstrated earlier, this criterion does not guarantee that a given comet was relatively close to the Sun for the first time. In other words, using only the original semimajor axis criterion we still have to deal with comets quite different dynamically and physically.

As a result of long time intervals of data, orbits of these comets are of highest quality of 1a+, and previous and next swarms of these orbits are very compact. Therefore dynamical status of each of these comets is firmly determined. It turns out, that C/2006~S3 having previous semimajor axis of about 100\,thousand au ($1/a_{\rm prev}=9.5\pm 0.3$ in units of $10^{-6}$\,au$^{-1}$)  is a dynamically new comet, extremely weakly bound to the Solar system (entire previous swarm of VCs is escaping). In contrast to that, C/2005~L3 is a dynamically old comet since it was in previous perihelion inside a planetary zone, therefore its surface was exposed to the Solar radiation. Additionally, during its previous passage inside the planetary system an orbit of C/2005~L3 might have been changed.  In particular, we know nothing about the semimajor axis of its orbit just before the previous entrance to the planetary zone. We can only state that it had semimajor axis of about 16.3\,thousand au ($1/a_{\rm prev}=61.7\pm 0.2$) and passed its previous perihelion within a distance of $7.68\pm 0.02$\,au from the Sun in the assumed model of Galactic and stellar perturbations and neglecting planetary perturbations during the previous perihelion passage.

Future dynamics of these comets is also quite different. C/2006~S3 is now leaving the Solar system in a slightly hyperbolic orbit. In contrast, C/2005~L3 is moving on a more tight orbit than before with a semimajor axis of about 3.4\,thousand au ($1/a_{\rm prev}=292.1\pm 0.2$ in units of $10^{-6}$\,au$^{-1}$) and will pass next perihelion at a similar distance  as  during the observed one.

Summarizing, comet C/2006~S3 was only once inside planetary zone, while C/2005~L3 is an example of  a comet potentially observable in all three consecutive perihelion passages investigated here. 

Both these types of evolution are quite common in the analysed sample of LPCs. Table~\ref{tab:previous-returning} reveals 24~dynamically old comets (including C/2005~L3) with the next perihelion distances smaller than 10~au. These comets are potentially observable in at least three perihelia. On the other hand, 22~other comets visited the inner part of planetary zone only during the observed perihelion and they fully deserve to be called \textit{one time visitors}. This subgroup of dynamically new comets is rather peculiar because they are almost all leaving solar system  on  slightly hyperbolic orbits like C/2006~S3. We found only one exception of C/2008~S3 which passed its previous perihelion at the distance greater than 200\,au from the Sun, in the observed perihelion was as close as 8.02\,au and in the next perihelion passage will be further than 20\,au from the Sun (for more details see Section~\ref{sub:particular-cases}). It is an interesting coincidence that in our sample of LPCs another object, C/2009~P2 Boattini ($q_{\rm obs}=6.55$\,au), also discovered in the course of the Catalina Sky Survey, has almost the same $1/a_{\rm prev}$ of $20.96\pm 1.40$ and a very similar observed orbital inclination: both comets are moving on retrograde orbits with an inclination of 162.7\degr (C/2008~S3) and 163.5\degr (C/2009~P2). C/2009~P2 also has very similar  previous and next orbit evolution to C/2008~S3. Only its next perihelion distance will be smaller.

\subsection{Interesting long term dynamics of C/2010~S1~LINEAR}\label{sub:past-dynamics-C/2010 S1}

\begin{figure}
\includegraphics[clip,angle=270,width=1.0\columnwidth]{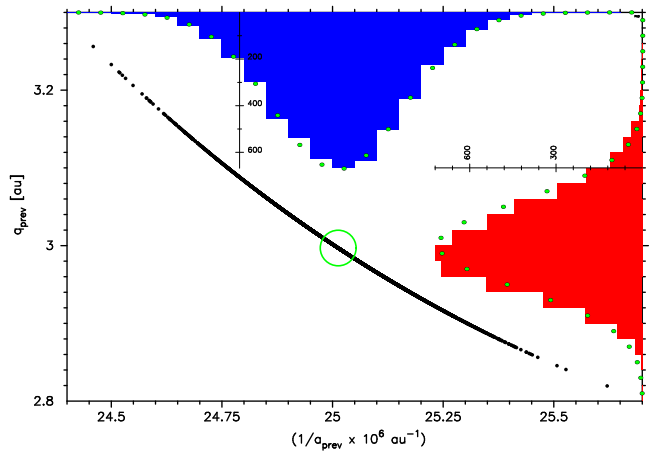}
\caption{\label{fig:hist-C/2010-S1}Two dimensional  dispersion of VCs for C/2010~S1 in a $1/a_{\textrm{prev}} \times  q_{\textrm{prev}}$ plane augmented with two marginal distributions of these parameters. Green dots present Gaussian fitting to these distributions. Nominal orbit parameters are in the centre of a large green circle.}
\end{figure}

\begin{figure}
\includegraphics[clip,angle=270,width=1.0\columnwidth]{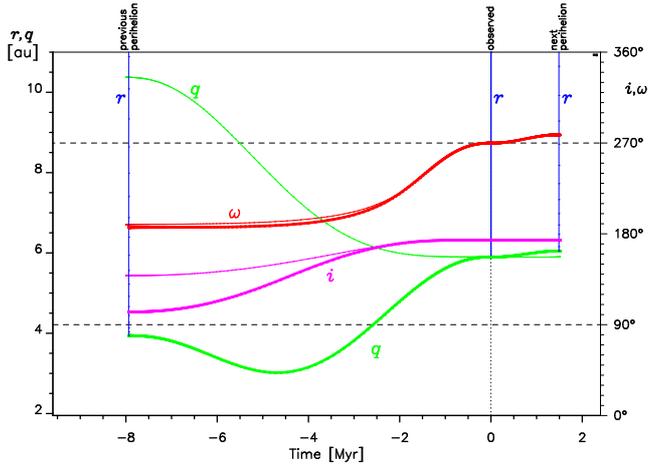}
\caption{\label{fig:tf11-gal-gal}Dynamical evolution of a nominal orbit of C/2010~S1 in two different models of Galactic perturbations: disc + centre (thick lines) and disk tide alone (thin lines). The horizontal time axis extends from the previous perihelion through the observed one up to the next perihelion passage. The left vertical axis is expressed in au and corresponds to the perihelion distance plot ($q$, green lines) as well as the heliocentric distance plots ($r$, thin, vertical blue lines). The right vertical axis is expressed in degrees and describes the evolution of the osculating inclination ($i$, magenta lines) and the argument of perihelion ($\omega$, red lines). Both these angular elements are expressed in the Galactic frame. }
\end{figure}

\begin{figure}
\includegraphics[clip,angle=270,width=1.0\columnwidth]{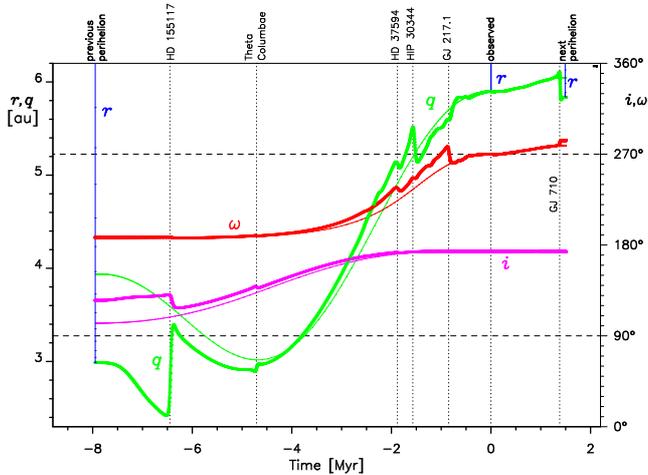}
\caption{\label{fig:tf11-gal-stars}Dynamical history and future of a nominal orbit of C/2010~S1. This figure is organized in a similar way that the previous one but now we compare the results obtained from a full force model (Galactic and stellar perturbations) with a simplified one, where stellar perturbations are omitted. One can observe small differences in angular elements and a remarkable change in  the   perihelion distance.}	
\end{figure}

This comet was discovered at a heliocentric distance of 8.85\,au from the Sun and 2.7\,yrs prior to the perihelion. From its discovery, C/2010~S1 was followed during 4.8 years through its perihelion (5.9\,au) up to 8.02\,au from the Sun, and more than 8.5\,thousands of positional measurements were obtained. Together with C/2005~L3 and C/2006~S3, this comet nucleus is expected to be large in size \citep{sarneczky-et-al:2016}.
\newline Such a rich data material covering several years allows to determine the orbit of the highest quality class of 1a+. Therefore, it is not surprising that previous and next orbits are so firmly fixed for a given dynamical model (see discussion below).

As it is shown in Table~\ref{tab:previous-returning} a whole swarm of C/2010~S1 VCs is returning. This swarm is also very compact but has a slightly non-Gaussian distribution of $q_{\textrm{prev}}$, as depicted in Fig.~\ref{fig:hist-C/2010-S1}. This figure shows a distribution of all 5001\,VCs of this comet in a $1/a_{\textrm{prev}} \times  q_{\textrm{prev}}$ plane (central black points) with a nominal orbit point in the centre of a green circle. Marginal distributions of  $1/a_{\textrm{prev}}$ and $q_{\textrm{prev}}$ are shown in blue and red, with the best fitting Gaussians depicted with small green points. The whole swarm of VCs of this comet is hidden under the nominal mark in Fig.~\ref{fig:fig13}.

The observed Galactic inclination of C/2010~S1 equals 174\degr ~-- its orbit is retrograde and lies almost in the Galactic disc plane. Such a configuration  causes the dynamics of this comet  to be  very sensitive to perturbations from the Galactic centre (in contrary to most of other comets, which evolution is mainly driven by a Galactic disc tide). Dynamical evolution of this orbit from the previous perihelion, through the observed one up to the next perihelion passage is presented in Figs.~\ref{fig:tf11-gal-gal}--\ref{fig:tf11-gal-stars}. The  first of these plots presents a remarkable difference between this orbit evolution when two different Galactic perturbations models are used: thin lines depict results of a disc tide action only while thick lines describe the dynamics under a full Galactic model (disc + centre). While the evolution of angular elements is rather similar in both cases, changes in the perihelion distance are completely different. If we take into account only a Galactic disc tide effect, the perihelion distance rises above 10\,au when going backward to the previous perihelion. After applying the full model of Galactic perturbations we observe a decreasing of the perihelion distance to $q_{\textrm{prev}}=3.9$\,au. The observed argument of perihelion (with respect to the Galactic plane) of this comet equals 269\degr (red lines in the figure). 
It is emphasized by horizontal dashed lines in Figs.~\ref{fig:tf11-gal-gal}--\ref{fig:tf11_2011l6a1_1} that the beginnings of the second and fourth quarter of $\omega$ (values of 90\degr and 270\degr) are important from the point of view of the Galactic disk perturbations , see for example \citet{byl:1983,matese-w:1992,breiter-dyb:1996} for a detailed explanation.
Typically, such a value coincides with the minimum in the curve representing perihelion distance evolution under Galactic perturbations, as it is depicted with the thin green line. When a full model is applied (thick green line) the situation might be quite different as in this particular case.	

Fig.~\ref{fig:tf11-gal-stars} also presents a dynamical evolution of C/2010~S1 but now we compare  the orbital evolution under the Galactic perturbation with and without passing stars included into a dynamical model.
We use here the same symbols and colour meanings as in Fig.~\ref{fig:tf11-gal-gal} but now thick lines describe the evolution under the simultaneous Galactic and stellar perturbations while thin lines were obtained by omitting the action from the stars. We also added several vertical, dashed lines depicting moments of the closest approach of certain stars to the comet. Names of these perturbers are placed at the top of this picture. One can notice that due to the combined action of Theta Columbae (HD\,42167, over 4 solar masses) and relatively strong perturbation from HD\,155117 (1.3 solar masses but coming much closer), the previous perihelion distance of C/2010~S1 is reduced to $q_{\textrm{prev}}=3.0$\,au. This figure also shows many smaller and rather  short changes in orbital elements caused by stars but only a few most visible are identified by dotted vertical lines.

Such a type of dynamical evolution where comet is deeply inside planetary zone in all three perihelion passages ($q < 10$\,au) can be observed in more than 20 per cent of comets from the investigated sample. However, particular inclination of C/2010~S1 makes its evolution rather unique due to a strong influence of the Galactic bulge.

\subsection{Comments on a few other particular cases}\label{sub:particular-cases}

\begin{figure}
\includegraphics[clip,angle=270,width=1.0\columnwidth]{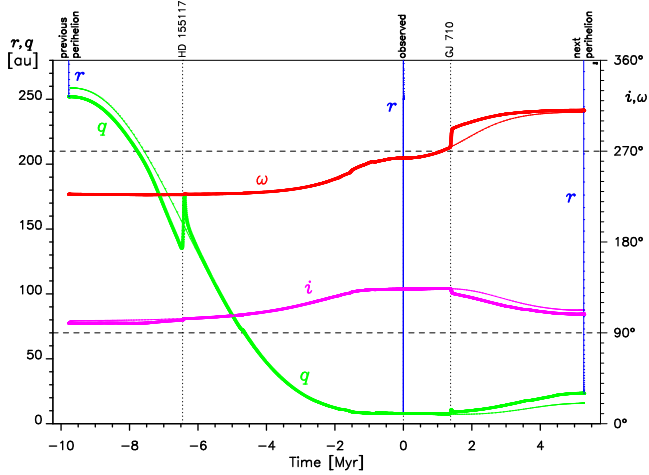}
\caption{\label{fig:tf11_2008s3a5_1}15 Myrs of past and future motion of C/2008~S3 -- nominal orbit 
evolution is presented. This plot is prepared in a similar way as Fig. \ref{fig:tf11-gal-gal}. A 	comparison of the results of calculations with (thick lines) and without (thin lines) stellar 	perturbations reveal a series of moderate comet-star interactions in this case.}
\end{figure}

\begin{figure}
\includegraphics[clip,angle=270,width=1.0\columnwidth]{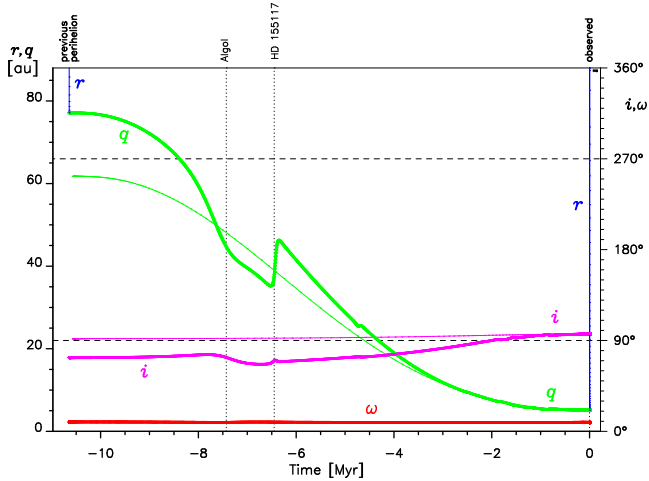}
\caption{\label{fig:tf11_2002a3a2_1}Over 10 Myrs of a past dynamical evolution of a nominal orbit of C/2002~A3 presented here in the same manner as in the previous figure. We compare here the results of a 
full force model (thick lines) with the simplified model, where only a Galactic disc tide is included.}
\end{figure}

C/2008~S3 LINEAR ($q_{\rm obs}=8.015$\,au, orbital quality class: 1a+) is an example of a comet that 
certainly will return in the next perihelion but at a large distance of 23.5\,au from the Sun. The uncertainty of this determination is very small, of about 0.3\,au in a given dynamical model. Such a big value of $q_{\textrm{next}}$ is only partially produced by a weak perturbation from Gliese~710, a star rather small but passing very close to the Sun in the next 1.4\,Myr. This is clearly depicted in Fig.~\ref{fig:tf11_2008s3a5_1}. However, a more pronounced stellar perturbation happened in the past motion of this comet. The same massive perturber that disturbed C/2010~S1's   motion, namely HD\,155117 also left a trace here. Without its action this comet would have $q_{\textrm{prev}}=260$\,au, a little bit more than the nominal value of about 253\,au given in Table~\ref{tab:previous-returning}. It seems that C/2008~S3 is a rare example of a comet that was only detectable (and deeply inside the planetary zone) in the observed perihelion, however does not leave the Solar system in the next perihelion (similarly as C/1999~F1, C/2009~P1 or C/2003~A2).

\begin{figure}
\includegraphics[clip,angle=270,width=1.0\columnwidth]{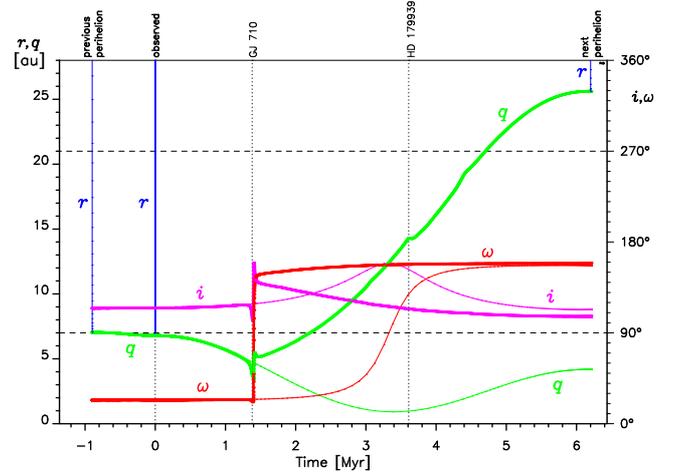}
\caption{\label{fig:tf11_2011l6a1_1}Long term dynamical evolution of a nominal orbit of C/2011~L6. 
Presented is a comparison between the results of a numerical integration of full equations of motion 
(thick lines) and the restricted case, where stellar perturbations are omitted. The arrangement of this 
figure, meaning of colours and symbols are the same as in previous picture.}
\end{figure}

Two interesting features can be found in the long term motion of another comet, C/2002~A3 LINEAR ($q_{\rm obs}=5.151$\,au, orbital quality class: 1a). The  first fact, already mentioned earlier is that, having  an  original orbit of $a_{\textrm{ori}}\cong$50\,000\,au it was captured by planetary perturbations into the shortest future orbit among studied in this paper -- this comet passed Jupiter at the distance of 0.502\,au on 2003~January~22. Its future semimajor axis is as small as 162\,au (notice that $a$-uncertainty is below 1\,au in the assumed model of motion) with $q_{\textrm{next}}=5.15$\,au and a period of only 2\,026~years.
The second interesting effect is shown in Fig.~\ref{fig:tf11_2002a3a2_1}. Due to the observed large inclination with respect to the Galactic plane this comet is also sensitive to the perturbation from the Galactic centre (as C/2010~S1). This part of disturbing forces caused a change of its orbit from prograde at previous perihelion to retrograde when this comet was observed. Such a reverse of the sense of motion with respect to the Galactic disk plane is impossible without Galactic centre action. 

\noindent To illustrate this effect we present in Fig.~\ref{fig:tf11_2002a3a2_1} a comparison of the backward numerical integration of this comet motion with two different force models: only disc tide model is shown with thin lines whereas thick lines describe the full model of simultaneous Galactic disc and centre action augmented with stellar perturbations. One can notice again an effect of HD\,155117 but also a wide in time disturbance caused by the Algol system. This perturber never comes too close (it passed farther than 3\,pc from the sun) but its total mass of six solar masses and a very small velocity of 4\,km\,s$^{-1}$ causes significant orbit changes for certain comets. In this case, stellar perturbations from HD\,155117 and Algol system reveal their substantial action only in a perihelion distance evolution while angular elements are almost unperturbed. There  are   also a dozen of weak stellar perturbations from nearby stars during the first two million years of backward motion of this comet but they are almost invisible due to the wide range of vertical scale of this plot. The future motion of C/2002~A3 in a very short orbit is also impossible to show with the time scale used in this picture.

Another interesting case is C/2011~L6 Boattini ($q_{\rm obs}=6.788$\,au, orbital quality class: 1b) due to its violent future motion. It has the largest future orbit among returning LPCs analysed here and its next orbit is described in the first row of Table~\ref{tab:next-mixed}. The orbital evolution of swarm of 5001\,VCs gives  more than 90 per cent of returning clones, only 483~VCs cross the assumed escape limit of 120\,000\,au and only eight of them have  an  eccentricity slightly greater than 1.0 at that distance. C/2011~L6 will pass farther than 20\,au from the Sun during the next perihelion but a detailed analysis showed that this is the result of a strong stellar perturbation from Gliese~710 as it is shown in Fig.~\ref{fig:tf11_2011l6a1_1} where the same colour coding is used as in Fig~\ref{fig:tf11_2002a3a2_1}. This small star  has  a mass of 0.6\,solar masses but it will pass very close to the Sun in 1.4\,Myr so  an  arbitrary close passage is possible for LPCs. A more detailed analysis of Gliese~710's  future approach to the Sun based on preliminary results of  the   Gaia mission was recently published by \citet{berski-dyb:2016}.
	
\noindent In the present calculations Gliese~710  will not change the semimajor axis of C/2011~L6's  orbit but will significantly increase its next perihelion distance from 3.7\,au up to over 20\,au. Notable local changes in angular elements are also visible in Fig.~\ref{fig:tf11_2011l6a1_1}. Several other stars will weakly perturb the  motion of this comet in the future but only the action of HD\,179939 can be yet recognized in this plot.

\section{Summary and conclusions}\label{sec:summary}

We have constructed a complete and fairly large sample of 94~distant LPCs ($q_{\rm obs}>3.1$\,au) discovered in the period 1901--2010 and having original semimajor axes longer than 5\,000\,au; six more comets detected more recently were added for richer statistics. 

Investigation presented here starts from the orbit determination using homogeneous methods of data treatment for each   analysed comet. In addition, by limiting only to large-perihelion LPCs, the impact of the NG-effects on orbital solutions was minimized although they are  still recognizable in some cases. All these make the presented statistics meaningful.

It might be surprising, but despite the large (or very large) perihelion distances of analysed LPCs, NG-effects were successfully determined for sixteen LPCs from their positional data. Generally, we used the  standard g(r)-function for water sublimation, only in two cases we deduced that the formula based on CO sublimation is more appropriate, though the adequacy of different formulae of g(r)-like functions   were usually barely testable on the positional data of these LPCs. All 100~LPCs orbits were than propagated through the planetary system to the past and to the future to obtain \textit{original} and \textit{future} barycentric orbit elements, recorded at 250~au from the Sun.

Next, we have studied the long-term dynamics of LPCs (including three perihelion passages) under the influence of both the radial and the vertical components of the Galactic tidal field as well as all currently known passing stars. We decided not to use recently published preliminary Gaia mission results \citep{GaiaDR1:2016} since they are significantly incomplete from the point of view of potential stellar perturbers of cometary motion. 

Dynamical investigation of each comet was based on a swarm of 5001\,VCs, including the nominal orbit. This allows us to include the uncertainties of orbital elements to statistical analysis at each step of orbital evolution. During the dynamical evolution back in time we focused on two stages: at the moments when LPCs reached  a  distance of 250\,au before entering  the planetary zone (\textit{original} orbits) and the moments of previous perihelion passages (\textit{previous} orbits). Similarly, for the future dynamical evolution we analysed orbits at the distance of 250\,au after leaving the planetary zone (\textit{future} orbits) and at the moments of next perihelion passages (\textit{next} orbits). 

\vspace{0.2cm}

Our most important conclusions are:

	\begin{itemize}
		\item The observed distribution of planetary perturbations, $\delta (1/a)$, has a spectacular  
decrease around zero in the range of $-0.000025$\,au$^{-1} < \delta (1/a) <  +0.000025$\,au$^{-1}$. In this paper, we widely analysed possible sources of such a phenomenon. Perhaps, it arose from the coincidence of a particular cometary stream with specific planetary configurations or it is an extraordinary statistical fluctuation. 
		\item Despite this deficit of negligible perturbations, we observe statistically significant 
		percentage of comets suffering small planetary perturbations: 30 per cent of analysed LPCs have 
		$|\delta (1/a)| < 0.000075$\,au$^{-1}$, and 49 per cent have  $|\delta (1/a)| < 0.000125$\,au$^{-1}$. 
		It shows that the Jupiter-Saturn barrier is leaky for this population of LPCs.
		\item Study of the past dynamics of LPCs to the previous perihelion passage clearly shows that 
		dynamically new comets may appear only when $1/a_{\rm ori}<50\cdot 10^{-6}$\,au$^{-1}$ 
		($a>20\,000$\,au). On the other hand, dynamically old comets are completely not present
		only when $1/a_{\rm ori}<25\cdot 10^{-6}$\,au$^{-1}$ ($a>40\,000$\,au).
		\item The detailed comparison of the dynamical status derived for each individual comet in this
		 paper with that obtained in Paper~I  shows that our dynamical conclusions are fully
		 confirmed. In Paper~I we analysed orbits of 64~large perihelion LPCs. Only 59 of them are in common with
		 the present paper since we restricted here to comets with $q>3.1$\,au. This condition
		 excluded 5~comets with slightly smaller perihelion distances (C/1974~F1, C/1992~J1, C/1997~J2, C/1999~Y1
		 and C/2001~K3).
		 Considering these 59 comets, we fully confirmed the dynamical status of all
		 comets classified by us as dynamically old in Paper~I. From a group of 28 LPCs recognized
		 as dynamically new, four comets (C/1978~A1, C/2003~S3, C/2004~T3 and C/2007~Y1) are now
		 shifted to a group of comets having an uncertain status, mainly due to a slightly more restrictive criterion used here. Only in the case of C/1978~A1 stellar perturbations
		 changed slightly orbits of a few VCs, what decreased a percentage of dynamically new VCs from 50.9 in
		 Paper~I to 48 in this paper, causing a mentioned reclassification. 
		\item Fifty of LPCs studied here are dynamically old, 33 comets are classified as dynamically new and 
		the remaining 17 comets have uncertain dynamical status. Thus, the overall statistics 
		is noticeably changed in comparison to Paper~I: fraction of dynamically old comets increased from 41 to 50 per cent, of dynamically new comets decreased from 48 to 33 per cent, and the percentage of uncertain cases increased from  10 to 17. This latter change is mainly a result of a slightly more
		restrictive criterion used in the present paper.
		\item Every third dynamically old comet is leaving the Solar system on hyperbolic orbits while 
		the same happens with a half of the remaining part of the sample, containing 
		dynamically new and uncertain comets.
		\item Statistical shortening of semimajor axes for comets with the returning swarms in the next 
		perihelion is observed: as many as 23~comets have semimajor axes shorter than 5\,000\,au. The same 
		number of 12~comets have semimajor axes between 10\,000\,au and 5\,000\,au as during the observed 
		perihelion passages. 
		\item The observed distribution of $1/a_{\rm ori}$ reveals a local minimum separating dynamically new comets from dynamically old comets. As far as we know this and the first one of our findings were never discussed before. 
		\item In \cite{dyb-kroli:2015} we summarized our results for 108 Oort spike comets studied by us at that time. This paper adds another  30 comets to our sample but the statistics remain almost unchanged: now we have 60 dynamically old comets (43.5 per cent),  53 dynamically new ones (38.5 per cent) and 25 objects with an  uncertain status (18 per cent). But this statistics from the point of view of all known Oort spike comets is still incomplete. However, among them we have complete sub-sample with $q > 3.1$\,au (89 comets) consisting of 43 per cent of dynamically old, 42 per cent of dynamically new and 15 per cent of an uncertain status.
		\item Since among comets with recognisable dynamical status we obtain more or less a half of dynamically old objects it is necessary to decrease by 50 per cent the stream of dynamically new objects with respect to the whole Oort spike objects. This should also decrease the estimated number of the Oort cloud comets by a factor of two. 
	\end{itemize}

\vspace{0.1cm}

Additionally, long-term dynamical studies reveal a large variety of orbital behaviour. Generally, we can draw a few more comments about long-term evolution and the role of passing stars in this evolution:

\begin{itemize}
			\item As many as 22~comets are certainly visiting the  interior of our planetary system during all three 
			consecutive perihelion passages investigated here (previous--observed--next). 
			We described in detail two interesting examples of such evolution for comets C/2005~L3 and C/2010~S1.
			\item More than 20~other comets visited the inner part of the  planetary zone only in the
			observed perihelion and these are of course dynamically new comets. 
			This subgroup of \textit{one time visitors} is very peculiar because they are almost all leaving solar
			system in a slightly hyperbolic orbits except the one case of C/2008~S3.
            \item This research fully confirmed our conclusion drawn in several our earlier papers: among known stars there is no significant perturber of past motion of any of investigated
             comets. Since all known potential stellar perturbers were included into our dynamical model this study
             shows noticeable but qualitatively insignificant influence of the passing stars on the motion 
             of observed LPCs. Stellar perturbations do not change the dynamical status of these comets. 
             This conclusion might be revised of course in the future after a complete data release from the Gaia
             mission.
			\item During the future evolution to the next perihelion perturbations of Gliese~710
			might be very important. We noticed several interesting cases among analysed LPCs which require more
			specific research with new data taken from Gaia mission. 
			Generally, we can expected that gravitational influence of Gliese~710 on the motion of LPCs 
			can be more spectacular than we showed in this study \citep{berski-dyb:2016}.
		\end{itemize}

All our orbital results (osculating heliocentric orbits, original and future barycentric orbits, and past and next solutions) are tabulated in a structured form and are included as supplementary material.

\section*{Acknowledgements}

The orbital calculation was partially performed using the numerical orbital package developed by
the Solar System Dynamics and Planetology Group at SRC PAS. This research has made use of NASA's Astrophysics Data System Bibliographic Services and was partially supported from the project 2015/17/B/ST9/01790 founded by National Science Centre in Poland.

{We are very grateful to the reviewer, Luke Dones, for his valuable comments and very detailed suggestions which allowed us to improve this paper. We also thank Andrzej M. So{\l}tan for his suggestions connected with statistical approach to some our results}.

\bibliographystyle{mnras}

\bibliography{moja23b.bib}


{\onecolumn{
		
		\appendix
		
		\begin{landscape}
			\section{Description of observational material and orbital quality assessment for comets with recalculated orbits in this paper}
			{\setlength\LTcapwidth{1.00\textwidth} 
				\setlength\LTleft{0pt}               
				\setlength\LTright{0pt}              
				\setlength{\tabcolsep}{1.8pt}
				{\footnotesize {
\begin{longtable}{lcccrrcccccccc}
							\caption{\label{tab:observational_material}
								Characteristics of an observational material and the new quality assessment for 31~near-parabolic comets (see references in the last col.). The observational material of comets taken for osculating orbit determination is
								described in Cols.\,$[4]-[8]$ and $[13]$, whereas the new orbit quality assessment resulting from the osculating orbit based on these data sets is shown in Cols.\,$[10]-[12]$. The second and third cols. show an osculating
								perihelion distance and perihelion time. The data distribution relative to a perihelion passage is presented in
								Cols.\,$[7]$ \& $[8]$, where 'pre' ('post') means that all observations were taken before (after) perihelion passage; 'pre+' means that significantly more pre-perihelion measurements were available,  Col.\,$[9]$
								shows the type of the best model possible to determine from the full interval of data; in the case of C/2003 WT$_{42}$ LINEAR the model better dedicated for past evolution is given in the second row.
								Cols.\,$[10]$ and $[11]$ give the Q$^*$ for GR and NG~models, respectively. The orbital class resulting from the model described in Col.\,$[9]$ and determined according to the new quality assessment is given in Col.\,$[12]$. Reference  Cat1 (last columns) refers to \citet{krolikowska:2014}. } \\
							\hline\hline
							Comet & q$_{\rm osc}$ & T            & Observational arc        & No  & Data      & Heliocentric   & Data  & Type & \multicolumn{2}{c}{Q$^*$} & New     & rms [arcsec]  & Ref  \\
							name  &            &                 & dates                    & of  & arc span  & distance span  & type  & of   &   GR    & NG              & orbital & /no of &      \\
							& {[}au{]}   & {[}yyyymmdd{]}  & {[}yyyymmdd--yyyymmdd{]} & obs & {[}yr{]}  & {[}au{]}       &       & model&   model & model           & class   & res.  &      \\
							$[1]$ & $[2]$  & $[3]$     & $[4]$              & $[5]$ & $[6]$& $[7]$     & $[8]$& $[9]$&$[10]$&$[11]$&$[12]$& $[13]$& $[14]$\\
							\hline
							\endfirsthead
							\caption{continued.}\\
							\hline\hline
							%
							%
							Comet & q$_{osc}$  & T               & Observational arc        & No  & Data      & Heliocentric   & Data  & Type & \multicolumn{2}{c}{Q$^*$} & New     & rms [arcsec]  & Ref  \\
							name  &            &                 & dates                    & of  & arc span  & distance span  & type  & of   &   GR    & NG              & orbital & /no of & \&   \\
							& {[}au{]}   & {[}yyyymmdd{]}  & {[}yyyymmdd--yyyymmdd{]} & obs & {[}yr{]}  & {[}au{]}       &       & model&   model & model           & class   & res.  & Notes \\
							$[1]$                            & $[2]$  & $[3]$     & $[4]$              & $[5]$ & $[6]$& $[7]$     & $[8]$& $[9]$&$[10]$&$[11]$&$[12]$& $[13]$& $[14]$\\
							\hline
							\endhead
							\hline 
							\endfoot
							\\
							\multicolumn{14}{c}{\bf New orbital solutions for Oort spike comets based on updated data set in comparison to  Sample B of \citet{krolikowska:2014} } \\  \\
							C/2000 SV$_{74}$ LINEAR          & 3.542  & 20020430  & 20000905--20050512  & 2207 & 4.7  & 6.26--9.51  & full & NG  & 9.0 & 9.0 & 1a+& 0.67/4403 & Cat1, new \\
							C/2003 WT$_{42}$ LINEAR          & 5.191  & 20060410  & 20031030--20080525  & 2564 & 4.6  & 8.26--7.66  & full & GR  & 9.0 & --  & 1a+& 0.44/5071 & Cat1, new \\
							&        &           & 20031030--20060804  & 1692 & 2.8  & 8.26--5.26  & pre  & GR  & 9.0 & --  & 1a+& 0.39/3339 & ~~~~~~~~~new \\
							C/2006 S3   LONEOS               & 5.131  & 20120416  & 19991013--20160505  & 6480 &16.6  & 26.1--11.4  & full & NG (CO) & 9.0 & 9.0  & 1a+& 0.50/10589& Cat1, new \\
							C/2008 FK$_{75}$ Lemmon-Siding S.& 4.511  & 20100829  & 20080331--20131106  & 3702 & 5.6  & 8.24--9.51  & full & NG (CO) & 9.0 & 9.0  & 1a+& 0.48/7376 & Cat1, new \\
\bottomrule
						\end{longtable}
				}}
			}
		\end{landscape}

		\newpage
		
		\section{Osculating orbital elements (heliocentric) }
		
		{\setlength\LTcapwidth{1.00\textwidth} 
			\setlength\LTleft{0pt}               
			\setlength\LTright{0pt}              
			\setlength{\tabcolsep}{2.5pt}{
				\begin{longtable}{ccrrrrrrr}
					\caption{\label{tab:orbit_osculating} Orbital elements of osculating heliocentric orbits for all 
						comets described in Table~\ref{tab:observational_material}.
						The successive columns contain :
						$[1]$ -- Comet designation, $[2]$ -- Epoch, i.e. osculation date,
						$[3]$ -- perihelion time [TT], $[4]$ -- perihelion distance, $[5]$ -- eccentricity, $[6]$ -- argument of perihelion,  $[7]$ -- longitude of the ascending node, $[8]$ -- inclination,    $[9]$ -- reciprocal semi-major axis in units of $10^{-6}$\,AU$^{-1}$. The angular elements refer to the equator and equinox of J2000. 
					}\\
					\hline\hline
					Comet      & Epoch         & T                   & q            &  e         & $\omega$      & $\Omega$    &  i         & $1/a_{\rm osc}$ \\
					& $[$yyyymmdd$]$&$[$yyyymmdd.dddddd$]$& $[$AU$]$     &            &  $[$\degr$]$  & $[$\degr$]$ & $[$\degr$]$& $[10^{-6}$AU$^{-1}]$   \\
					$[1]$      & $[2]$       & $[3]$                 & $[4]$        & $[5]$      & $[6]$         & $[7]$       & $[8]$      & $[9]$          \\
					\hline
					\endfirsthead
					\caption{continued.}\\
					\hline\hline
					Comet      & Epoch         & T                   & q            &  e         & $\omega$      & $\Omega$    &  i         & $1/a_{\rm osc}$ \\
					& $[$yyyymmdd$]$&$[$yyyymmdd.dddddd$]$& $[$AU$]$     &            &  $[$\degr$]$  & $[$\degr$]$ & $[$\degr$]$& $[10^{-6}$AU$^{-1}]$   \\
					$[1]$      & $[2]$       & $[3]$                 & $[4]$        & $[5]$      & $[6]$         & $[7]$       & $[8]$      & $[9]$          \\
					\hline
					\endhead
					\hline 
					\endfoot
					%
					\\
					\multicolumn{9}{c}{\bf New orbital solutions for Oort spike comets based on (the updated data set in comparison to Sample B} \\ 
					\multicolumn{9}{c}{\bf of \citet{krolikowska:2014}} \\  \\
					\multicolumn{9}{c}{Orbits based on the entire data sets} \\ \\
					C/2000 SV$_{74}$ &  20020506 &  20020430.47430545 &    3.54159154 &    1.00479869 &   76.226371 &   24.185273 &   75.237137 & -1354.952 \\
					&           &     $\pm$0.00010264 &$\pm$.00000164 &$\pm$.00000203 &$\pm$.000018 &$\pm$.000005 &$\pm$.000008 &$\pm$     0.572 \\
\bottomrule
				\end{longtable}
		}}

		\newpage
		
		\section{Original barycentric orbital elements}
		
		{\setlength\LTcapwidth{1.00\textwidth} 
			\setlength\LTleft{0pt}               
			\setlength\LTright{0pt}              
			\setlength{\tabcolsep}{2.5pt}{
				\begin{longtable}{ccrrrrrrr}
					\caption{\label{tab:orbit_original} Orbital elements of original barycentric orbits, i.e. before entering the planetary zone for all comets described in Table~\ref{tab:observational_material}.
						The successive columns contain:
						$[1]$ -- Comet designation, $[2]$ -- Epoch, i.e. osculation date, $[3]$ -- perihelion time [TT], $[4]$ -- perihelion distance, $[5]$ -- eccentricity, $[6]$ -- argument of perihelion, $[7]$ -- longitude of the ascending node, $[8]$ -- inclination, $[9]$ -- reciprocal original semi-major axis in units of $10^{-6}$\,AU$^{-1}$. 
						The angular elements refer to the equator and equinox of J2000.} \\
					\hline\hline
					Comet      & Epoch         & T                   & q            &  e         & $\omega$      & $\Omega$    &  i         & $1/a_{\rm ori}$ \\
					& $[$yyyymmdd$]$&$[$yyyymmdd.dddddd$]$& $[$AU$]$     &            &  $[$\degr$]$  & $[$\degr$]$ & $[$\degr$]$& $[10^{-6}$AU$^{-1}]$   \\
					$[1]$      & $[2]$       & $[3]$                 & $[4]$        & $[5]$      & $[6]$         & $[7]$       & $[8]$      & $[9]$          \\
					\hline
					\endfirsthead
					\caption{continued.}\\
					\hline\hline
					Comet      & Epoch         & T                   & q            &  e         & $\omega$      & $\Omega$    &  i         & $1/a_{\rm ori}$ \\
					& $[$yyyymmdd$]$&$[$yyyymmdd.dddddd$]$& $[$AU$]$     &            &  $[$\degr$]$  & $[$\degr$]$ & $[$\degr$]$& $[10^{-6}$AU$^{-1}]$   \\
					$[1]$      & $[2]$       & $[3]$                 & $[4]$        & $[5]$      & $[6]$         & $[7]$       & $[8]$      & $[9]$          \\
					\hline
					\endhead
					\hline 
					\endfoot
					%
					\\
					\multicolumn{9}{c}{\bf New orbital solutions for Oort spike comets based on the updated data set in comparison to Sample B} \\ 
					\multicolumn{9}{c}{\bf of \citet{krolikowska:2014}} \\  \\
					\multicolumn{9}{c}{Orbits based on the entire data sets} \\ \\
					C/2000 SV$_{74}$ &  16980827 &  20020430.08118337 &    3.52665909 &    0.99967965 &   76.492790 &   24.152992 &   75.360945 &    90.835 \\
					&           &     $\pm$0.00031199 &$\pm$.00000215 &$\pm$.00000272 &$\pm$.000065 &$\pm$.000010 &$\pm$.000012 &$\pm$     0.770 \\
\bottomrule
				\end{longtable}
		}}
		
		
		\newpage
		
		\section{Future barycentric orbital elements}
		
		{\setlength\LTcapwidth{1.00\textwidth} 
			\setlength\LTleft{0pt}               
			\setlength\LTright{0pt}              
			{\setlength{\tabcolsep}{2.5pt}{
					\begin{longtable}{ccrrrrrrr}
						\caption{\label{tab:orbit_future} Orbital elements of future barycentric orbits, i.e. after suffering the planetary perturbations for all comets described in Table~\ref{tab:observational_material}.
							The successive columns contain:
							$[1]$ -- Comet designation, $[2]$ -- Epoch, i.e. osculation date, $[3]$ -- perihelion time [TT], $[4]$ -- perihelion distance, $[5]$ -- eccentricity, $[6]$ -- argument of perihelion, $[7]$ -- longitude of the ascending node, $[8]$ -- inclination, $[9]$ -- reciprocal original semi-major axis in units of $10^{-6}$\,AU$^{-1}$. 
							The angular elements refer to the equator and equinox of J2000.} \\
						\hline\hline
						Comet      & Epoch         & T                   & q            &  e         & $\omega$      & $\Omega$    &  i         & $1/a_{\rm fut}$ \\
						& $[$yyyymmdd$]$&$[$yyyymmdd.dddddd$]$& $[$AU$]$     &            &  $[$\degr$]$  & $[$\degr$]$ & $[$\degr$]$& $[10^{-6}$AU$^{-1}]$   \\
						$[1]$      & $[2]$       & $[3]$                 & $[4]$        & $[5]$      & $[6]$         & $[7]$       & $[8]$      & $[9]$          \\
						\hline
						\endfirsthead
						\caption{continued.}\\
						\hline\hline
						Comet      & Epoch         & T                   & q            &  e         & $\omega$      & $\Omega$    &  i         & $1/a_{\rm fut}$ \\
						& $[$yyyymmdd$]$&$[$yyyymmdd.dddddd$]$& $[$AU$]$     &            &  $[$\degr$]$  & $[$\degr$]$ & $[$\degr$]$& $[10^{-6}$AU$^{-1}]$   \\
						$[1]$      & $[2]$       & $[3]$                 & $[4]$        & $[5]$      & $[6]$         & $[7]$       & $[8]$      & $[9]$          \\
						\hline
						\endhead
						\hline 
						\endfoot
						%
						%
						\\
						\multicolumn{9}{c}{\bf New orbital solutions for Oort spike comets based on the updated data set in comparison to Sample B} \\ 
						\multicolumn{9}{c}{\bf of \citet{krolikowska:2014}} \\  
						\multicolumn{9}{c}{Orbits based on the entire data sets} \\ \\
						C/2000 SV$_{74}$ &  23040413 &  20020429.65912770 &    3.53676654 &    1.00018991 &   76.072968 &   24.187876 &   75.352548 &   -53.695 \\
						&           &     $\pm$0.00029332 &$\pm$.00000209 &$\pm$.00000213 &$\pm$.000061 &$\pm$.000010 &$\pm$.000008 &$\pm$     0.602 \\
						C/2003 WT$_{42}$ &  23140220 &  20060410.40770193 &    5.18237935 &    0.99895846 &   92.354993 &   48.405146 &   31.423892 &   200.977 \\
\bottomrule
					\end{longtable}
			}}
		}
		
		
		\newpage
		
		\begin{landscape}
			
			\section{Past orbital elements}

			{\setlength\LTcapwidth{1.00\textwidth} 

				\begin{longtable}{lcccccccl}
					\caption{\label{tab:previous-returning}
						Previous orbits of 66 comets from the studied sample which have a swarm of VCs fully returning in 
						the past. In columns {[}2{]}--{[}5{]} presented are inverse of the semimajor axis ($1/a_{{\rm prev}}$), 
						perihelion and aphelion distance ($q_{{\rm prev}}$, $Q_{{\rm prev}}$), and the epoch ($Time_{{\rm prev}}$) of 
						the previous perihelion passage. These elements are described either by a mean value (with the $1\sigma$ 
						uncertainty) for the normal distributions or by three deciles at 10, 50 (median) and 90 per cent. Columns 
						{[}6{]}--{[}8{]} inform how many VCs in each swarm have $q_{{\rm prev}}$ below 10\,au (column {[}6{]}), in 
						the range of 10--20\,au (column {[}7{]}), and larger than 20\,au (column {[}12{]}). In the last column the dynamical status of the  comet is given, see text for a detailed explanation.}
					\\
					\hline 
					\multirow{3}{*}{Comet name } & \multirow{3}{*}{$1/a_{{\rm prev}}$ $[10^{-6}$AU$^{-1}]$ } & \multirow{3}{*}{$q_{{\rm prev}}$ {[}au{]} } & \multirow{3}{*}{$Q_{{\rm prev}}$ {[}$10^{3}$au{]} } & \multirow{3}{*}{$Time_{{\rm prev}}$ {[}$10^{6}$yr{]} } & \multicolumn{3}{l}{Percentage of VCs } & \multirow{3}{*}{Status }\\
					&  &  &  &  &  \multicolumn{3}{l}{with $q_{{\rm prev}}$ {[}au{]}:}  & \\
					&  &  &  &  & \textless{}10  & 10--20  & \textgreater{}20  & \\
					$[1]$  & $[2]$  & $[3]$  & $[4]$  & $[5]$  & $[6]$  & $[7]$  & $[8]$  & $[9]$ \\
					\hline 
					&  &  &  &  &  &  &  & \\
					\endfirsthead
					\caption{continued.}\\
					\\
					\hline 
					\multirow{3}{*}{Comet name } & \multirow{3}{*}{$1/a_{{\rm prev}}$ $[10^{-6}$AU$^{-1}]$ } & \multirow{3}{*}{$q_{{\rm prev}}$ {[}au{]} } & \multirow{3}{*}{$Q_{{\rm prev}}$ {[}$10^{3}$au{]} } & \multirow{3}{*}{$Time_{{\rm prev}}$ {[}$10^{6}$yr{]} } & \multicolumn{3}{l}{Percentage of VCs } & \multirow{3}{*}{Status }\\
					&  &  &  &  &  \multicolumn{3}{l}{with $q_{{\rm prev}}$ {[}au{]}:}  & \\
					&  &  &  &  & \textless{}10  & 10--20  & \textgreater{}20  & \\
					$[1]$  & $[2]$  & $[3]$  & $[4]$  & $[5]$  & $[6]$  & $[7]$  & $[8]$  & $[9]$ \\
					\hline
					&  &  &  &  &  &  &  & \\
					\endhead
					C/2001 B2          & $199.29\pm1.10$             & $5.3290\pm0.0004$          & $10.03\pm0.06$        & $0.3485\pm0.0029$     & 100  &   0 &   0 & DO+\\
					C/2006 X1          & 131.84 -- 185.72 -- 237.73  & 6.13 -- 6.135 -- 6.14      & 8.41 -- 10.8 -- 15.2  & 0.27 -- 0.39 -- 0.65  & 100  &   0 &   0 & DO+\\
					C/2003 O1          & $180.99\pm0.81$             & $6.8856\pm0.0006$          & $11.04\pm0.05$        & $0.4037\pm0.0028$     & 100  &   0 &   0 & DO+\\
					C/2010 L3          & $172.76\pm2.12$             & $9.909\pm0.001$            & $11.57\pm0.14$        & $0.433\pm0.008$       & 100  &   0 &   0 & DO+\\
					C/1991 C3          & $172.65\pm22.27$            & 4.79 -- 4.80 -- 4.83       & 10.0 -- 11.6 -- 13.9  & 0.35 -- 0.43 -- 0.57  & 100  &   0 &   0 & DO+\\
					C/2007 U1          & $165.93\pm1.17$             & $3.3681\pm0.0008$          & $12.05\pm0.09$        & $0.461\pm0.005$       & 100  &   0 &   0 & DO+\\
					C/2009 UG$_{89}$   & $163.61\pm0.48$             & $3.9854\pm0.0004$          & $12.22\pm0.04$        & $0.471\pm0.002$       & 100  &   0 &   0 & DO+\\
					C/2008 P1          & $149.98\pm0.55$             & $3.8297\pm0.0009$          & $13.33\pm0.05$        & $0.537\pm0.003$       & 100  &   0 &   0 & DO+\\
\hline 
				\end{longtable}
				{\setlength{\tabcolsep}{4.0pt}{
						\begin{longtable}{lrrrcccccccl}
							\caption{\label{tab:previous-mixed}Previous orbits of 34 comets with their past swarms of VCs mixed or fully escaping. The 
								table is arranged in a similar manner as Table~\ref{tab:previous-returning} but after the first column with comet name we added three new columns containing numbers of returning {[}R{]} and escaping {[}E{]} VCs, and additionally the number of hyperbolas {[}H{]} among escaping VCs. The term 'synchronous' placed in column {[}8{]} means that the whole swarm of VCs has been stopped  at the same time. {[}R{]} added in columns {[}5{]} -- {[}8{]} means, that the given result is obtained from the returning part of the swarm of VCs. An additional horizontal line separates  the first 11~comets with more than 95~per cent of VCs returning and included in Fig.13.}
							\\
							\hline 
							\multirow{3}{*}{Comet name } & & & & \multirow{3}{*}{$1/a_{{\rm prev}}$ $[10^{-6}$AU$^{-1}]$ } & \multirow{3}{*}{$q_{{\rm prev}}$ {[}au{]} } & \multirow{3}{*}{$Q_{{\rm prev}}$ {[}$10^{3}$au{]} } & \multirow{3}{*}{$Time_{{\rm prev}}$ {[}$10^{6}$yr{]} } & \multicolumn{3}{l}{Percentage of VCs} & \multirow{3}{*}{Status }\\
							&        \multicolumn{3}{c}{Number of VCs}       &  &  &  &  &  \multicolumn{3}{l}{with $q_{{\rm prev}}$ {[}au{]}:} & \\
							& {[}R{]}  & {[}E{]}  & {[}H{]}  &  &  &  &  & \textless{}10  & 10--20  & \textgreater{}20  & \\
							$[1]$  & $[2]$  & $[3]$  & $[4]$  & $[5]$  & $[6]$  & $[7]$  & $[8]$  & $[9]$  & $[10]$  & $[11]$  & $[12]$ \\
							\hline 
							\endfirsthead
							\caption{continued.}\\
							\\
							\hline 
							\multirow{3}{*}{Comet name } & & & & \multirow{3}{*}{$1/a_{{\rm prev}}$ $[10^{-6}$AU$^{-1}]$ } & \multirow{3}{*}{$q_{{\rm prev}}$ {[}au{]} } & \multirow{3}{*}{$Q_{{\rm prev}}$ {[}$10^{3}$au{]} } & \multirow{3}{*}{$Time_{{\rm prev}}$ {[}$10^{6}$yr{]} } & \multicolumn{3}{l}{Percentage of VCs} & \multirow{3}{*}{Status }\\
							&        \multicolumn{3}{c}{Number of VCs}       &  &  &  &  &  \multicolumn{3}{l}{with $q_{{\rm prev}}$ {[}au{]}:} & \\
							& {[}R{]}  & {[}E{]}  & {[}H{]}  &  &  &  &  & \textless{}10  & 10--20  & \textgreater{}20  & \\
							$[1]$  & $[2]$  & $[3]$  & $[4]$  & $[5]$  & $[6]$  & $[7]$  & $[8]$  & $[9]$  & $[10]$  & $[11]$  & $[12]$ \\
							\hline 
							&  &  &  &  &  &  &  &  &   &   &   \\
							\endhead
							\\
							C/1997 A1         & 5000{*} &       1 &       0 & $21.80\pm1.63${[}R{]}              & 55.7 -- 110 -- 232{[}R{]}         & 83.8 -- 91.6 -- 101{[}R{]}   & 8.53 -- 9.75 -- 11.4{[}R{]}  &   0  &   0 & 100 & DN+\\
							C/2013 L2         & 5000{*} &       1 &       0 & $90.79\pm18.92${[}R{]}             & 4.47 -- 4.75 -- 4.82{[}R{]}       & 17.4 -- 22.0 -- 30.1{[}R{]}  & 0.80 -- 1.14 -- 1.84{[}R{]}  & 100  &   0 &   0 & DO+\\
							C/2010 D3         & 4997{*} &       4 &       0 & $20.11\pm1.19${[}R{]}              & 11.4 -- 20.7 -- 36.8{[}R{]}       & 92.3 -- 99.5 -- 108{[}R{]}   & 9.81 -- 11.0 -- 12.3{[}R{]}  &   6  &  41 &  53 & DN \\
							C/2006 YC         & 4965{*} &      36 &       0 & 30.91 -- 45.49 -- 60.97{[}R{]}     & 6.46 -- 9.89 -- 34.2{[}R{]}       & 32.8 -- 44.0 -- 64.7{[}R{]}  & 2.09 -- 3.24 -- 5.75{[}R{]}  &  51  &  30 &  19 & DO \\
							C/1925 F1         & 4950{*} &      51 &       0 & 24.55 -- 35.04 -- 45.62{[}R{]}     & 0.78 -- 2.00 -- 58.2{[}R{]}       & 43.8 -- 57.1 -- 81.5{[}R{]}  & 3.22 -- 4.75 -- 7.94{[}R{]}  &  74  &   7 &  19 & DO \\
							C/1960 M1         & 4943{*} &      58 &       1 & $43.01\pm10.53${[}R{]}             & 5.59 -- 8.19 -- 24.8{[}R{]}       & 35.3 -- 46.8 -- 68.2{[}R{]}  & 2.33 -- 3.56 -- 6.29{[}R{]}  &  64  &  23 &  13 & DO \\
							C/2002 A3         & 4940{*} &      61 &       0 & $20.88\pm1.75${[}R{]}              & 40.8 -- 75.8 -- 153{[}R{]}        & 86.2 -- 95.7 -- 107{[}R{]}   & 9.01 -- 10.6 -- 12.7{[}R{]}  &   0  &   0 & 100 & DN+\\
\hline 
						\end{longtable}
						
						{\setlength\LTcapwidth{1.00\textwidth} 

							\begin{longtable}{lrrrcccccccl}
								\caption{\label{tab:all-previous} All parameters of the previous orbits for all 100~LPCs arranged chronologically. The structure of the table is identical to that of Table~\ref{tab:previous-mixed}.} 
								\\
								\hline 
								& & & & \multirow{3}{*}{$1/a_{{\rm prev}}$ $[10^{-6}$AU$^{-1}]$ } & \multirow{3}{*}{$q_{{\rm prev}}$ {[}au{]} } & \multirow{3}{*}{$Q_{{\rm prev}}$ {[}$10^{3}$au{]} } & \multirow{3}{*}{$Time_{{\rm prev}}$ {[}$10^{6}$yr{]} } & \multicolumn{3}{l}{Percentage of VCs} & \multirow{3}{*}{Status }\\
								Comet &        \multicolumn{3}{c}{Number of VCs}       &  &  &  &  &  \multicolumn{3}{l}{with $q_{{\rm prev}}$ {[}au{]}:} & \\
								name & {[}R{]}  & {[}E{]}  & {[}H{]}  &  &  &  &  & \textless{}10  & 10--20  & \textgreater{}20  & \\
								$[1]$  & $[2]$  & $[3]$  & $[4]$  & $[5]$  & $[6]$  & $[7]$  & $[8]$  & $[9]$  & $[10]$  & $[11]$  & $[12]$ \\
								\hline 
								\endfirsthead
								\caption{continued.}\\
								\\
								\hline 
								& & & & \multirow{3}{*}{$1/a_{{\rm prev}}$ $[10^{-6}$AU$^{-1}]$ } & \multirow{3}{*}{$q_{{\rm prev}}$ {[}au{]} } & \multirow{3}{*}{$Q_{{\rm prev}}$ {[}$10^{3}$au{]} } & \multirow{3}{*}{$Time_{{\rm prev}}$ {[}$10^{6}$yr{]} } & \multicolumn{3}{l}{Percentage of VCs} & \multirow{3}{*}{Status }\\
								Comet &        \multicolumn{3}{c}{Number of VCs}       &  &  &  &  &  \multicolumn{3}{l}{with $q_{{\rm prev}}$ {[}au{]}:} & \\
								name & {[}R{]}  & {[}E{]}  & {[}H{]}  &  &  &  &  & \textless{}10  & 10--20  & \textgreater{}20  & \\
								$[1]$  & $[2]$  & $[3]$  & $[4]$  & $[5]$  & $[6]$  & $[7]$  & $[8]$  & $[9]$  & $[10]$  & $[11]$  & $[12]$ \\
								\hline 
								&  &  &  &  &  &  &  &  &   &   &   \\
								\endhead
								\\
								C/1914 M1         & 3901{*} &    1100 &     377 & 22.88 -- 43.75 -- 71.32{[}R{]}     & 3.83 -- 4.54 -- 13.1{[}R{]}       & 28.0 -- 45.7 -- 87.4{[}R{]}  & 1.65 -- 3.45 -- 9.23{[}R{]}  &  83  &  14 &   3 & DO \\
								C/1925 F1         & 4950{*} &      51 &       0 & 24.55 -- 35.04 -- 45.62{[}R{]}     & 0.78 -- 2.00 -- 58.2{[}R{]}       & 43.8 -- 57.1 -- 81.5{[}R{]}  & 3.22 -- 4.75 -- 7.94{[}R{]}  &  74  &   7 &  19 & DO \\
								C/1935 Q1         &    1306 & 3695{*} &      24 & 13.30 $\pm$ 5.17            & 23.0 -- 105 -- 248{[}R{]}  & 99.9 -- 150 -- 294    & 2.90 (synchronous)    &  --  &  -- &  -- & DN+\\
								C/1942 C2         &       1 &    5000 & 4933{*} & -29.13 $\pm$ 13.45          & 0.46 -- 0.54 -- 0.76       & --                    & 1.66 (synchronous)    &  --  &  -- &  -- & DN+\\
								C/1948 T1         & 4815{*} &     186 &       0 & 23.01 -- 34.76 -- 47.58     & 5.50 -- 12.9 -- 79.6       & 42.0 -- 57.5 -- 86.9  & 3.04 -- 4.86 -- 8.97  &  39  &  26 &  35 & DU \\
								C/1954 O2         & 5001{*} &       0 &       0 & $47.06\pm3.94$              & 1.62 -- 2.06 -- 2.47       & 38.4 -- 42.5 -- 47.6  & 2.65 -- 3.09 -- 3.66  & 100  &   0 &   0 & DO+\\
								C/1954 Y1         & 5001{*} &       0 &       0 & $38.19\pm2.66$              & 10.3 -- 13.1 -- 18.1       & 48.2 -- 52.4 -- 57.5  & 3.72 -- 4.21 -- 4.84  &   8  &  86 &   6 & DU \\
\hline 
							\end{longtable}
						}
					}

					\vfill\eject
					\newpage

					\section{Next orbital elements}
					
					{\setlength\LTcapwidth{1.00\textwidth} 
						\begin{longtable}{lccccccccl}
							\caption{\label{tab:next-returning}Orbital elements recorded at the next perihelion passage for 49 comets 
								that have a swarm of VCs fully returning. The arrangement of this table is identical to Table 
								\ref{tab:previous-returning} except the last column. Here we simply copied  the respective dynamical status from Table~\ref{tab:all-previous}.}
							\\
							\hline 
							\multirow{3}{*}{Comet name } & \multirow{3}{*}{$1/a_{{\rm next}}$ $[10^{-6}$AU$^{-1}]$ } & \multirow{3}{*}{$q_{{\rm next}}$ {[}au{]} } & \multirow{3}{*}{$Q_{{\rm next}}$ {[}$10^{3}$au{]} } & \multirow{3}{*}{$Time_{{\rm next}}$ {[}$10^{6}$yr{]} } & \multicolumn{3}{l}{Percentage of VCs } & \multirow{3}{*}{Status }\\
							&  &  &  &  &  \multicolumn{3}{l}{with $q_{{\rm next}}$ {[}au{]}:}  & \\
							&  &  &  &  & \textless{}10  & 10--20  & \textgreater{}20  & \\
							$[1]$  & $[2]$  & $[3]$  & $[4]$  & $[5]$  & $[6]$  & $[7]$  & $[8]$  & $[9]$ \\
							\hline 
							\endfirsthead
							\caption{continued.}\\
							\\
							\hline 
							\multirow{3}{*}{Comet name } & \multirow{3}{*}{$1/a_{{\rm next}}$ $[10^{-6}$AU$^{-1}]$ } & \multirow{3}{*}{$q_{{\rm next}}$ {[}au{]} } & \multirow{3}{*}{$Q_{{\rm next}}$ {[}$10^{3}$au{]} } & \multirow{3}{*}{$Time_{{\rm next}}$ {[}$10^{6}$yr{]} } & \multicolumn{3}{l}{Percentage of VCs } & \multirow{3}{*}{Status }\\
							&  &  &  &  &  \multicolumn{3}{l}{with $q_{{\rm next}}$ {[}au{]}:}  & \\
							&  &  &  &  & \textless{}10  & 10--20  & \textgreater{}20  & \\
							$[1]$  & $[2]$  & $[3]$  & $[4]$  & $[5]$  & $[6]$  & $[7]$  & $[8]$  & $[9]$ \\
							\hline 
							&  &  &  &  &  &  &  &   \\
							\endhead
							\\
							C/2002 A3          & 6173.58 $\pm$ 1.93           & 5.147430 $\pm$ 0.000008     & 0.3188 $\pm$ 0.0001   & 0.008775 $\pm$ 0.000001  & 100 &   0 &   0 & DN+ \\
							C/1991 C3          & 1418.66 $\pm$ 22.32          & 4.77130 $\pm$ 0.00005       & 1.41 $\pm$ 0.02       & 0.0254 $\pm$ 0.0004      & 100 &   0 &   0 & DO+ \\
							C/2007 D1          & 738.36 $\pm$ 0.95            & 8.78764 $\pm$ 0.00001       & 2.700 $\pm$ 0.004     & 0.0565 $\pm$ 0.0001      & 100 &   0 &   0 & DU  \\
							C/1993 K1          & 594.76 $\pm$ 7.66            & 4.84531 $\pm$ 0.00001       & 3.30 -- 3.36 -- 3.41  & 0.074 -- 0.076 -- 0.077  & 100 &   0 &   0 & DU  \\
							C/2000 CT$_{54}$   & 585.35 $\pm$ 2.53            & 3.152563 $\pm$ 0.000007     & 3.41 $\pm$ 0.02       & 0.0773 $\pm$ 0.0005      & 100 &   0 &   0 & DO+ \\
							C/1974 V1          &  574.60 $\pm$ 12.12          & 6.00978 $\pm$ 0.00007       & 3.48 $\pm$ 0.07       & 0.079 $\pm$ 0.002        & 100 &   0 &   0 & DN  \\
							C/1999 U1          & 537.63 $\pm$ 3.00            & 4.12809 $\pm$ 0.00002       & 3.72 $\pm$ 0.02       & 0.0869 $\pm$ 0.0007      & 100 &   0 &   0 & DO  \\
\hline 
						\end{longtable}
						
						\begin{longtable}{lcccccccl}
							\caption{\label{tab:next-mixed} Next orbit elements for 15~comets with their VCs swarms mixed or fully escaping.
								The arrangement of this table is identical to that of Table~\ref{tab:previous-mixed} but columns containing the distribution of the perihelion distance are not included here.}
							\\
							\hline 
							\multirow{2}{*}{Comet name } & \multicolumn{3}{c}{Number of VCs} & \multirow{2}{*}{$1/a_{{\rm next}}$ 
								$[10^{-6}$AU$^{-1}]$ } & \multirow{2}{*}{$q_{{\rm next}}$ {[}au{]} } & \multirow{2}{*}{$Q_{{\rm next}}$ 
								{[}$10^{3}$au{]} } & \multirow{2}{*}{$Time_{{\rm next}}$ {[}$10^{6}$yr{]} } & \multirow{2}{*}{Status }\\
							& {[}R{]}  & {[}E{]}  & {[}H{]}  &  &  &  &  & \\
							$[1]$  & $[2]$  & $[3]$  & $[4]$  & $[5]$  & $[6]$  & $[7]$  & $[8]$  & $[9]$ \\
							\hline 
							\endfirsthead
							\caption{continued.}\\
							\\
							\hline 
							\multirow{2}{*}{Comet name } & \multicolumn{3}{c}{Number of VCs} & \multirow{2}{*}{$1/a_{{\rm next}}$ 
								$[10^{-6}$AU$^{-1}]$ } & \multirow{2}{*}{$q_{{\rm next}}$ {[}au{]} } & \multirow{2}{*}{$Q_{{\rm next}}$ 
								{[}$10^{3}$au{]} } & \multirow{2}{*}{$Time_{{\rm next}}$ {[}$10^{6}$yr{]} } & \multirow{2}{*}{Status }\\
							& {[}R{]}  & {[}E{]}  & {[}H{]}  &  &  &  &  & \\
							$[1]$  & $[2]$  & $[3]$  & $[4]$  & $[5]$  & $[6]$  & $[7]$  & $[8]$  & $[9]$\\ 
							\hline 
							&  &  &  &  &  &  &  \\
							\endhead
							\\ 
							C/2011 L6         & 4518{*} &     483 &      8  & 20.56 -- 30.80 -- 43.05{[}R{]}      & 3.70 -- 20.2 -- 239 {[}R{]} & 46.5 -- 65.0 -- 97.0{[}R{]}  & 3.54 -- 5.85 -- 10.7{[}R{]} & DO+\\
							C/1914 M1         &   1854  & 3147{*} &   1855  & 8.11 $\pm$ 25.03             & 0.16 -- 1.95 -- 19.2 {[}R{]}& 43.08 -- 101 -- 480   & 1.65 (synchronous)   & DO \\
							C/2006 X1         &   1825  & 3176{*} &   2379  & 2.14 $\pm$ 41.19             & 6.12 -- 10.1 -- 126 {[}R{]} & 29.1 -- 71.3 -- 379   & 1.37 (synchronous)   & DO+\\
							C/1977 D1         &   1644  &    3357 & 2718{*} & -5.19 $\pm$ 45.24            & 3.16 -- 6.00 -- 435 {[}R{]} & 28.6 -- 68.8 -- 367   & 1.18 (synchronous)   & DO \\
							C/1960 M1         &    370  & 4631{*} &   2365  & 0.74 $\pm$ 10.97             & 0.77 -- 1.06 -- 1.45        & 110 -- 260 -- 1388    & 2.06 (synchronous)   & DO \\
							C/2006 S2         &    311  &    4690 & 3626{*} & -10.81 $\pm$ 18.07           & 2.01 -- 2.21 -- 2.68        & 82.6 -- 223 -- 1338   & 1.64 (synchronous)   & DO+\\
							C/2002 R3         &    153  & 4848{*} &   1231  & 4.30 $\pm$ 6.25              & 17.6 -- 21.6 -- 26.4        & 152 -- 318 -- 1475    & 2.45 (synchronous)   & DO+\\
\hline 
						\end{longtable}
					}

					{\setlength\LTcapwidth{1.00\textwidth} 
						{\setlength{\tabcolsep}{4.0pt}{
								
								\begin{longtable}{lrrrcccccccl}
									\caption{\label{tab:all-next}{The next orbit descriptions for all 100 comets arranged chronologically for easier browsing.  The  construction of this table is identical to that of Table~\ref{tab:all-previous}.}}
									\\
									\hline 
									& & & & \multirow{3}{*}{$1/a_{{\rm next}}$ $[10^{-6}$AU$^{-1}]$ } & \multirow{3}{*}{$q_{{\rm next}}$ {[}au{]} } & \multirow{3}{*}{$Q_{{\rm next}}$ {[}$10^{3}$au{]} } & \multirow{3}{*}{$Time_{{\rm next}}$ {[}$10^{6}$yr{]} } & \multicolumn{3}{l}{Percentage of VCs} & \multirow{3}{*}{Status }\\
									Comet &        \multicolumn{3}{c}{Number of VCs}       &  &  &  &  &  \multicolumn{3}{l}{with $q_{{\rm next}}$ {[}au{]}:} & \\
									name  & {[}R{]}  & {[}E{]}  & {[}H{]}  &  &  &  &  & \textless{}10  & 10--20  & \textgreater{}20  & \\
									$[1]$  & $[2]$  & $[3]$  & $[4]$  & $[5]$  & $[6]$  & $[7]$  & $[8]$  & $[9]$  & $[10]$  & $[11]$  & $[12]$ \\
									\hline 
									\endfirsthead
									\caption{continued.}\\
									\\
									\hline 
									& & & & \multirow{3}{*}{$1/a_{{\rm next}}$ $[10^{-6}$AU$^{-1}]$ } & \multirow{3}{*}{$q_{{\rm next}}$ {[}au{]} } & \multirow{3}{*}{$Q_{{\rm next}}$ {[}$10^{3}$au{]} } & \multirow{3}{*}{$Time_{{\rm next}}$ {[}$10^{6}$yr{]} } & \multicolumn{3}{l}{Percentage of VCs} & \multirow{3}{*}{Status }\\
									Comet &        \multicolumn{3}{c}{Number of VCs}       &  &  &  &  &  \multicolumn{3}{l}{with $q_{{\rm next}}$ {[}au{]}:} & \\
									name & {[}R{]}  & {[}E{]}  & {[}H{]}  &  &  &  &  & \textless{}10  & 10--20  & \textgreater{}20  & \\
									$[1]$  & $[2]$  & $[3]$  & $[4]$  & $[5]$  & $[6]$  & $[7]$  & $[8]$  & $[9]$  & $[10]$  & $[11]$  & $[12]$ \\
									\hline 
									&  &  &  &  &  &  &  &  &   &   &   \\
									\endhead
									\\
									C/1914 M1         &   1854  & 3147{*} &   1855  & 8.11 $\pm$ 25.03             & 0.16 -- 1.95 -- 19.2{[}R{]} & 43.08 -- 101 -- 480   & 1.65 (synchronous)       &  -- &  -- &  -- & DO \\
									C/1925 F1         & 5001{*} &       0 &       0 &  127.97 $\pm$ 8.33           & 4.28 -- 4.31 -- 4.35        & 14.4 -- 15.6 -- 17.1  & 0.618 -- 0.696 -- 0.795  & 100 &   0 &   0 & DO \\
									C/1935 Q1         &       0 &    5001 & 5001{*} &   -285.43$\pm$ 5.18          &         4.54$\pm$ 0.01      &         --            & 1.01 (synchronous)       &  -- &  -- &  -- & DN+\\
									C/1942 C2         &       0 &    5001 & 5001{*} &   -278.14$\pm$13.46          &        10.29$\pm$ 0.28      &         --            & 0.97 (synchronous)       &  -- &  -- &  -- & DN+\\
									C/1948 T1         & 5001{*} &       0 &       0 &  225.20 $\pm$ 10.05          & 3.262 -- 3.264 -- 3.266     & 8.40 -- 8.88 -- 9.42  & 0.279 -- 0.302 -- 0.330  & 100 &   0 &   0 & DU \\
									C/1954 O2         & 5001{*} &       0 &       0 &  255.57 $\pm$ 3.94           & 3.8708 $\pm$ 0.0002         & 7.82 $\pm$ 0.12       & 0.251 $\pm$ 0.006        & 100 &   0 &   0 & DO+\\
									C/1954 Y1         &       0 &    5001 & 5001{*} &   -228.41$\pm$ 2.65          &         1.42$\pm$ 0.01      &         --            & 1.11 (synchronous)       &  -- &  -- &  -- & DU \\
\hline 
								\end{longtable}
								
							}
						}
				}}
			}
		\end{landscape}


\bsp	
\label{lastpage}
\end{document}